\documentclass[a4paper, 12pt]{article}

\usepackage[symbol]{footmisc}
\usepackage{algorithm}
\usepackage{algpseudocode}
\usepackage{amsmath}

\usepackage{titlesec}
\usepackage{amssymb}
\usepackage{bm}
\usepackage{physics}
\usepackage{bigints}
\usepackage{bbm}
\usepackage{comment}

\usepackage{fullpage}% set margins

\newcommand{\lowerromannumeral}[1]{\romannumeral#1\relax} %print roman lower case numbers
\newcommand{\upperRomannumeral}[1]{\uppercase\expandafter{\romannumeral#1}}

\usepackage{graphicx} % enable graphics, images and resizebox
\usepackage{multicol} % enable multicolumn
\usepackage{lipsum}
\usepackage{natbib}
\usepackage{xcolor}

\usepackage{mathtools}

\usepackage{booktabs}   
\definecolor{whitesmoke}{rgb}{0.96, 0.96, 0.96}
\definecolor{shadecolor}{named}{whitesmoke}

\definecolor{powderblue(web)}{rgb}{0.69, 0.88, 0.9}
\definecolor{antiquewhite}{rgb}{0.98, 0.92, 0.84}
\definecolor{navajowhite2}{RGB}{ 238,207,161}
\usepackage{caption}
\usepackage{gensymb} % for \degree
\usepackage[export]{adjustbox}

\usepackage{hyperref}

\usepackage{color}
\usepackage{tikz}

\usetikzlibrary{shapes,decorations,arrows,calc,arrows.meta,fit,positioning}
\tikzset{
    -Latex,auto,node distance =1 cm and 1 cm,semithick,
    state/.style ={ellipse, draw, minimum width = 0.7 cm},
    point/.style = {circle, draw, inner sep=0.04cm,fill,node contents={}},
    bidirected/.style={Latex-Latex,dashed},
    el/.style = {inner sep=2pt, align=left, sloped}
}

\begin{document}

\begin{center}

{\Large \bfseries A Zero-Inflated Poisson Latent Position Cluster Model}

\vspace{5mm}

{\large Chaoyi Lu\footnote[1]{Address for correspondence: \texttt{chaoyi.lu.stat@gmail.com}}$^{\dagger\ddagger}$, Riccardo Rastelli$^{\dagger}$, Nial Friel$^{\dagger\ddagger}$} \\
$^{\dagger}$School of Mathematics and Statistics, University College Dublin, Dublin, Ireland\\
$^{\ddagger}$Insight Research Ireland Centre for Data Analytics, University College Dublin, Dublin, Ireland

\vspace{5 mm}

\end{center}

%----------------------------------------------------------------------------------------------------------------------------------------------------------------------------------------------------------------------------------------------------------------------------------------
%----------------------------------------------------------------------------------------------------------------------------------------------------------------------------------------------------------------------------------------------------------------------------------------
%----------------------------------------------------------------------------------------------------------------------------------------------------------------------------------------------------------------------------------------------------------------------------------------
%----------------------------------------------------------------------------------------------------------------------------------------------------------------------------------------------------------------------------------------------------------------------------------------

\begin{abstract}

\noindent The latent position network model (LPM) is a popular approach for the statistical analysis of network data. A central aspect of this model is that it assigns nodes to random positions in a latent space, such that the probability of an interaction between each pair of individuals or nodes is determined by their distance in this latent space. A key feature of this model is that it allows one to visualize nuanced structures via the latent space representation.
The LPM can be further extended to the Latent Position Cluster Model (LPCM), 
to accommodate the clustering of nodes by assuming that the latent positions
are distributed following a finite mixture distribution.
In this paper, we extend the LPCM to accommodate missing network data and apply this to non-negative discrete weighted social networks.
By treating missing data as ``unusual'' zero interactions, we propose a combination of the LPCM with the zero-inflated Poisson distribution. 
Statistical inference is based on a novel partially collapsed Markov chain Monte Carlo algorithm, where a Mixture-of-Finite-Mixtures (MFM) model is adopted to automatically determine the number of clusters and optimal group partitioning. 
Our algorithm features a truncated absorb-eject move, which is a novel adaptation of an idea commonly used in collapsed samplers, within the context of MFMs. 
Another aspect of our work is that we illustrate our results on 3-dimensional latent spaces, maintaining clear visualizations while achieving more flexibility than 2-dimensional models.
The performance of this approach is illustrated via three carefully designed simulation studies, as well as four different publicly available real networks, where some interesting new perspectives are uncovered.
\end{abstract}


%----------------------------------------------------------------------------------------------------------------------------------------------------------------------------------------------------------------------------------------------------------------------------------------
%----------------------------------------------------------------------------------------------------------------------------------------------------------------------------------------------------------------------------------------------------------------------------------------
%----------------------------------------------------------------------------------------------------------------------------------------------------------------------------------------------------------------------------------------------------------------------------------------
%----------------------------------------------------------------------------------------------------------------------------------------------------------------------------------------------------------------------------------------------------------------------------------------

\section{Introduction}

The Latent Position Model (LPM) was introduced for social network analysis by \citet{hoff2002latent}. An important feature of the LPM is that it facilitates network visualization via latent space modeling.
The LPM assumes that the data distribution depends on the stochastic latent positions of the nodes of the network. 
Usually these latent positions are assumed to lie in a Euclidean space. 
We refer the reader to \citet{LPNM_2023,rastelli2016properties,salter2012review} for detailed reviews of this model.
Clustering of nodes plays a central role in statistical network analysis: the LPM can be extended to this context following the pioneering work of \citet{handcock2007model}.
This leads us to the well-known Latent Position Cluster Model (LPCM) whereby the latent positions of the nodes are assumed to follow a mixture of multivariate normal distributions. 
With these assumptions, the network model can better support the presence of communities and assortative mixing.

A limitation of the original LPM and LPCM models is that they are only defined for the networks with binary interactions between the individuals. 
An interesting extension of these models to the weighted setting is provided by \citet{sewell2015latent,sewell2016latent}, where the authors extend the LPM to dynamic unipartite networks with weighted edges and propose a general link function for the expectation of the edge weights.
In this paper, we focus on non-dynamic unipartite networks with non-negative discrete weighted edges, corresponding to networks where the edge values are typically integer counts.

Missing data is a common issue in statistical data analysis which generally leads to an overabundance of zeros.
In this paper, we classify zero entries in a dataset either as ``true'' zeros or as ``unusual'' zeros.
A zero-inflated model proposed by \citet{lambert1992zero} assumes a probability model for the occurrence of unusual zeros, which can be paired with the Poisson distribution as a canonical choice for count weighted data.
In this situation, an unusual zero may correspond to an edge that although present, the corresponding edge weight is not recorded. 

Zero-inflation is well-explored in the area of regression models with many extensions, for example, \citet{hall2000zero}, \citet{ridout2001score}, \citet{ghosh2006bayesian}, \citet{lemonte2019zero} to cite a few.
By contrast, zero-inflated models are not widely applied in the statistical analysis of network data. 
Zero-inflation is mentioned in \citet{sewell2016latent}: while their dynamic latent space model may be extended to the zero-inflated case for sparse networks, their applications do not cover this extension.
Some advancements in this line of literature include \citet{marchello2024deep},
which incorporate the zero-inflated model within the Latent Block Model (LBM) \citep{govaert2003clustering} for dynamic bipartite networks, and \citet{lu2025zero}, which adopts the zero-inflation to the Stochastic Block Model (SBM) for unipartite networks.
The SBM is widely used clustering model for network analysis that shares some similarities with the LPM, since both models are based on a latent variable framework. Differently from the LPM, the SBM is capable of representing disassortative patterns, but, 
on the other hand, the LPM can provide clearer and more interpretable network visualizations.

In this paper, we propose to incorporate a Zero-Inflated Poisson (ZIP) distribution within the LPCM leading to the Zero-Inflated Poisson Latent Position Cluster Model (ZIP-LPCM), and, in doing so, we create a simultaneous framework that can be used to characterize zero inflation, clustering and the latent space visualization.
Similar model assumptions can be found in \citet{ma2024statistical} where a Latent Space Zero-Inflated Poisson (LS-ZIP) model is proposed.
However, differently from our model structure, their LS-ZIP embeds an inner-product latent space model \citep{hoff2003random,ma2020universal} and proposes two different sets of latent positions for the Poisson rate and for the probability of unusual zeros, respectively.
Furthermore, their model does not account for the clustering of nodes, which is a central feature of our proposed model.

We employ a Bayesian framework to infer the model parameters and the latent variables of our ZIP-LPCM. Our inferential framework takes inspiration from the literature on the LPCM, the Mixture of Finite Mixtures (MFM) model \citep{miller2018mixture, geng2019probabilistic}, and collapsed Gibbs sampling, and we combine some key ideas from the available literature to create our own original procedure.
As regards the LPCM, commonly used inference methods include a variational expectation maximization algorithm \citep{salter2013variational}, and Markov chain Monte Carlo methods \citep{handcock2007model}. 
A contribution close to our own is that of \citet{ryan2017bayesian} where the authors exploit conjugate priors to calculate a marginalized posterior in analytic form, and then target this distribution using a so-called ``collapsed'' sampler. 
Similar parameter-collapsing ideas are also employed in \citet{mcdaid2012clustering,wyse2012block}.
In this paper, we propose to follow a similar strategy as that presented in \citet{lu2025zero} for the inference task leveraging the partially collapsed Gibbs sampler introduced by \citet{van2008partially,park2009partially}.

As regards the choice of the number of clusters, we also make an original contribution by combining some approaches and ideas available from the literature.
We highlight \citet{nobile2007bayesian}, where the authors introduced an Absorb-Eject (AE) move to automatically choose the number of clusters. Here we propose a variant of this move to better match our framework.
Indeed, as a prior distribution for the clustering variables and number of clusters, we adopt a MFM model, along with the supervision idea introduced by \citet{legramanti2022extended}.
In this case, the AE move can further facilitate the estimation of clusters, but such a step can only be defined if the framework allows for the existence of empty clusters. Unfortunately, this is not the case for MFMs, and so the move and the model are incompatible.
To address this impasse, we propose a new Truncated Absorb-Eject (TAE) move which allows us to efficiently explore the sample space thus obtaining good estimates of the clustering variables and of the number of groups.

This paper is organized as follows. Section~\ref{The model} provides a detailed introduction of our proposed zero-inflated Poisson latent position cluster model.
Section~\ref{Bayesian inference} explains how we design the Bayesian inference process of our model where the incorporation of a mixture of finite mixtures model as well as its supervised version is included in Section~\ref{MFM_Sup}.
The idea of partially collapsing the model parameters in the posterior distribution, and the newly proposed truncated absorb-eject move, are introduced in Section~\ref{Collapsing_Inference} and Section~\ref{TAE}, respectively.
The detailed steps and designs of the partially collapsed Metropolis-within-Gibbs algorithm for the inference are illustrated in Section~\ref{PCG}.
In Section~\ref{SS}, we show the performance of our strategy via three carefully designed simulation studies within which different scenarios are proposed to tackle different real world situations.
Applications on four different real social networks with different network sizes are included in Section~\ref{RDA}.
Finally, Section~\ref{Conclusion} concludes this paper and provides a few possible future directions.

%----------------------------------------------------------------------------------------------------------------------------------------------------------------------------------------------------------------------------------------------------------------------------------------
%----------------------------------------------------------------------------------------------------------------------------------------------------------------------------------------------------------------------------------------------------------------------------------------
%----------------------------------------------------------------------------------------------------------------------------------------------------------------------------------------------------------------------------------------------------------------------------------------
%----------------------------------------------------------------------------------------------------------------------------------------------------------------------------------------------------------------------------------------------------------------------------------------

\section{Model}
\label{The model}

In this paper we focus on weighted networks with non-negative discrete edges, and denote $N$ as the total number of individuals in the network. This framework is particularly common in observed real datasets, most notably whereby edges indicate the intensity or frequency of interactions. Examples include some types of friendship and professional networks, email networks, phone call networks, proximity networks. In this paper we consider several real datasets, including a summit co-attendance network which records the number of times that two criminal suspects co-attended summits within a specified period. Since the edges indicate the number of co-attendances, these are observed as non-negative integers.
The model introduced in this section is designed for directed networks, however it is straightforward to apply it to the undirected case.
A network is usually observed by an $N \times N$ adjacency matrix denoted as $\boldsymbol{Y}$, where each element $y_{ij}$ is a non-negative integer indicating the interaction from node $i$ to node $j$, and reflecting the corresponding interaction strength.
An element $y_{ij}=0$ corresponds to a non-interaction or a zero interaction.
Self-loops are not allowed.

%----------------------------------------------------------------------------------------------------------------------------------------------------------------------------------------------------------------------------------------------------------------------------------------
%----------------------------------------------------------------------------------------------------------------------------------------------------------------------------------------------------------------------------------------------------------------------------------------

\subsection{Zero-inflated Poisson model}

In real networks which are observed in practice, there is often an overabundance of zeros.
It is possible that these zeros are due to the nature of the network's sparse architecture, but they may also be due to missing data or misreported data.
Since practitioners are typically not aware of the presence of unusual zeros, we aim to provide a model-based solution that helps detecting any unusual zeros in the observed data.

We consider the Zero-Inflated Poisson (ZIP) model \citep{lambert1992zero}, which is a commonly used framework to deal with an excessive number of zeros in $\boldsymbol{Y}$.
The ZIP model assumes that each observed interaction $y_{ij}$ follows
\begin{equation}
\label{ZI}
\text{P}(y_{ij}|\lambda_{ij},p_{ij}) = \begin{cases} 
p_{ij} + (1-p_{ij})f_{\text{Pois}}(0|\lambda_{ij}),&\text{if } y_{ij}=0;\\  
(1-p_{ij})f_{\text{Pois}}(y_{ij}|\lambda_{ij}),&\text{if } y_{ij}=1,2,\dots, % 
\end{cases}
\end{equation}
for $i,j=1,2,\dots,N;\; i\neq j$, where $f_{\text{Pois}}(\cdot|\lambda_{ij})$ is the probability mass function of the Poisson distribution with parameter $\lambda_{ij}$.
Here, the zeros assigned with probability mass $p_{ij} + (1-p_{ij})f_{\text{Pois}}(0|\lambda_{ij})$ can be classified into two types: ``structural'' zeros, which are observed with probability $p_{ij}$, and Poisson zeros, which naturally arise from the Poisson distribution, and are observed with probability $(1-p_{ij})f_{\text{Pois}}(0|\lambda_{ij})$, respectively.
If we treat the Poisson zeros as the conventional zeros, then we refer to the other observed zeros as the ``structural'' or ``unusual'' zeros, since these are not determined according to the natural distribution that we assume on the data.
We formalize the two possibilities by augmenting the model in Eq.~\eqref{ZI} with a new indicator variable,
$\nu_{ij}$, indicating whether the corresponding $y_{ij}$ is a structural zero or not, and thus the data distribution is determined separately for the two cases below:
\begin{equation}
\label{Aug_ZI}
y_{ij}|\lambda_{ij},\nu_{ij} \sim \begin{cases} 
\mathbbm{1}(y_{ij}=0),&\text{if } \nu_{ij}=1;\\  
\text{Pois}(\lambda_{ij}),&\text{if } \nu_{ij}=0, 
\end{cases}
\end{equation}
where, for every $i$ and $j$, the collection of $\nu_{ij}\sim \text{Bernoulli}(p_{ij})$ constitutes the $N \times N$ structural zero indicator matrix $\boldsymbol{\nu}$.
The function $\mathbbm{1}(y_{ij}=0)$ is an indicator function returning $1$ if $y_{ij}=0$ and returning $0$ otherwise.

As far as $\nu_{ij}=1$, the observed $y_{ij}$ is a structural zero with probability $1$.
Here, we interpret such a ``structural'' zero as an ``unusual'' zero or missing data that replaces a true interaction weight 
which follows the corresponding $\text{Pois}(\lambda_{ij})$ distribution.
A zero arising from $f_{\text{Pois}}(\cdot|\lambda_{ij})$ is thus treated as a ``true'' zero.
We denote the covert true interaction as $x_{ij}$: we assume that this value is not observed when $\nu_{ij}=1$, and that it follows the same Poisson distribution of $y_{ij}|\lambda_{ij},\nu_{ij}=0$. 
Thus, based on the augmented zero-inflated model in Eq.~\eqref{Aug_ZI} as well as the observed $\boldsymbol{Y}$, the augmented data $\boldsymbol{X}$, which is a $N \times N$ matrix with entries $\{x_{ij}\}$, is in the form
\begin{equation}
\label{Aug_X}
\begin{cases} 
x_{ij}\sim \text{Pois}(\lambda_{ij}),&\text{if } \nu_{ij}=1;\\
x_{ij}=y_{ij},&\text{if } \nu_{ij}=0.
\end{cases}
\end{equation}
Here, the case $x_{ij}=y_{ij}$ is equivalent to $x_{ij} \sim \mathbbm{1}(x_{ij}=y_{ij})$ when $\nu_{ij}=0$.
A similar data augmentation framework has appeared in other works, for example, 
\citet{tanner1987calculation,ghosh2006bayesian}.
The augmented $\boldsymbol{X}$ is known as the missing data imputed adjacency matrix.

%----------------------------------------------------------------------------------------------------------------------------------------------------------------------------------------------------------------------------------------------------------------------------------------
%----------------------------------------------------------------------------------------------------------------------------------------------------------------------------------------------------------------------------------------------------------------------------------------

\subsection{Zero-inflated Poisson latent position cluster model}

To characterize the Pois($\lambda_{ij}$) distribution under the $\nu_{ij}=0$ case of the augmented ZIP model in Eq.~\eqref{Aug_ZI}, we employ the Latent Position Cluster Model (LPCM) \citep{handcock2007model}, which is an extended version of the Latent Position Model (LPM) \citep{hoff2002latent}.
Each node $i:\; i=1,2,\dots,N$ in the network is assumed to have a latent position $\boldsymbol{u}_i\in \mathbbm{R}^d$, and we denote the collection of all latent positions as $\boldsymbol{U}:=\{\boldsymbol{u}_i\}$.
A generalization of the latent position model without considering covariates can be expressed in the form $g[\mathbbm{E}(y_{ij})]=h(\boldsymbol{u}_i,\boldsymbol{u}_j)$, where $g(\cdot)$ is some link function, and $h:\mathbbm{R}^d\times \mathbbm{R}^d\rightarrow \mathbbm{R}$ is a function of two nodes' latent positions.
Here, we make standard assumptions on the functions $g(\cdot)$ and $h(\cdot,\cdot)$, which link the Poisson rate $\lambda_{ij}$ in Eq.~\eqref{Aug_ZI} with the latent positions $\boldsymbol{U}$ as follows:
\begin{equation}
\label{LPM}
\text{log}(\lambda_{ij})=\beta-||\boldsymbol{u}_i-\boldsymbol{u}_j||.
\end{equation}
In the equation above, $||\cdot||$ is the Euclidean distance, while $\beta\in\mathbbm{R}$ can be interpreted as an intercept term where higher values bring larger interaction weights as well as lower chance of a true zero.
Note that the model characterization in Eq.\eqref{LPM} is symmetric for directed edges in directed networks, that is, $\log(\lambda_{ij}) = \log(\lambda_{ji})$.
This may be seen as a limitation of the framework, but it can be overcome when additional information, such as covariates, are available and thus are characterized in Eq.\eqref{LPM}. 
We note that also other variations of latent space models, such as the projection model discussed in \cite{hoff2002latent} or the random effects model of \cite{krivitsky2009representing}, can be formulated in ways that allow for non-symmetric patterns.
However, in this paper, we do not pursue this characteristic directly, but our model can certainly be adapted to deal with non-symmetric patterns.

The LPCM further assumes that each latent position $\boldsymbol{u}_i$ is drawn from a finite mixture of $\bar{K}$ multivariate normal distributions, each corresponding to a different group:
\begin{equation}
\label{LPCM}
\boldsymbol{u}_i\sim\sum_{k=1}^{\bar{K}}\pi_k\text{MVN}_d(\boldsymbol{\mu}_k,1/\tau_k\mathbbm{I}_d).
\end{equation}
Here, each $\pi_k$ is the probability that a node is clustered into group $k$, and $\sum_{k=1}^{\bar{K}}\pi_k=1$.
The combination of Eqs.~\eqref{Aug_ZI}, \eqref{LPM} and \eqref{LPCM} defines our Zero-Inflated Poisson Latent Position Cluster Model (ZIP-LPCM).\\

Letting $z_i \in \{1,2,\dots,\bar{K}\}$ denote the group membership of node $i$, the mixture in Eq.~\eqref{LPCM} can be augmented as
\begin{equation*}
\label{Aug_LPCM}
\boldsymbol{u}_i|z_i=k\sim\text{MVN}_d(\boldsymbol{\mu}_k,1/\tau_k\mathbbm{I}_d),
\end{equation*}
where a multinomial$(1,\boldsymbol{\Pi})$ distribution is assumed for each $\boldsymbol{z_i}$, and $\boldsymbol{\Pi}:=(\pi_1,\pi_2,\dots,\pi_{\bar{K}})$.
The notation $\boldsymbol{z_i}:=(z_{i1},z_{i2},\dots,z_{i\bar{K}})$ is equivalent to $z_i=k$ if we let $z_{ig}=\mathbbm{1}(g=k)$ 
for $g=1,2,\dots,\bar{K}$, and thus $f_{\text{multinomial}}(\boldsymbol{z_i}|1,\boldsymbol{\Pi}) = \pi_{z_i}$.
Note that $\bar{K}$ here denotes the total number of possible groups that the network is assumed to have, even though some groups may be empty. 
Further, we denote $\boldsymbol{z}:=(z_1,z_2,\dots,z_N)$ as a vector of group membership indicators for all $\{z_i\}$, while $\boldsymbol{\mu}:=\{\boldsymbol{\mu}_k:k=1,2,\dots,\bar{K}\}$, and $\boldsymbol{\tau}:=(\tau_1,\tau_2,\dots,\tau_{\bar{K}})$.

The indicator of unusual zero $\nu_{ij}$ in Eq.~\eqref{Aug_ZI} is proposed to instead follow a $\text{Bernoulli}(p_{z_iz_j})$, that is, a typical Stochastic Block Model (SBM) structure is further assumed for $\bm{\nu}$, where we replace the probability of unusual zero $p_{ij}$ with a cluster-dependent counterpart $p_{z_iz_j}$, as in a network block-structure.
One intuition behind this is that the probability of an unusual zero observed between two nodes is expected to vary depending on their respective groups.
Taking the summit co-attendance real network for criminal suspects as an example here, the occurrences of unusual zeros possibly correspond to different group-level secrecy strategies applied by the group leaders where some summits are intentionally hidden to the law enforcement investigations \citep{lu2025zero}, leading to different levels of unusual zero probability within and between different groups.

There are other possible ways to model the probability of unusual zeros. 
A homogeneous setup where the probability is the same for all pairs of nodes is also possible, albeit naive. 
This would correspond to a special case of our framework. 
Alternatively, one may characterize the probability of unusual zeros using the latent distances.
However, we eventually decided not to pursue this type of model formulation in this paper.
One could argue: should the unusual zero probability be larger or smaller when the nodes are close to each other? 
Intuitively one could choose that a large distance would make it more likely that there is an unusual zero. 
However, it turns out that the Poisson distribution would also be more likely to give a zero in that case. 
So, the logic behind this assumption does not seem particularly convincing. 
On the other hand, a SBM structure can give disassortative patterns which can be more flexible for this particular task.
In any case, we emphasize again that there is no one solution to this, and alternative approaches to ours may be perfectly valid.
We denote $\boldsymbol{P}$ as a $\bar{K} \times \bar{K}$ matrix with each entry $p_{gh}$ denoting the probability of unusual zeros for the interactions from group $g$ to group $h$, where $g,h=1,2,\dots,\bar{K}$.

The following Directed Acyclic Graph (DAG) provides a clear visualization of the relationships between the observed adjacency matrix data $\bm{Y}$ and all the model latent variables and parameters as well as the augmented missing data imputed adjacency matrix $\bm{X}$ for our proposed ZIP-LPCM:
\begin{center}
\begin{tikzpicture}

\node[state] (yij) at (-2,0) {$y_{ij}$};
\node[state] (xij) at (-5,0) {$x_{ij}$};
\node[state,rectangle] (beta) at (-2,2) {$\beta$};
\node[state,rectangle] (uiuj) at (1,-2) {$\bm{u}_i,\bm{u}_j$};
\node[state,rectangle] (nuij) at (1,2) {$\nu_{ij}$};
\node[state,rectangle] (zizj) at (1,0) {$\bm{z}_i,\bm{z}_j$};
\node[state,rectangle] (mutau) at (4,-2) {$\bm{\mu},\bm{\tau}$};
\node[state,rectangle] (P) at (4,2) {$\bm{P}$};
\node[state,rectangle] (Pi) at (4,0) {$\bm{\Pi}$};

\path (yij) edge (xij);
\path (beta) edge (xij);
\path (beta) edge (yij);
\path (uiuj) edge (xij);
\path (uiuj) edge (yij);
\path (nuij) edge (yij);
\path (nuij) edge (xij);
\path (zizj) edge (nuij);
\path (zizj) edge (uiuj);
\path (mutau) edge (uiuj);
\path (P) edge (nuij);
\path (Pi) edge (zizj);

\end{tikzpicture}
\end{center}
The complete likelihood of the ZIP-LPCM is thus written as:
\begin{equation}
\label{CompleteLikelihood}
\begin{split}
& f(\boldsymbol{Y}, \boldsymbol{\nu},\boldsymbol{U},\boldsymbol{z}|\beta, \boldsymbol{P}, \boldsymbol{\mu}, \boldsymbol{\tau}, \boldsymbol{\Pi}) 
= f(\boldsymbol{Y}|\beta, \boldsymbol{U},\boldsymbol{\nu})f(\boldsymbol{\nu}|\boldsymbol{P},\boldsymbol{z})f(\boldsymbol{U}|\boldsymbol{\mu}, \boldsymbol{\tau}, \boldsymbol{z})f(\boldsymbol{z}|\boldsymbol{\Pi})\\
&= \prod^N_{\substack{i,j: i \neq j, \\ \nu_{ij} = 0}}f_{\text{Pois}}(y_{ij}|\text{exp}(\beta-||\boldsymbol{u}_i-\boldsymbol{u}_j||))
\prod^N_{i,j:i \neq j} f_{\text{Bern}}(\nu_{ij}|p_{z_iz_j}) \times\\
&\hspace{1em}\times \prod^N_{i =1} f_{\text{MVN}_d}(\boldsymbol{u}_i|\boldsymbol{\mu}_{z_i},1/\tau_{z_i}\mathbbm{I}_d)
\prod^N_{i =1} f_{\text{multinomial}}(\boldsymbol{z_i}|1,\boldsymbol{\Pi}),\\
\end{split}
\end{equation}
which can be calculated analytically and efficiently for all choices of parameter values.

%----------------------------------------------------------------------------------------------------------------------------------------------------------------------------------------------------------------------------------------------------------------------------------------
%----------------------------------------------------------------------------------------------------------------------------------------------------------------------------------------------------------------------------------------------------------------------------------------

\subsection{Mixture of finite mixtures and supervision}
\label{MFM_Sup}

Instead of the multinomial distribution, we further consider the Mixture-of-Finite-Mixtures (MFM) model for the clustering $\bm{z}$, because such a model is actually an extension of the Dirichlet-multinational conjugacy and allows the corresponding inference procedure to automatically choose the number of clusters.

A natural conjugate prior for the multinomial parameters is 
$$\bm{\Pi} \sim \text{Dirichlet}(\alpha,\dots,\alpha).$$
By further proposing a prior $\bar{K}\sim \pi_{\bar{K}}(\cdot)$, the MFM marginalizes both $\boldsymbol{\Pi}$ and $\bar{K}$, leading to the probability mass function of a MFM, which is defined for the unlabeled clustering and reads as follows:
\begin{equation}
\label{MFM_PMF_C}
\begin{split} 
f(\boldsymbol{C}(\boldsymbol{z}))= \sum_{k=1}^{\infty} \frac{k_{(K)}}{(k\alpha)^{(N)}}\pi_{\bar{K}}(k)\prod_{G\in \boldsymbol{C}(\boldsymbol{z})} \alpha^{(|G|)},
\end{split}
\end{equation}
where $\boldsymbol{C}(\boldsymbol{z}):=\{G_k: k=1,2,\dots,\bar{K};\;|G_k|>0\}$ is a set of non-empty unordered collections of nodes with each collection $G_k:=\{i:i=1,2,\dots,N;\;z_i=k\}$ containing all the nodes from group $k$.
Here, $|G_k|$ denotes the number of nodes inside the collection, and the number of non-empty collections/groups $K := |\boldsymbol{C}(\boldsymbol{z})|$.
In the case that $G_k$ is the empty set, we have $|G_k|=0$.
The ascending factorial notation $\alpha^{(n)}:=\alpha(\alpha+1)\cdots(\alpha+n-1)$, while the descending factorial notation $k_{(K)}:=k(k-1)\cdots(k-(K-1))$.
The parameter $\bar{K}$ is collapsed by summing over all the possible $\bar{K}=k$ values from $k=1$ to $k=\infty$.
Note that $\boldsymbol{C}(\boldsymbol{z})$ is invariant under any relabeling of $\boldsymbol{z}$.
The $k_{(K)}\equiv\binom{k}{K}K!$ term in Eq.~\eqref{MFM_PMF_C} is the number of ways to relabel a specific $\boldsymbol{z}$ or to label the unlabeled partition $\boldsymbol{C}(\boldsymbol{z})$ provided with $\bar{K}=k$. 
The natural choice of $\bar{K}$ prior is a zero-truncated Poisson(1) distribution \citep{geng2019probabilistic,nobile2005bayesian,mcdaid2012clustering} that is assumed throughout this paper.

To ensure that the clustering $\boldsymbol{z}$ is invariant under any relabeling of it, we adopt a particular labeling method following the procedure used in 
\citet{rastelli2018optimal}. 
We assign node $1$ to group $1$ by default, and then iteratively assign the next node either to a new empty group or to an existing group.
In this way, the defined $\boldsymbol{z}$ only contains $K$ occupied groups and is one-to-one correspondence to $\boldsymbol{C}(\boldsymbol{z})$ regardless of whether the clustering before label-switching has empty groups or not.
The clustering dependent parameters, $\boldsymbol{\mu},\boldsymbol{\tau}$ and $\boldsymbol{P}$ are relabeled accordingly and the entries relevant to empty groups are treated as redundant.
The probability mass function of the MFM in Eq.~\eqref{MFM_PMF_C} may be rewritten as
\begin{equation}
\label{MFM_PMF}
\begin{split} 
f(\boldsymbol{z})= \mathcal{W}_{N,K}\prod^K_{g=1} \alpha^{(n_g)},
\end{split}
\end{equation}
where $n_g$ is the number of nodes in group $g$. 
The non-negative weight
$$
\mathcal{W}_{N,K}:= \sum_{k=1}^{\infty} \frac{k_{(K)}}{(k\alpha)^{(N)}}\pi_{\bar{K}}(k)
$$
satisfies the recursion $\mathcal{W}_{N,K}:= (N+K\alpha)\mathcal{W}_{N+1,K}+\alpha\mathcal{W}_{N+1,K+1}$ with $\mathcal{W}_{1,1}=1/\alpha$, and we refer to \citet{miller2018mixture} for the details of the computation of $\mathcal{W}_{N,K}$.

One could proceed to determine a generative/predictive urn scheme by checking the formulae differences between the $f(\boldsymbol{z})$ and each $f(\{\boldsymbol{z},z_{N+1}\})$ for $z_{N+1}=1,2,\dots,K,K+1$. 
Then, sampling from Eq.~\eqref{MFM_PMF} can be performed using the following procedure:
assuming that the first node labeled as node $1$ is assigned to group $1$ by default, then the probability of the subsequent node being assigned to existing non-empty groups or to a new empty group is defined as:
\begin{equation}
\label{MFM_urn}
\text{P}(z_{N+1}=k|\boldsymbol{z}) \propto \begin{cases} 
n_k+\alpha, & \text{for}\;k = 1,2,\dots,K;\\
\frac{\mathcal{W}_{N+1,K+1}}{\mathcal{W}_{N+1,K}}\alpha, & \text{for}\;k = K+1,
\end{cases}
\end{equation}
that is conditional on the current clustering $\boldsymbol{z}$ with parameters $N$ and $K$, where $n_k$ is defined as the number of nodes in group $k$.
We refer to Theorem 4.1 of \citet{miller2018mixture} for a more detailed discussion on this generative procedure.
This generative scheme 
also belongs to the Ewens-Pitman two-parameter family of Exchangeable Partitions \citep{gnedin2009characterizations,pitman2006combinatorial}.\\

Some exogenous node attributes may be available when analyzing real datasets, and, in this case, we leverage the idea of supervised priors proposed by \citet{legramanti2022extended} to account for such information in the modeling.
We use $\boldsymbol{c} = \{c_1,c_2,\dots,c_N\}$ to denote the categorical node attributes in our context where each $c_i\in \{1,2,\dots,C\}$, and propose that
\begin{equation}
\label{MFM_PMF_Sup}
\begin{split}
f(\boldsymbol{z}|\boldsymbol{c})
&\propto \mathcal{W}_{N,K}\prod^{K}_{g=1}\text{P}(\boldsymbol{c}_g)\alpha^{(n_g)},\\
\end{split}
\end{equation}
where $\boldsymbol{c}_g := \{c_i: z_i = g\}$ and where the distribution $\text{P}(\boldsymbol{c}_g)$ is chosen as the Dirichlet-multinomial cohesion of \citet{muller2011product}, which is given by:
\begin{equation}
\label{Sup_DMCohesion}
\begin{split}
\text{P}(\boldsymbol{c}_g) = \frac{\prod^C_{c=1}\Gamma\left(n_{g,c}+\omega_c\right)}{\Gamma\left(n_g+\omega_0\right)}\frac{\Gamma\left(\omega_0\right)}{\prod^C_{c=1}\Gamma\left(\omega_c\right)}.
\end{split}
\end{equation}
Here, $n_{g,c}$ denotes the number of nodes in group $g$ that have the node attribute $c$, whereas $\omega_c$ is the cohesion parameter for each level $c$ of the node attributes and $\omega_0 = \sum^C_{c=1}\omega_c$.
Similar to Eq.~\eqref{MFM_urn}, we can determine an urn scheme also for the supervised MFM as:
\begin{equation}
\label{MFM_urn_Sup}
\text{P}(z_{N+1}=k|\boldsymbol{z},\boldsymbol{c},c_{N+1})  \propto \begin{cases} 
\frac{n_{k,c_{N+1}}+\omega_{c_{N+1}}}{n_k+\omega_0}(n_k+\alpha), & \text{for } k=1,2,\dots,K; \\ 
\frac{\omega_{c_{N+1}}}{\omega_0}\frac{\mathcal{W}_{N+1,K+1}}{\mathcal{W}_{N+1,K}}\alpha, & \text{for } k = K+1,
\end{cases}
\end{equation}
which, compared to Eq.~\eqref{MFM_urn}, is inflated or deflated by a term that favors allocating the new node $N+1$ to the group with higher fraction of the same node attribute as the $c_{N+1}$, that is, the allocation is supervised by $\bm{c}$.

Note that, in the case that some categorical node attributes $\bm{c}$ are available, these can be considered as a reference clustering under the MFM framework.
For example, individuals in a social network can be clustered into different groups based on their affiliations, while they can also be clustered in a different way according to their job categories.
However, none of these reference clusterings might be the ``true'' clustering or the ``best'' clustering for the network.
Our supervised MFM framework can be treated as a type of supervised prior, in the sense of \citet{legramanti2022extended}, with a reference clustering $\bm{c}$: this should not be confused with the general idea of supervised learning, which is often used in similar contexts.
We refer to \citet{legramanti2022extended} for more details on the use and terminology of supervised priors.

%----------------------------------------------------------------------------------------------------------------------------------------------------------------------------------------------------------------------------------------------------------------------------------------
%----------------------------------------------------------------------------------------------------------------------------------------------------------------------------------------------------------------------------------------------------------------------------------------
%----------------------------------------------------------------------------------------------------------------------------------------------------------------------------------------------------------------------------------------------------------------------------------------
%----------------------------------------------------------------------------------------------------------------------------------------------------------------------------------------------------------------------------------------------------------------------------------------

\section{Partially collapsed Bayesian inference}
\label{Bayesian inference}

In this section, we illustrate our original inference processes which aim to jointly infer the intercept $\beta$, the indicator of unusual zeros $\boldsymbol{\nu}$, the latent positions $\boldsymbol{U}$, the latent clustering indicator $\boldsymbol{z}$ and the number of occupied groups $K$.
The model includes other parameters, such as $\boldsymbol{\mu}$ and $\boldsymbol{\tau}$, which are dealt with via marginalization, as shown in Section~\ref{Collapsing_Inference}.
We emphasize that $K$ indicates the number of occupied groups for the clustering $\boldsymbol{z}$, and thus can be calculated directly from $\boldsymbol{z}$. This should not be confused with the total number of groups $\bar{K}$.
This distinction is crucial since we leverage mixture-of-finite-mixtures \citep{miller2018mixture,geng2019probabilistic} to marginalize $\bar{K}$ in Section~\ref{MFM_Sup}.
As a result, we only need to focus on non-empty groups during the inference procedures.

%----------------------------------------------------------------------------------------------------------------------------------------------------------------------------------------------------------------------------------------------------------------------------------------
%----------------------------------------------------------------------------------------------------------------------------------------------------------------------------------------------------------------------------------------------------------------------------------------

\subsection{Inference and collapsing}
\label{Collapsing_Inference}

In the case that the exogenous node attributes $\boldsymbol{c}$ are available, adopting the supervised MFM introduced in Eq.~\eqref{MFM_PMF_Sup} as well as the missing data imputation shown in Eq.~\eqref{Aug_X} leads to the posterior distribution of our ZIP-LPCM being written as:
\begin{equation}
\label{PosteriorX}
\resizebox{.925\hsize}{!}{$
\begin{split}
\pi(\boldsymbol{X}, \boldsymbol{\nu}, \boldsymbol{U}, \boldsymbol{z},\beta,\boldsymbol{P},\boldsymbol{\mu},\boldsymbol{\tau}|\boldsymbol{Y},\boldsymbol{c})
&\propto f(\boldsymbol{X}|\boldsymbol{Y}, \beta, \boldsymbol{U},\boldsymbol{\nu})f(\boldsymbol{Y}|\beta, \boldsymbol{U},\boldsymbol{\nu})f(\boldsymbol{\nu}|\boldsymbol{P},\boldsymbol{z})f(\boldsymbol{U}|\boldsymbol{\mu},\boldsymbol{\tau},\boldsymbol{z}) f(\boldsymbol{z}|\boldsymbol{c})\\
&\hspace{1em}\times\pi(\boldsymbol{P})\pi(\boldsymbol{\mu})\pi(\boldsymbol{\tau})\pi(\beta)\\
&=f(\boldsymbol{X}|\beta, \boldsymbol{U})f(\boldsymbol{\nu}|\boldsymbol{P},\boldsymbol{z})f(\boldsymbol{U}|\boldsymbol{\mu},\boldsymbol{\tau},\boldsymbol{z})f(\boldsymbol{z}|\boldsymbol{c})\pi(\boldsymbol{P})\pi(\boldsymbol{\mu})\pi(\boldsymbol{\tau})\pi(\beta),\\
\end{split}$}
\end{equation}
where the $f(\boldsymbol{Y}|\beta, \boldsymbol{U},\boldsymbol{\nu})$, the $f(\boldsymbol{\nu}|\boldsymbol{P},\boldsymbol{z})$ and the $f(\boldsymbol{U}|\boldsymbol{\mu},\boldsymbol{\tau},\boldsymbol{z})$ are exactly the same as the ones shown in the complete likelihood in  Eq.~\eqref{CompleteLikelihood}.
Furthermore, similar to the $f(\boldsymbol{Y}|\beta, \boldsymbol{U},\boldsymbol{\nu})$, the $f(\boldsymbol{X}|\boldsymbol{Y}, \beta, \boldsymbol{U},\boldsymbol{\nu})$ above can be obtained based on its sampling process in Eq.~\eqref{Aug_X}.
The combination of the $f(\boldsymbol{Y}|\beta, \boldsymbol{U},\boldsymbol{\nu})$ and the $f(\boldsymbol{X}|\boldsymbol{Y}, \beta, \boldsymbol{U},\boldsymbol{\nu})$ in Eq.~\eqref{PosteriorX} is equivalent to the $f(\boldsymbol{X}|\beta, \boldsymbol{U})$ which reads as follows:
\begin{equation*}
\label{Posterior_X}
\begin{split}
f(\boldsymbol{X}|\beta, \boldsymbol{U})&=\prod^N_{i,j:i \neq j}f_{\text{Pois}}(x_{ij}|\text{exp}(\beta-||\boldsymbol{u}_i-\boldsymbol{u}_j||)).
\end{split}
\end{equation*}
However, in our context, not all the real networks that we work on provide exogenous node attributes.
In the case that $\boldsymbol{c}$ is not available, the unsupervised prior $f(\boldsymbol{z})$ from Eq.~\eqref{MFM_PMF} is instead proposed in Eq.~\eqref{PosteriorX} to replace the $f(\boldsymbol{z}|\boldsymbol{c})$. \\

We leverage conjugate prior distributions to marginalize a number of model parameters from the posterior distribution in (\ref{PosteriorX}).
This methodology is also known as ``collapsing'' and has already been exploited in, for example, \citet{mcdaid2012clustering,wyse2012block,ryan2017bayesian,rastelli2018choosing,legramanti2022extended,lu2025zero}.
By proposing the conjugate prior distributions:
\begin{equation}
\label{tau_prior}
\tau_k\sim\text{Gamma}(\alpha_1,\alpha_2/2)\ \text{for}\ k=1,\dots,K,
\end{equation}
\begin{equation}
\label{mu_prior}
\boldsymbol{\mu}_k|\tau_k \sim \text{MVN}_d(\boldsymbol{0},1/(\omega\tau_k)\mathbbm{I}_d)\ \text{for}\ k=1,\dots,K,
\end{equation}
where $(\alpha_1,\alpha_2,\omega)$ are hyperparameters to be specified a priori, the collapsed posterior distribution of the ZIP-LPCM is obtained as:
\begin{equation}
\label{CollapsedPosteriorX}
\begin{split}
\pi(\boldsymbol{X}, \boldsymbol{\nu}, \boldsymbol{U}, \boldsymbol{z},\beta,\boldsymbol{P}|\boldsymbol{Y},\boldsymbol{c}) &\propto f(\boldsymbol{X}|\beta, \boldsymbol{U})f(\boldsymbol{\nu}|\boldsymbol{P},\boldsymbol{z})f(\boldsymbol{U}|\boldsymbol{z})f(\boldsymbol{z}|\boldsymbol{c})\pi(\boldsymbol{P})\pi(\beta),\\
\end{split}
\end{equation}
where the $f(\boldsymbol{U}|\boldsymbol{z})$ is calculated according to the methodology explored in \citet{ryan2017bayesian}:
\begin{equation}
\label{Collpased_U}
\resizebox{.925\hsize}{!}{$
\begin{split}
f(\boldsymbol{U}|\boldsymbol{z}) = \prod^{K}_{k=1} \left\{\frac{\alpha_2^{\alpha_1}}{\Gamma(\alpha_1)}\frac{\Gamma(\alpha_1+\frac{d}{2}n_k)}{\pi^{\frac{d}{2}n_k}}\left(\frac{\omega}{\omega+n_k}\right)^{\frac{d}{2}}\left[\alpha_2-\frac{\norm{\sum_{i:z_i=k}\boldsymbol{u}_i}^2}{n_k+\omega}+\sum_{i:z_i=k}\norm{\boldsymbol{u}_i}^2\right]^{-(\frac{d}{2}n_k+\alpha_1)}\right\}.
\end{split}$}
\end{equation}\\

Now that we have our target collapsed posterior distribution, we apply a partially collapsed Markov chain Monte Carlo approach \citep{van2008partially,park2009partially} aiming to infer the latent clustering $\boldsymbol{z}$, the latent indicators of unusual zeros $\boldsymbol{\nu}$, the intercept $\beta$ and the latent positions $\boldsymbol{U}$ from the posterior in Eq.~\eqref{CollapsedPosteriorX}.
This is accomplished by constructing a sampler which consists of 
multiple steps for each of the target variables and parameters, and for the number of occupied groups $K$ simultaneously.
The sampling of the imputed adjacency matrix $\boldsymbol{X}$ straightforwardly follows from Eq.~\eqref{Aug_X}, and we leverage ideas similar to those in \citet{lu2025zero} to infer $\boldsymbol{\nu}$ and $\boldsymbol{z}$.
Another layer of conjugacy is proposed for the inference procedure of the probability of unusual zeros $\boldsymbol{P}$ that is required by the sampling step of $\boldsymbol{\nu}$, and we further develop a new truncated absorb-eject move tailored for the clustering without empty groups to facilitate the clustering inference.
The sampling of $\boldsymbol{U}$ and $\beta$ are performed via two standard Metropolis-Hastings steps.
More details of all these sampling steps are carefully discussed and provided next.

\subsubsection[Inference for $\nu$]{Inference for $\boldsymbol{\nu}$}
\label{nu_Inference_Step}

Recall that each of the latent indicators of unusual zeros $\{\nu_{ij}\}$ is assumed to follow the $\text{Bern}(p_{z_iz_j})$ distribution. 
However, note that each $\nu_{ij}$ must be zero by assumption when the corresponding observed $y_{ij}>0$, so only those $\{\nu_{ij}:y_{ij}=0\}$ are required to be inferred during the inference.
Conditional on that the observed interaction $y_{ij}$ is a zero interaction, the probability of such an interaction being an unusual zero is:
\begin{equation}
\label{P_m0}
\text{P}(\nu_{ij}=1|y_{ij}=0,p_{z_iz_j},\beta,\boldsymbol{u}_i,\boldsymbol{u}_j)=\frac{p_{z_iz_j}}{p_{z_iz_j}+(1-p_{z_iz_j})f_{\text{Pois}}(0|\text{exp}(\beta-||\boldsymbol{u}_i-\boldsymbol{u}_j||)},
\end{equation}
and, on the contrary, the conditional probability of it not being an unusual zero is $\text{P}(\nu_{ij}=0|y_{ij}=0,p_{z_iz_j},\beta,\boldsymbol{u}_i,\boldsymbol{u}_j)=1-\text{P}(\nu_{ij}=1|y_{ij}=0,p_{z_iz_j},\beta,\boldsymbol{u}_i,\boldsymbol{u}_j)$.
This motivates the sampling of each $\nu_{ij}$ to follow:
\begin{equation}
\label{nu_Inference}
\begin{split}
\nu_{ij}|y_{ij},p_{z_iz_j},\beta,\boldsymbol{u}_i,\boldsymbol{u}_j
&\sim
\begin{cases} 
 \text{Bern}\left(\frac{p_{z_iz_j}}{p_{z_iz_j}+(1-p_{z_iz_j})f_{\text{Pois}}(0|\text{exp}(\beta-||\boldsymbol{u}_i-\boldsymbol{u}_j||)}\right), &  \text{if}\;y_{ij}=0;\\ 
\mathbbm{1}(\nu_{ij}=1), & \text{if}\; y_{ij} > 0.
\end{cases}
\end{split}
\end{equation}
This distribution is actually the full-conditional distribution of $\nu_{ij}$ under the posterior:
\begin{equation}
\label{CollapsedPosteriorY}
\pi(\boldsymbol{\nu}, \boldsymbol{U}, \boldsymbol{z},\beta, \boldsymbol{P}|\boldsymbol{Y},\boldsymbol{c}) \propto f(\boldsymbol{Y}|\beta, \boldsymbol{U},\boldsymbol{\nu})f(\boldsymbol{\nu}|\boldsymbol{P},\boldsymbol{z})f(\boldsymbol{U}|\boldsymbol{z})f(\boldsymbol{z}|\boldsymbol{c})\pi(\boldsymbol{P})\pi(\beta),\\
\end{equation}
which further marginalizes the augmented missing data imputed elements from Eq.~\eqref{CollapsedPosteriorX}, that is, collapsing the $f(\boldsymbol{X}|\boldsymbol{Y}, \beta, \boldsymbol{U},\boldsymbol{\nu})$ therein.
The sampling in Eq.~\eqref{nu_Inference} requires the inference of the probability of unusual zeros defined for the interactions between and within those occupied groups.
This can be accomplished by further proposing the conjugate prior distributions: 
\begin{equation}
\label{p_prior}
p_{gh}\sim\text{Beta}(\beta_1,\beta_2)\ \text{for}\ g,h=1,\dots,K,
\end{equation}
leading to a typical Gibbs sampling step for each non-redundant probability of unusual zeros:
\begin{equation}
\label{P_Inference}
p_{gh}|\boldsymbol{\nu},\boldsymbol{z} \sim \text{Beta}\left(\boldsymbol{\nu}_{gh} + \beta_1,n_{gh}-\boldsymbol{\nu}_{gh}+\beta_2\right),\;\text{for}\; g,h=1,2,\dots,K,
\end{equation}
where $(\beta_1,\beta_2)$ are hyperparameters, and $n_{gh} := \sum_{i,j:i\neq j}^N\mathbbm{1}(z_i=g, z_j=h)$.
The notation $\boldsymbol{\nu}_{gh}$ denotes the sum of all the $\nu_{ij}|z_i=g, z_j=h, i \neq j$.
Note that the conditional probability of unusual zero provided that the corresponding observed interaction is a zero interaction shown in Eq.~\eqref{P_m0} is of key interest for practitioners to explore when observing a zero interaction.
We refer to \citet{lu2025zero} for a detailed discussion of the $\boldsymbol{\nu}$ sampling in Eq.~\eqref{nu_Inference}.

\subsubsection[Inference for $z$]{Inference for $\boldsymbol{z}$}

Carrying out inference for the clustering variable $\boldsymbol{z}$ based on the supervised MFM prior in Eq.~\eqref{MFM_urn_Sup} simultaneously infers the clustering allocations and automatically chooses the number of groups.
However, the dimension of the model parameter matrix $\boldsymbol{P}$ becomes problematic when the nodes are proposed to be assigned to a new empty group.
Thus, following the prior distribution introduced in Eq.~\eqref{p_prior}, we can marginalize the target posterior in Eq.~\eqref{CollapsedPosteriorX} in a different way compared to the posterior in Eq.~\eqref{CollapsedPosteriorY}.
Here, we collapse the parameter $\boldsymbol{P}$ from Eq.~\eqref{CollapsedPosteriorX} following, for example, \citet{mcdaid2012clustering,lu2025zero}, and this leads to a posterior:
\begin{equation}
\label{CollapsedPosteriorP}
\begin{split}
\pi(\boldsymbol{X}, \boldsymbol{\nu}, \boldsymbol{U}, \boldsymbol{z},\beta|\boldsymbol{Y},\boldsymbol{c}) &\propto f(\boldsymbol{X}|\beta, \boldsymbol{U})f(\boldsymbol{\nu}|\boldsymbol{z})f(\boldsymbol{U}|\boldsymbol{z})f(\boldsymbol{z}|\boldsymbol{c})\pi(\beta),\\
\end{split}
\end{equation}
where the collapsed likelihood function $f(\boldsymbol{\nu}|\boldsymbol{z})$ reads as follows:
\begin{equation}
\label{Collpased_nu}
\begin{split}
f(\boldsymbol{\nu}|\boldsymbol{z}) 
&= \prod^{K}_{g=1,h=1}\left[\frac{\text{B}\left(\boldsymbol{\nu}_{gh} + \beta_1,n_{gh}-\boldsymbol{\nu}_{gh}+\beta_2\right)}{\text{B}(\beta_1,\beta_2)}\right].
\end{split}
\end{equation}
Here, B($\cdot,\cdot$) is the beta function.
The sampling of each $z_i$ is based on the normalized probability proportional to its full-conditional distribution of the posterior in Eq.~\eqref{CollapsedPosteriorP} \citep{legramanti2022extended}, that is,
\begin{equation}
\label{zi_FullCondLike}
\begin{split}
\text{P}(z_i = k|\boldsymbol{\nu}, \boldsymbol{U},\boldsymbol{c}, \boldsymbol{z}^{-i}) 
&\propto \text{P}(z_i = k|\boldsymbol{c}, \boldsymbol{z}^{-i})f(\boldsymbol{\nu}|z_i = k, \boldsymbol{z}^{-i})f(\boldsymbol{U}|z_i = k, \boldsymbol{z}^{-i}),\\
\end{split}
\end{equation}
where the notation $\boldsymbol{z}^{-i} := \boldsymbol{z}\backslash\{z_i\}$ 
contains all the clustering indicators except $z_i$, and the $\text{P}(z_i = k|\boldsymbol{c}, \boldsymbol{z}^{-i})$ follows the supervised MFM urn scheme in Eq.~\eqref{MFM_urn_Sup} by assuming that the node $i$ is removed from the network and then is treated as a new node to be assigned a group in the network. 
More specifically, we have
\begin{equation}
\label{zi_Inference}
\text{P}(z_i = k|\boldsymbol{c}, \boldsymbol{z}^{-i})  \propto \begin{cases} 
\frac{n^{-i}_{k,c_i}+\omega_{c_i}}{n^{-i}_k+\omega_0}(n^{-i}_k+\alpha), & \text{for}\; k=1,2,\dots,K^{-i}; \\ 
\frac{\omega_{c_i}}{\omega_0}\frac{\mathcal{W}_{N^{-i}+1,K^{-i}+1}}{\mathcal{W}_{N^{-i}+1,K^{-i}}}\alpha& \text{for}\; k = K^{-i}+1,
\end{cases}
\end{equation}
where $(\cdot)^{-i}$ denotes the corresponding statistics obtained after removing node $i$ from the network.
If removing node $i$ makes one of the groups empty, the remaining non-empty groups in $\boldsymbol{z}^{-i}$ should be relabeled in ascending order by letting $z_j=z_j-1$ for all the $\{z_j:j=1,2,\dots,N; j\neq i; z_j>z_i\}$ during the inference procedures.
If $\boldsymbol{c}$ is not available, the $\text{P}(z_i = k|\boldsymbol{c}, \boldsymbol{z}^{-i})$ should be replaced with the $\text{P}(z_i = k|\boldsymbol{z}^{-i})$ which instead follows the unsupervised MFM urn scheme in Eq.~\eqref{MFM_urn}, that is, the specific form can be obtained by removing the $(n^{-i}_{k,c_i}+\omega_{c_i})/(n^{-i}_k+\omega_0)$ and the $\omega_{c_i}/\omega_0$ terms in Eq.~\eqref{zi_Inference}.
We refer to \citet{miller2018mixture,geng2019probabilistic,legramanti2022extended,lu2025zero} for more details of the inference procedure of $\boldsymbol{z}$.

%----------------------------------------------------------------------------------------------------------------------------------------------------------------------------------------------------------------------------------------------------------------------------------------
%----------------------------------------------------------------------------------------------------------------------------------------------------------------------------------------------------------------------------------------------------------------------------------------

\subsection{Truncated absorb-eject move}
\label{TAE}

Since the latent clustering variable $\boldsymbol{z}$ is updated one element at a time based on Eq.~\eqref{zi_FullCondLike}, the inference algorithm is susceptible to getting stuck in local posterior modes. 
Thus, we propose to leverage an Absorb-Eject (AE) move proposed by \citet{nobile2007bayesian} to facilitate the clustering inference and to deal with such a sampling issue.
However, the AE move, as suggested by \citet{nobile2007bayesian}, may create empty groups, and this does not work well with our method, which requires non-empty groups.
Thus we instead propose a Truncated Absorb-Eject (TAE) move which specifically addresses this issue by no longer creating empty groups, as we describe more in detail here below.

Similar to a typical AE move, we have two reversible moves in each iteration of the inference algorithm: a truncated $eject$ move, denoted as $eject^T$, and an $absorb$ move.
In general, with probability $\text{P}(eject^T)$, the $eject^T$ move is applied and, with probability $1-\text{P}(eject^T)$, the $absorb$ move is applied.
As an exception, the $eject^T$ move is applied with probability $1$ if $K=1$, while the $absorb$ move is applied with probability $1$ when $K=N$.

\begin{itemize}
\item $Eject^T$ move: first randomly pick one of the $K$ non-empty groups, say group $g$. 
Then, we sample an ejection probability from a prior distribution, $p_e\sim \text{Beta}(a,a)$, and each node in group $g$ has probability $p_e$ to be reallocated to the new group labeled as $K+1$. 
Thus, on the contrary, each node stays in group $g$ with probability $1-p_e$.
The proposed state is denoted as $(\boldsymbol{z}',K'=K+1)$ after the reallocation.
If this process creates an empty group, that is, either the proposed group $g$ or the proposed group $K+1$ in $\boldsymbol{z}'$ is an empty group, we propose to abandon this truncated AE move and remains at the current state, $(\boldsymbol{z},K)$.

If the reallocation does not create an empty group, we propose:
$$\{\boldsymbol{z},K\}\rightarrow\{\boldsymbol{z}',K'=K+1\},$$
with the proposal probability
\begin{equation}
\label{eject_prop}
\begin{split}
&\text{P}(\{\boldsymbol{z},K\}\rightarrow\{\boldsymbol{z}',K'\})=\frac{\int^1_0 p_e^{n'_{K'}}(1-p_e)^{n'_g}\pi_{\text{beta}}(p_e|a,a)\text{d}p_e}{1-p_0}\text{P}(eject^T)\frac{1}{K},
\end{split}
\end{equation}
where $n'_{g}$ and $n'_{K'}$, respectively, denotes the number of nodes in group $g$ and in group $K'$ of the proposed clustering $\boldsymbol{z}'$.
Here, the reallocation probability $p_e$ is collapsed leading to:
\begin{equation}
\label{eject_prop_collapsed_p_e}
\resizebox{.855\hsize}{!}{$
\begin{split}
&\hspace{1em}\int^1_0 p_e^{n'_{K'}}(1-p_e)^{n'_g}\pi_{\text{beta}}(p_e|a,a)\text{d}p_e = \int^1_0 p_e^{n'_{K'}}(1-p_e)^{n'_g}\frac{\Gamma(2a)}{\Gamma(a)^2}(1-p_e)^{a-1}p_e^{a-1}\text{d}p_e\\
&=\frac{\Gamma(2a)}{\Gamma(a)^2}\int^1_0p_e^{n'_{K'}+a-1}(1-p_e)^{n'_g+a-1}=\frac{\Gamma(2a)}{\Gamma(a)^2}\frac{\Gamma(a+n'_g)\Gamma(a+n'_{K'})}{\Gamma(2a+n_g)},\\
\end{split}$}
\end{equation}
where $n_g$ is the number of nodes in group $g$ of the current clustering $\boldsymbol{z}$, and note here that $n_g = n'_{K'} + n'_{g}$.
The $p_0$ in Eq.~\eqref{eject_prop} is the probability that all the selected nodes are reallocated in one group leaving another group empty, and is calculated following the similar way as Eq.~\eqref{eject_prop_collapsed_p_e}, that is,
\begin{equation*}
\resizebox{.925\hsize}{!}{$
\begin{split} 
\text{P}(n'_{K'}=0)=\text{P}(n'_{g}=0)=\frac{p_0}{2}= \int_0^1p_e^0(1-p_e)^{n_{g}}\pi_{\text{beta}}(p_e|a,a)\text{d}p_e=\frac{\Gamma(2a)}{\Gamma(a)}\frac{\Gamma(a+n_{g})}{\Gamma(2a+n_{g})}.
\end{split}$}
\end{equation*}
The prior parameter $a$ for the reallocation probability $p_e$ is set by checking the pre-computed look-up table of $p_0$ with respect to $a$ and $n_{g}$.
According to our simulation studies, since we do not expect too many moves to be abandoned due to creating empty groups, we propose to set $p_0=0.02$ in our experiments.

On the contrary, the reverse proposal probability is
\begin{equation*}
\begin{split}
&\text{P}(\{\boldsymbol{z}',K'\}\rightarrow\{\boldsymbol{z},K\})=\frac{\text{P}(absorb)}{K}.
\end{split}
\end{equation*}

Thus this $eject^T$ move is accepted 
with probability
\begin{equation}
\label{AE_AcceptRatio}
\begin{split}
&\text{min}\left(1,\frac{f(\boldsymbol{\nu}|\boldsymbol{z}')f(\boldsymbol{U}|\boldsymbol{z}')f(\boldsymbol{z}'|\boldsymbol{c})}{f(\boldsymbol{\nu}|\boldsymbol{z})f(\boldsymbol{U}|\boldsymbol{z})f(\boldsymbol{z}|\boldsymbol{c})}\frac{\text{P}(\{\boldsymbol{z}',K'\}\rightarrow\{\boldsymbol{z},K\})}{\text{P}(\{\boldsymbol{z},K\}\rightarrow\{\boldsymbol{z}',K'\})}\right),
\end{split}
\end{equation}
where the details of the $f(\boldsymbol{\nu}|\boldsymbol{z}), f(\boldsymbol{U}|\boldsymbol{z})$ and $f(\boldsymbol{z}|\boldsymbol{c})$ terms are, respectively, provided in Eqs.~\eqref{Collpased_nu}, \eqref{Collpased_U} and \eqref{MFM_PMF_Sup}.
Otherwise, we remain at the current state $(\boldsymbol{z},K)$.

\item $Absorb$ move: first randomly select two groups, say groups $g,h$ with $g<h$, from the $K$ groups and merge them together into cluster $g$.  
Then relabel all the groups in ascending order from group $1$ to group $K':=K-1$. 
We accept this $absorb$ move with the same probability scheme in Eq.~\eqref{AE_AcceptRatio} but with different proposal probability:
\begin{equation*}
\begin{split}
&\text{P}(\{\boldsymbol{z},K\}\rightarrow\{\boldsymbol{z}',K'\})=\frac{\text{P}(absorb)}{K(K-1)},
\end{split}
\end{equation*}
while the reverse proposal probability is
\begin{equation*}
\begin{split}
\text{P}(\{\boldsymbol{z}',K'\}\rightarrow\{\boldsymbol{z},K\})=\frac{\frac{\Gamma(2a)}{\Gamma(a)^2}\frac{\Gamma(a+n_g)\Gamma(a+n_h)}{\Gamma(2a+n'_g)}}{1-2\frac{\Gamma(2a)}{\Gamma(a)}\frac{\Gamma(a+n'_{g})}{\Gamma(2a+n'_{g})}}\frac{\text{P}(eject^T)}{K'(K'+1)}.
\end{split}
\end{equation*}
The proof of the above formula is similar to Eq.~\eqref{eject_prop_collapsed_p_e}.
If the absorb move is not accepted, we remain at the current state.
\end{itemize}

Note that our proposed TAE move is a Metropolis-Hastings move that extends the typical AE move proposed in \citet{nobile2007bayesian}, wherein a complete study and discussion on these moves are provided.

%----------------------------------------------------------------------------------------------------------------------------------------------------------------------------------------------------------------------------------------------------------------------------------------
%----------------------------------------------------------------------------------------------------------------------------------------------------------------------------------------------------------------------------------------------------------------------------------------

\subsection{Partially collapsed Metropolis-within-Gibbs}
\label{PCG}

Inferring the indicators of unusual zeros $\boldsymbol{\nu}$ and the latent clustering indicators $\boldsymbol{z}$ from different posteriors does not guarantee the convergence to the same target distribution. 
However, note that both posteriors in Eq.~\eqref{CollapsedPosteriorY} and in Eq.~\eqref{CollapsedPosteriorP} are different partially collapsed forms of the posterior in Eq.~\eqref{CollapsedPosteriorX} which is further a partially collapsed form of the posterior:
\begin{equation}
\label{unCollapsedPosteriorX}
\resizebox{.925\hsize}{!}{$
\pi(\boldsymbol{X}, \boldsymbol{\nu}, \bar{\boldsymbol{P}}, \boldsymbol{U}, \boldsymbol{z},\beta|\boldsymbol{Y},\boldsymbol{c}) \propto f(\boldsymbol{X}|\boldsymbol{Y}, \beta, \boldsymbol{U},\boldsymbol{\nu})f(\boldsymbol{Y}|\beta, \boldsymbol{U},\boldsymbol{\nu})f(\boldsymbol{\nu}|\bar{\boldsymbol{P}},\boldsymbol{z})f(\boldsymbol{U}|\boldsymbol{z})f(\boldsymbol{z}|\boldsymbol{c})\pi(\beta)\pi(\bar{\boldsymbol{P}}),$}
\end{equation}
where $\bar{\boldsymbol{P}}$ is defined as a $N \times N$ matrix of which $\boldsymbol{P}$ is a submatrix.
The joint prior distribution $\pi(\bar{\boldsymbol{P}}) \propto \prod_{g=1,h=1}^Nf_{\text{Beta}}(\bar{p}_{gh}|\beta_1,\beta_2)$ is proposed to be the product up to the maximum reachable value of $K$, that is, $N$, and the $\{\bar{p}_{gh}\}$ here are the entries of $\bar{\boldsymbol{P}}$.
Thus we leverage the partially collapsed Gibbs approach proposed by \citet{van2008partially,park2009partially} 
to ensure the correct target stationary distribution.
Our Algorithm~\ref{PCMwG} further adopts the Metropolis-Hastings (M-H) steps of inferring $\beta$ and $\boldsymbol{U}$ leading to the Partially Collapsed Metropolis-within-Gibbs (PCMwG) algorithm.
Note that the full-conditional distribution of $\beta$ remains the same under either the posterior in Eq.~\eqref{CollapsedPosteriorX} or the posterior in Eq.~\eqref{unCollapsedPosteriorX}, that is,
\begin{equation}
\label{Beta_fullCond}
\begin{split}
p(\beta|\boldsymbol{X},\boldsymbol{U}) \propto f(\boldsymbol{X}|\beta, \boldsymbol{U})\pi(\beta) = \left[\prod^N_{i,j:i \neq j}f_{\text{Pois}}(x_{ij}|\text{exp}(\beta-||\boldsymbol{u}_i-\boldsymbol{u}_j||))\right]\pi(\beta),
\end{split}
\end{equation}
while the full-conditional distribution of each $\boldsymbol{u}_i$ is also invariant under the two different posteriors:
\begin{equation}
\label{U_fullCond}
\resizebox{.925\hsize}{!}{$
\begin{split}
&\hspace{1.2em}p(\boldsymbol{u}_i|z_i=k, \beta,\boldsymbol{X},\boldsymbol{U}\backslash\{\boldsymbol{u}_i\}) \propto f(\boldsymbol{X}|\beta, \boldsymbol{U})f(\boldsymbol{U}|\boldsymbol{z}^{-i},z_i=k)\\
& \propto \left[\prod^N_{j:j \neq i}f_{\text{Pois}}(x_{ij},x_{ji}\mid\text{exp}(\beta-||\boldsymbol{u}_i-\boldsymbol{u}_j||))\right] \left[\alpha_2-\frac{\norm{\sum_{j:z_j=k}\boldsymbol{u}_j}^2}{n_k+\omega}+\sum_{j:z_j=k}\norm{\boldsymbol{u}_j}^2\right]^{-(\frac{d}{2}n_k+\alpha_1)}.
\end{split}$}
\end{equation}
Here we set normal proposal distribution with variance $\sigma^2_{\beta}$ for the M-H step of $\beta$, while we propose a multivariate normal proposal distribution for each latent position $\boldsymbol{u}_i$ with covariance $\sigma^2_{\boldsymbol{U}}\mathbbm{I}_d$.
Both proposal distributions are centered at the previous state of the corresponding parameter or latent variables.

\begin{algorithm}[h!]
\caption{A partially collapsed Metropolis-within-Gibbs sampler for ZIP-LPCM.}
\label{PCMwG}
\begin{algorithmic} 
\State \textbf{Input}: $\boldsymbol{Y}, \boldsymbol{c}, \sigma^2_{\beta}, \sigma^2_{\boldsymbol{U}}, \alpha_1,\alpha_2, \omega, \alpha, \beta_1,\beta_2,\{\omega_c:c=1,2,\dots,C\}$.
\State Initialize $\boldsymbol{U}, \boldsymbol{z}, \beta, \boldsymbol{P},\boldsymbol{\nu}, \boldsymbol{X}$.
\For {$t=1$ to $T$}

\For {$i,j = 1,2,\dots,N$ and $i \neq j$}
\State \textcolor{red}{1.} Sample $\nu_{ij}$ from Eq.~\eqref{nu_Inference}.
\State \textcolor{red}{2.} Sample $x_{ij}$ from Eq.~\eqref{Aug_X} where $\lambda_{ij}$ is obtained from Eq.~\eqref{LPM}.
\EndFor

\begin{itemize}
\addtolength{\itemindent}{1em}
  \item[\textcolor{red}{3.1}] Propose $\beta' \sim \text{N}(\beta,\sigma^2_{\beta})$.
  \item[\textcolor{red}{3.2}] Based on Eq.~\eqref{Beta_fullCond}, accept $\beta = \beta'$ with probability $\text{min}\left(1,\frac{p(\beta'|\boldsymbol{X},\boldsymbol{U})}{p(\beta|\boldsymbol{X},\boldsymbol{U})}\right)$.
  \item[] Otherwise, set $\beta = \beta$.
\end{itemize}

\For {$i = 1,2,\dots,N$}
\State \textcolor{red}{4.1} Propose $\boldsymbol{u}_i' \sim \text{MVN}(\boldsymbol{u}_i,\sigma^2_{\boldsymbol{U}}\mathbbm{I}_d)$.
\State \textcolor{red}{4.2} Based on Eq.~\eqref{U_fullCond}, accept $\boldsymbol{u}_i = \boldsymbol{u}_i'$ with probability 
$$\text{min}\left(1,\frac{p\left(\boldsymbol{u}_i'|z_i, \beta,\boldsymbol{X},\boldsymbol{U}\backslash\{\boldsymbol{u}_i\}\right)}{p\left(\boldsymbol{u}_i|z_i, \beta,\boldsymbol{X},\boldsymbol{U}\backslash\{\boldsymbol{u}_i\}\right)}\right).$$
\State \hspace{1em} Otherwise, set $\boldsymbol{u}_i = \boldsymbol{u}_i$.
\EndFor

\vspace{0.5em}

\For {$i = 1,2,\dots,N$}
\State \textcolor{red}{5.} Each clustering variable $z_i$ is inferred to be $k$ with the normalized  
\State \hspace{1em} probability proportional to Eq.~\eqref{zi_FullCondLike} for $k=1,2,\dots,K^{-i}+1$.
\EndFor

\vspace{0.5em}
\State \textcolor{red}{6.} The truncated AE move is applied following Section~\ref{TAE}.
\vspace{0.5em}

\For {$g,h= 1,2,\dots,K$}
\State \textcolor{red}{7.} Infer $p_{gh}$ from Eq.~\eqref{P_Inference}.
\EndFor

\EndFor

\State \textbf{Output}: Posterior samples of $\boldsymbol{\nu}, \boldsymbol{X}, \beta, \boldsymbol{U}, \boldsymbol{z},\boldsymbol{P}, K$ for each iteration $t = 1,2,\dots,T$.
\end{algorithmic}
\end{algorithm}

According to \citet{van2008partially,park2009partially}, the ordering of the steps in Algorithm~\ref{PCMwG} matters in that the sampling of $\boldsymbol{X}$ must be performed after the sampling on $\boldsymbol{\nu}$; also, the inference on $\boldsymbol{P}$ must be performed after the step on $\boldsymbol{z}$ and after the truncated AE move.
If these requirements are not satisfied, the algorithm would yield a Markov chain targeting an unknown stationary distribution.\\

The posterior samples of interests are those of $\boldsymbol{\nu}, \beta, \boldsymbol{U}, \boldsymbol{z}$ and $K$.
Since the sampled values of each $\nu_{ij}$ are either 0s or 1s, the posterior mean of the $\nu_{ij}$, denoted as $\hat{\nu}_{ij}$, provides an approximation of the conditional probability in Eq.~\eqref{P_m0}.
Such an approximation takes into account the uncertainty generated by $\beta, \boldsymbol{U},\boldsymbol{z}$ and $\boldsymbol{P}$,
where $p_{z_iz_j}$ varies along with different posterior clustering $\boldsymbol{z}$ throughout the posterior samples.
Similarly, the posterior mean is also calculated for the intercept $\beta$ to obtain its summary statistic, $\hat{\beta}$.

As concerns the latent clustering variable, we may obtain a point estimate via
\begin{equation}
\label{hat_z}
\hat{\boldsymbol{z}}=\underset{\boldsymbol{z}'}{\arg\min}\;\mathbbm{E}_{\text{post}}[\mathcal{L}_{VI}(\boldsymbol{z},\boldsymbol{z}')|\boldsymbol{Y},\boldsymbol{c}],
\end{equation}
within which $\mathcal{L}_{VI}$ is the Variation of Information (VI) loss function \citep{meilua2007comparing,wade2018bayesian} defined as $\mathcal{L}_{VI}(\boldsymbol{z},\boldsymbol{z}') = 2H(\boldsymbol{z},\boldsymbol{z}')-H(\boldsymbol{z})-H(\boldsymbol{z}')$, measuring the log-ratio of the mutual and the individual entropies of the clusterings.
Here, $H(\boldsymbol{z})$ is the entropy of $\boldsymbol{z}$, and $H(\boldsymbol{z},\boldsymbol{z}')$ is the joint entropy of $\boldsymbol{z}$ and $\boldsymbol{z}'$.
Such a method is widely used in the literature and is shown to perform well for obtaining a point estimate of the clustering from a set of posterior samples, for example, \citet{rastelli2018optimal,legramanti2022extended,lu2025zero}.
An estimated $\hat{K}$ can thus be obtained based on the number of non-empty groups in $\hat{\boldsymbol{z}}$.
A natural approach to summarize posterior samples of latent positions $\boldsymbol{U}$ is to first apply Procrustes transformation \citep{borg2005modern} on each posterior sample with respect to a reference $\boldsymbol{U}$, and then to calculate the posterior mean from the posterior samples of each $\boldsymbol{u}_i$.
However, we find that, under LPCMs, the latent positions can still keep rotating within each individual group, especially when the groups are well separated. 
This creates further complications that are not resolved by the standard Procrustes transformations. 
For this reason, we opt for a more pragmatic approach and resort to 
obtain a point estimate of $\boldsymbol{U}$ via 
\begin{equation}
\label{hat_U}
\hat{\boldsymbol{U}}=\underset{\boldsymbol{U}}{\arg{\max}}{}_{\text{post}}\left[f(\boldsymbol{X}|\beta, \boldsymbol{U})f(\boldsymbol{\nu}|\boldsymbol{P},\boldsymbol{z})f(\boldsymbol{U}|\boldsymbol{z})f(\boldsymbol{z}|\boldsymbol{c})\mathbbm{1}(\boldsymbol{z}=\hat{\boldsymbol{z}})\right],
\end{equation}
which is the posterior latent positions from the posterior state which maximizes the complete likelihood function of the posterior in Eq.~\eqref{unCollapsedPosteriorX} among those states whose corresponding posterior clustering is identical to $\hat{\boldsymbol{z}}$.
In the case that no posterior clustering agrees with $\hat{\boldsymbol{z}}$, the posterior states after burn-in process are all considered.
However, this is not the case of any experiments we illustrate in the next sections.

%----------------------------------------------------------------------------------------------------------------------------------------------------------------------------------------------------------------------------------------------------------------------------------------
%----------------------------------------------------------------------------------------------------------------------------------------------------------------------------------------------------------------------------------------------------------------------------------------
%----------------------------------------------------------------------------------------------------------------------------------------------------------------------------------------------------------------------------------------------------------------------------------------
%----------------------------------------------------------------------------------------------------------------------------------------------------------------------------------------------------------------------------------------------------------------------------------------

\subsection{Choice of the number of latent dimensions}
\label{dimensions}
In our simulations and applications, we generally propose $d=3$ for the latent positions, aiming to provide 3-d visualizations of the complex networks.
Latent spaces with higher dimensions cannot be visualized in practice and are generally not required to represent most typical real networks, so $d=2$ or $d=3$ dimensions are usually more than sufficient.
In fact, in \citet{hoff2002latent}, the authors defined (for binary networks) a concept of representability that determines by which dimension $d$ of the latent positions a network is representable.
However, it is not easy to theoretically evaluate the value $d$, especially for complex networks, regardless of whether such a concept can be well extended to weighted networks. 
In any case, the authors argue that a small value of $d$ can be sufficient to represent a large variety of common social networks.
In this spirit, we perform our simulation studies in three dimensions, and, in the real data applications of Section~\ref{RDA}, we compare the $d=3$ performance to the corresponding $d=2$ performance for each real network we focus on, giving some nuanced views for the different cases.
We note that model choice for LPMs remains a critical research question. Our work contributes in this research direction by providing additional examples of the typical results that can be obtained in two and three dimensional LPMs.

\section{Simulation studies}
\label{SS}
In this section, we show the performance of the newly proposed ZIP-LPCM + MFM model in both a supervised and unsupervised setting.
For clarity, the inference algorithm of the unsupervised case is obtained by simply replacing Eq.~\eqref{MFM_urn_Sup} with Eq.~\eqref{MFM_urn} in Eq.~\eqref{zi_FullCondLike} within the Algorithm~\ref{PCMwG} Step 5, and by replacing Eq.~\eqref{MFM_PMF_Sup} with Eq.~\eqref{MFM_PMF} in Step 6 of the algorithm.
Further, we also make comparisons with the binary LPCM explored in \citet{ryan2017bayesian}, which we suitably extend to the non-negative weighted networks framework.
The extension of the binary LPCM to the weighted case is done by replacing the Bernoulli logistic link function with the Poisson distribution equipped with the link function in Eq.~\eqref{LPM}, leading to the Poisson LPCM (Pois-LPCM).
Note that the Pois-LPCM is a specific case of the ZIP-LPCM if we let the probability of unusual zeros for each pair of nodes be zero, corresponding to the situation that no unusual zero or missing data is assumed for the model.
Both supervised and unsupervised MFM priors are also proposed for the Pois-LPCM along with the truncated AE move to achieve a more fair performance comparison.
Thus the corresponding inference processes for the Pois-LPCM also follow Algorithm~\ref{PCMwG}, the only differences being the removal of Steps 1, 2, 7, as well as all the $\boldsymbol{\nu}$ terms in Steps 5, 6, and also treating $\boldsymbol{X}$ as $\boldsymbol{Y}$ instead.\\

We propose three simulation studies.
In simulation study 1, we randomly generate two artificial networks: one from a ZIP-LPCM (scenario 1) and one from a Pois-LPCM (scenario 2).
We implement the supervised and the unsupervised versions of ZIP-LPCM and of Pois-LPCM on the artificial network in each scenario, aiming to show that our newly proposed ZIP-LPCM is able to perform well when it is mis-specified to networks without unusual zeros or missing data.
We also aim to illustrate that the capablility of modeling unusual zeros plays a key role during the inference, leading to better performance of the ZIP-LPCM compared to the Pois-LPCM when we fit them to networks with unusual zeros.

Instead, in simulation study 2, we focus on synthetic networks which are generated from the Zero-Inflated Poisson Stochastic Block Model (ZIP-SBM) explored in \citet{lu2025zero}.
Also for this second simulation study, we consider two scenarios. 
In the first scenario we have a basic community structure, whereas in the second we also include hubs and thus disassortative patterns.
Our goal is to analyze whether the ZIP-LPCM is able to fit well to networks that generated from a different model, that is, the ZIP-SBM.
Indeed, it is known that the ZIP-SBM is able to characterize specific structures which are not possible to represent with the latent positions of the ZIP-LPCM (due to the inherent transitivity of the latent space), and thus may be more flexible when compared to the ZIP-LPCM.
However, the ZIP-SBM cannot capture much variability within the blocks, and cannot provide the latent space views of the networks, thus making the ZIP-LPCM a viable modeling choice.
Note that the simulation settings of all the models in this section mimic the structures of the real networks from our real data applications, thus, the simulation study performance provide valuable benchmarks for the analyses of real networks.

In simulation study 3, we further explore the robustness of our ZIP-LPCM approach performance by replicated experiments on different sampled networks with the same model settings.
We also investigate the generalization of our results to networks with larger network sizes.

\subsection{Simulation study 1}
\label{SS1}

\begin{figure}[h!]
\centering
\includegraphics[scale=0.335]{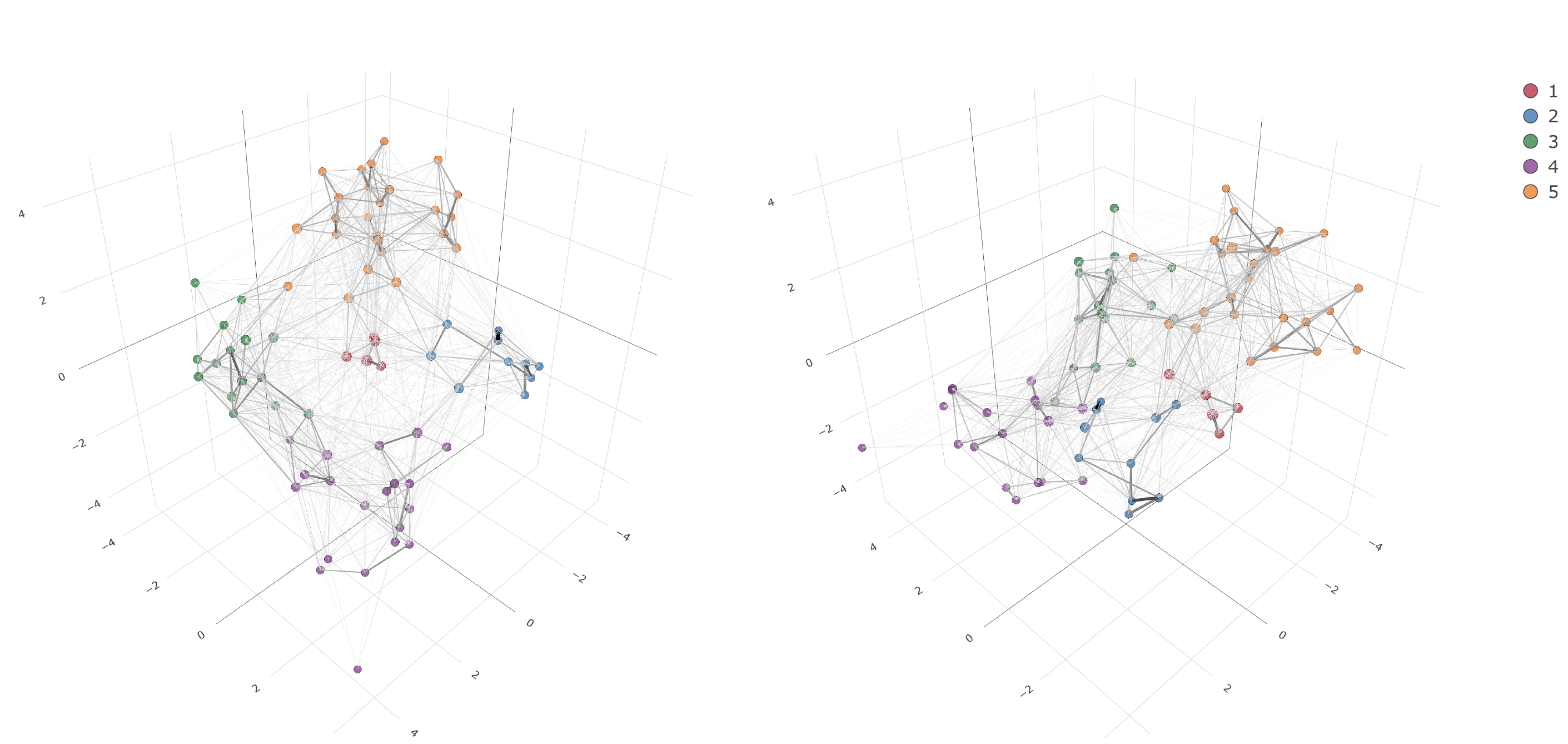}
\includegraphics[scale=0.33]{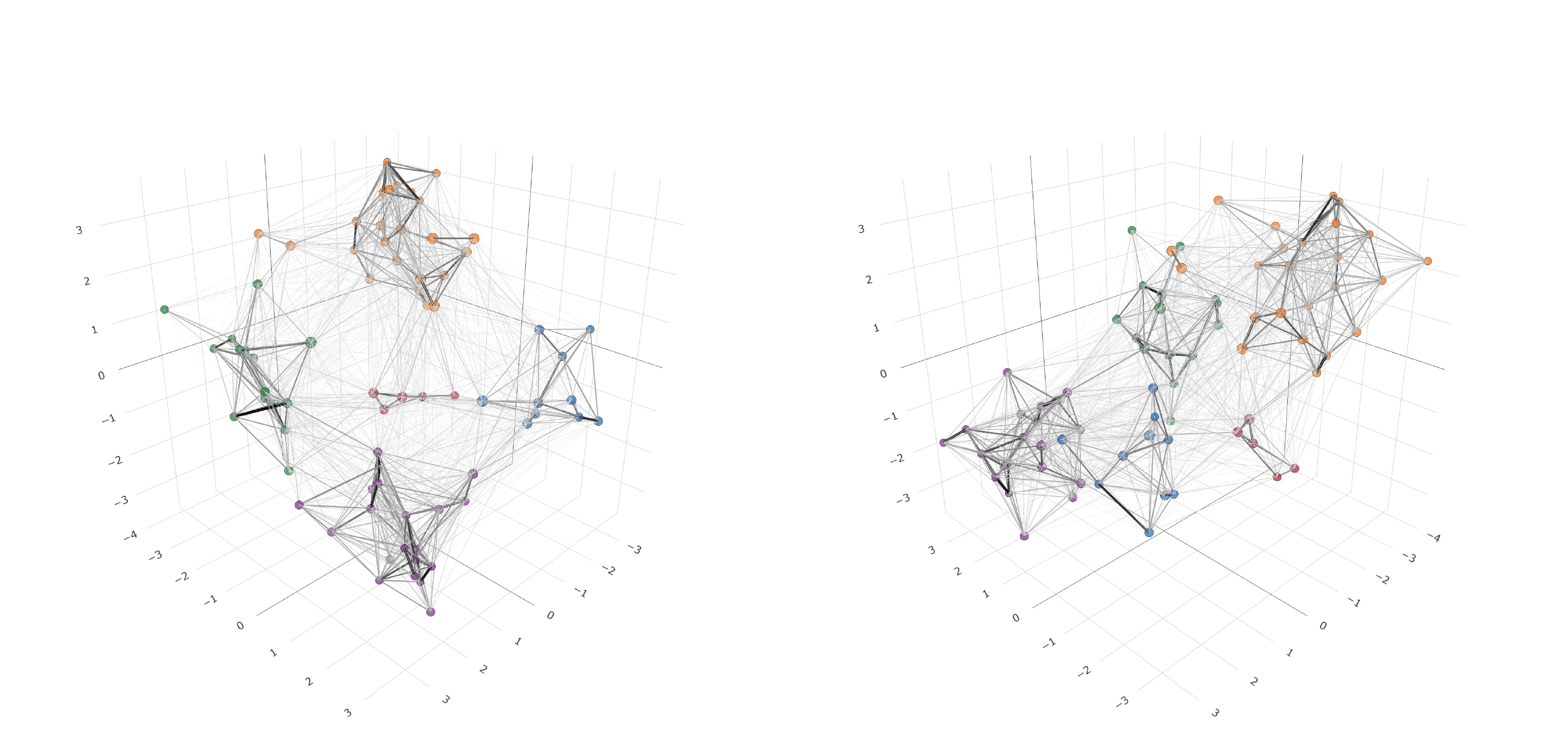}
\caption{Simulation study 1 synthetic networks. The 1st row plots correspond to the scenario 1 network, while the 2nd row plots correspond to the scenario 2 network. The 1st column plots show the 3-dimensional plots of the latent positions with different node colors denoting the corresponding true clustering. Node sizes are proportional to node betweeness, whereas edge widths and colors are proportional to edge weights. The 2nd column plots are the rotated plots of the 1st column latent position plots that are rotated for $90\degree$ clockwise with respect to the vertical axis.}
\label{SS1Sce1Sce2obsY}
\end{figure}

\begin{figure}[h!]
\centering
\includegraphics[scale=0.55]{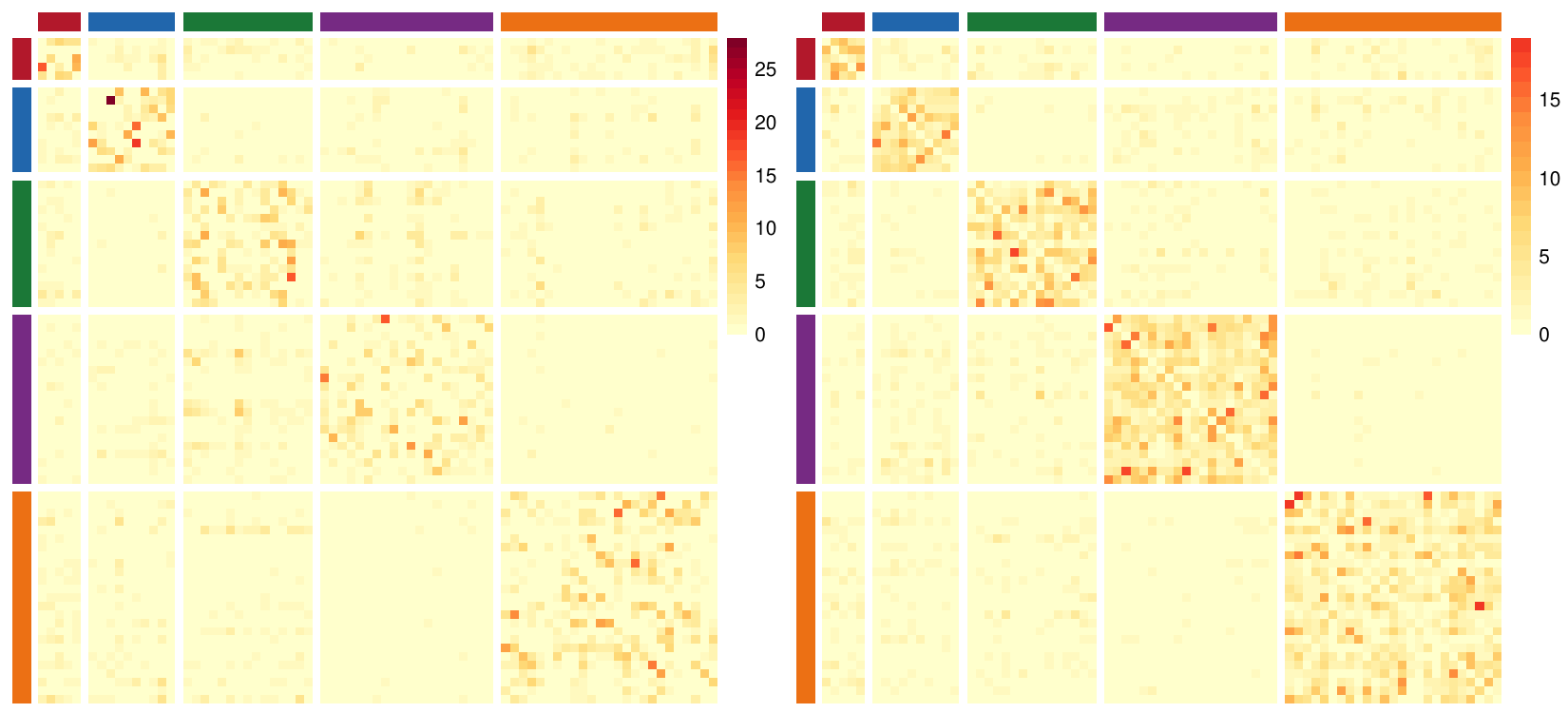}
\caption{Simulation study 1. Synthetic networks' adjacency matrix heatmap plots. Darker entries correspond to higher edge weights. The side-bars indicate the reference clustering $\boldsymbol{z}^*$. Left plot: scenario 1 network, generated from a ZIP-LPCM. Right plot: scenario 2 network, generated from a Pois-LPCM.}
\label{SS1obsYadj}
\end{figure}

We propose synthetic networks with $N=75$ nodes and $K=5$ clusters, where the true clustering $\boldsymbol{z}^*$ has group sizes: $n_1=5,n_2=10,n_3=15,n_4=20$ and $n_5=25$.
The network that we generate for scenario 1 is randomly simulated from a ZIP-LPCM with settings: $\beta=3, \boldsymbol{\tau} = (\frac{1}{0.25},\frac{1}{0.50},\frac{1}{0.75},1,\frac{1}{1.25})$ and
\begin{equation*}
\resizebox{.9\hsize}{!}{$
\boldsymbol{\mu} = \left[
\begin{pmatrix}
-1.5 \\-1.5\\-1.5
\end{pmatrix},
\begin{pmatrix}
 -2\\ 2\\0
\end{pmatrix},
\begin{pmatrix}
2\\-2\\0
\end{pmatrix},
\begin{pmatrix}
2\\2\\-2
\end{pmatrix},
\begin{pmatrix}
-2\\-2\\2
\end{pmatrix}\right],
\boldsymbol{P}=\begin{pmatrix}
0.40 & 0.05 & 0.10 & 0.05 & 0.10 \\
0.10 & 0.40 & 0.05 & 0.10 & 0.05 \\
0.05 & 0.10 & 0.40 & 0.05 & 0.10 \\
0.10 & 0.05 & 0.10 & 0.40 & 0.05 \\
0.05 & 0.10 & 0.05 & 0.10 & 0.40
\end{pmatrix}$}.
\end{equation*}
The network in scenario 2 is simulated from a Pois-LPCM with the same set of parameters shown above, but excluding the probability of unusual zeros $\boldsymbol{P}$.
In order to assess the performance of the estimation procedure, the above parameter values, which are used for generating the networks, are treated as the reference values to be compared to the corresponding posterior samples.
We further use $(\cdot)^*$ to denote these reference values, that is, $(K^*,\boldsymbol{z}^*,\beta^*,\boldsymbol{\tau}^*,\boldsymbol{\mu}^*,\boldsymbol{P}^*)$ indicate the true parameter values that have generated the data.
However, not all of these parameters are actually used, for example, the parameters $\boldsymbol{\tau}$ and $\boldsymbol{\mu}$ are marginalized out during the inference and thus they are not taken into account.
Figure~\ref{SS1Sce1Sce2obsY} illustrates the latent positions and the clustering used for generating the networks in two scenarios.
The pattern of the latent positions for the second scenario are similar to the first scenario one but with denser edges because no missing data is assumed. 
We refer to \url{ https://github.com/Chaoyi-Lu/ZIP-LPCM} for more details including the 3-d interactive plots of the latent positions as well as all the implementation code of the experiments in this paper.
Figure~\ref{SS1obsYadj} shows the heatmap plots of the adjacency matrices for both synthetic networks.

Prior settings similar to those of \citet{handcock2007model,ryan2017bayesian} are applied for all the experiments of this paper: $\alpha=3$ for the MFM and $\alpha_1=1,\alpha_2=0.103$ in Eq.~\eqref{tau_prior}.
The tuning parameter $\omega$ in Eq.~\eqref{mu_prior} is instead set as $0.01$, encouraging more split clusters in the latent space.
A common prior distribution of $\bar{K}$ in MFM is assumed as we discussed in Section~\ref{MFM_Sup}, that is, $\bar{K}\sim\text{Pois}(1)|\bar{K}>1$.
Based on the ZIP-SBM results illustrated in \citet{lu2025zero}, we also propose to set a $\text{Beta}(1,9)$ prior for the probability of unusual zeros by default in Eq.~\eqref{p_prior}, but we also check in the simulation studies the sensitivity of this prior setting by considering four other different options, that is, $\text{Beta}(1,1),\text{Beta}(1,3),\text{Beta}(1,19)$ and $\text{Beta}(1,99)$.

As regards the node attributes, a contaminated version of the true clustering $\boldsymbol{z}^*$ is used as the exogenous node attributes $\boldsymbol{c}$ where the clustering of 20 out of 75 nodes are randomly reallocated. 
This setting aims to check whether the output are robust when noise exists in the node attributes.
The prior of $\beta$ is proposed to be a continuous uniform distribution defined on a large enough interval centered around zero so that the corresponding probability density function can always be canceled in the acceptance ratio of Step 3.2 of Algorithm~\ref{PCMwG}.
The cohesion parameter of the supervised MFM is canonically set as $\omega_c=1$ for $c=1,2,\dots,C$.

With regard to the MCMC posterior samples, we run the algorithm for 12,000 iterations, where the first 2000 iterations are discarded as burn-in for each setting that we consider.
The convergence and mixing of the posterior samples of the ZIP-LPCM is monitored by checking the trace plots of the corresponding values of the complete likelihood function of the posterior in Eq.~\eqref{CollapsedPosteriorX}, that is, 
$$f(\boldsymbol{Y},\boldsymbol{X},\boldsymbol{\nu},\boldsymbol{U},\boldsymbol{z}|\beta,\boldsymbol{P},\boldsymbol{c})= f(\boldsymbol{X}|\beta, \boldsymbol{U})f(\boldsymbol{\nu}|\boldsymbol{P},\boldsymbol{z})f(\boldsymbol{U}|\boldsymbol{z})f(\boldsymbol{z}|\boldsymbol{c}).$$
The complete likelihood of the Pois-LPCM is obtained by simply removing the $f(\boldsymbol{\nu}|\boldsymbol{P},\bm{z})$ term and by replacing $\boldsymbol{X}$ with $\boldsymbol{Y}$ in the above equation.
In practice, we observe that the posterior samples of all implementations in simulation study 1 and simulation study 2 quickly converge to the stationary distribution within approximately 2000 iterations, however we conservatively use a much larger number of iteration to ensure satisfactory convergence.

The initial clustering is proposed to be the trivial clustering where each node occupies a different group.
This leads to higher computational burden required in early iterations of the implementations: on a laptop equipped with a Intel Core i7 1.80GHz CPU, the supervised ZIP-LPCM algorithm implemented for 2000 iterations in this simulation study 1 take around 5.5 mins to finish, the 6000-iteration implementations take approximately 14 mins, and the 12,000-iteration implementations that we mainly focus on in practice take around 24 mins.
The execution time of the unsupervised implementations is generally 20\% less than that of the supervised implementations, and the time spent by Pois-LPCM algorithm executions is approximately 12.5\% less than that of the ZIP-LPCM cases.

Latent positions are initialized by a natural approach where classical multidimensional scaling \citep{gower1966some} is applied on the geodesic distance matrix of the observed adjacency matrix, in the same style as \citet{hoff2002latent}.
The proposal variances of M-H steps of $\beta$ and $\boldsymbol{U}$ are tuned so that the corresponding acceptance rates are approximately 0.23.
The probability of applying an $eject^T$ move is set as $\text{P}(eject^T)=0.5$ by default.\\

Recall that a point estimate of the posterior clustering and of the latent positions is, respectively, obtained by Eq.~\eqref{hat_z} and Eq.~\eqref{hat_U}.
The set of distances $\{d_{ij}=||\boldsymbol{u}_i-\boldsymbol{u}_j||:i,j=1,2,\dots,N;i>j\}$ between each pair of nodes are invariant under any rotation or translation of $\boldsymbol{U}$, so that the set of $\{\hat{d}_{ij}\}$ can be obtained via posterior mean of each $d_{ij}$ in the posterior samples, accounting for the uncertainty of all other model parameters and latent variables.
We treat the distances obtained from $\boldsymbol{U}^*$ as the corresponding reference values to be compared to those $\{\hat{d}_{ij}\}$, that is, $\{d^*_{ij}=||\boldsymbol{u}_i^*-\boldsymbol{u}_j^*||:i,j=1,2,\dots,N;i>j\}$.
Similarly, the probability of unusual zeros $\boldsymbol{P}$ is, by definition, composed by $\{p_{gh}:g,h=1,2,\dots,K\}$ for each pair of non-empty clusters and is dependent on the clustering.
Due to the fact that the posterior clustering keeps mixing during the inference leading to different number of clusters, the identifiability problem occurs in the posterior samples of $\boldsymbol{P}$.
However, instead of such a group-level parameter, we focus on an individual-level $N \times N$ matrix, $\boldsymbol{p}$, within which each $ij$th entry is the probability of unusual zero $p_{z_iz_j}$, and we obtain the corresponding $\hat{p}_{z_iz_j}$ via posterior mean accounting for the uncertainty of the posterior clustering.
The $p^*_{z_iz_j}$ is proposed to be zero for each pair of nodes $i,j$ in scenario 2 by considering that the Pois-LPCM is a specific case of the ZIP-LPCM.

The performance of eight different cases are illustrated in Table~\ref{SS1table}.
The table shows that the supervised implementations are able to provide better clustering performance with lower uncertainty in both scenarios as expected, even though there is significant contamination existing in the exogenous node attributes.
The Pois-LPCM cases fail to recover the true clustering and are shown to be unable to fit well to the network generated from the ZIP-LPCM in scenario 1.
On the contrary, the ZIP-LPCM provides good performance in both scenario 1 and 2, especially when the parameter $\beta_2$ of the prior of the probability of unusual zeros is set to be around $9$.
In the case that $\beta_2=99$, the prior encourages very small probability of unusual zeros, and this makes the ZIP-LPCM become close to the Pois-LPCM leading to small $\{\hat{p}_{z_iz_j}\}$ shown as, for example, $\mathbbm{E}(\{|\hat{p}_{z_iz_j}-p^*_{z_iz_j}|\})${\scriptsize [\text{sd}]} where $p^*_{z_iz_j}=0$ in scenario 2 of Table~\ref{SS1table}.
Moreover, the inferred $\hat{\boldsymbol{z}}$ from the cases of ``ZIP-LPCM Sup \text{Beta}(1,99)'' and ``Pois-LPCM Sup'' is also very similar in scenario 1.
Though the estimated clustering is not detailed here, the materials can be provided upon request or following the provided code in the \textbf{Code and Data} section to reproduce the output.\\
\setlength{\tabcolsep}{12pt}
\begin{table}[h!]
\renewcommand{\arraystretch}{0.9}
\centering
\caption{\footnotesize{Simulation study 1. Performance of eight different implementations where (\lowerromannumeral{1}) $\hat{K}$: the number of clusters in $\hat{\boldsymbol{z}}$;  (\lowerromannumeral{2}) $\text{VI}(\hat{\boldsymbol{z}},\boldsymbol{z}^*)$: the VI distance between the point estimate $\hat{\boldsymbol{z}}$ and the true clustering $\boldsymbol{z}^*$; (\lowerromannumeral{3}) $\mathbbm{E}_{\boldsymbol{z}}[\text{VI}(\hat{\boldsymbol{z}},\boldsymbol{z}) \mid \boldsymbol{Y}]$: the minimized expected posterior VI loss of the clustering with respect to $\hat{\boldsymbol{z}}$. This statistic measures the uncertainty of the posterior clustering around the $\hat{\boldsymbol{z}}$; (\lowerromannumeral{4}) $\mathbbm{E}(\{|\hat{d}_{ij}-d^*_{ij}|\})${\scriptsize [\text{sd}]}: the mean of $\{|\hat{d}_{ij}-d^*_{ij}|: i,j=1,2,\dots,N; i>j\}$ with the corresponding standard deviation (sd) shown in the square bracket; (\lowerromannumeral{5}) $\hat{\beta}$: the posterior mean of $\beta$; (\lowerromannumeral{6}) $\mathbbm{E}(\{|\hat{p}_{z_iz_j}-p^*_{z_iz_j}|\})${\scriptsize [\text{sd}]}: the mean of $\{|\hat{p}_{z_iz_j}-p^*_{z_iz_j}|: i,j=1,2,\dots,N; i>j\}$ with sd in the square bracket. More details are included in Section~\ref{SS1}. The best performance within each column are highlighted in bold font.}}
\begin{adjustbox}{width=1.02\textwidth,center=\textwidth}
\begin{tabular}[c]{c|cc|cc|cc|cc|cc|cc}
\multicolumn{1}{c}{} & \multicolumn{2}{c}{$\hat{K}$} & \multicolumn{2}{c}{$\text{VI}(\hat{\boldsymbol{z}},\boldsymbol{z}^*)$}  &  \multicolumn{2}{c}{$\mathbbm{E}_{\boldsymbol{z}}[\text{VI}(\hat{\boldsymbol{z}},\boldsymbol{z}) \mid \boldsymbol{Y}]$}  & \multicolumn{2}{c}{$\mathbbm{E}(\{|\hat{d}_{ij}-d^*_{ij}|\})${\scriptsize [\text{sd}]}} &   \multicolumn{2}{c}{$\hat{\beta}$}  & \multicolumn{2}{c}{$\mathbbm{E}(\{|\hat{p}_{z_iz_j}-p^*_{z_iz_j}|\})${\scriptsize [\text{sd}]}}  \\
\midrule
\textsc{Scenario} & 1 & 2 & 1 & 2 & 1 & 2 & 1 & 2&1&2&1& 2\\
\midrule
ZIP-LPCM Sup \text{Beta}(1,1)& 7 & {\bf5} & 0.68 & {\bf0.00} &  0.551 & 0.031 &  {0.359{\scriptsize[0.266]}} &  {1.356{\scriptsize[1.097]}} &  2.95 & 2.93 &  {0.193{\scriptsize[0.158]}} &  {0.157{\scriptsize[0.127]}}\\
ZIP-LPCM Sup \text{Beta}(1,3)& {\bf5} & {\bf5} & {\bf0.00} & {\bf0.00} &  0.018 & {\bf0.008} &  {0.250{\scriptsize[0.203]}} &  {1.347{\scriptsize[1.088]}} &  3.03 & {\bf2.96} &  {0.084{\scriptsize[0.065]}} &  {0.116{\scriptsize[0.081]}}\\
\midrule
ZIP-LPCM Sup \text{Beta}(1,9)& {\bf5} & {\bf5} & {\bf0.00} & {\bf0.00} &  0.009 & 0.011 &  {\bf0.249{\scriptsize[0.201]}} &  {1.341{\scriptsize[1.081]}} &  {\bf3.02} & {\bf2.96} &  {0.033{\scriptsize[0.034]}} &  {0.074{\scriptsize[0.050]}}\\
ZIP-LPCM unSup \text{Beta}(1,9)& {\bf5} & {\bf5} & {\bf0.00} & {\bf0.00} &  0.072 & 0.052 &  {0.250{\scriptsize[0.202]}} &  {1.339{\scriptsize[1.080]}} &  {\bf3.02} & {\bf2.96} &  {0.032{\scriptsize[0.031]}} &  {0.069{\scriptsize[0.043]}}\\
\midrule
ZIP-LPCM Sup \text{Beta}(1,19)& {\bf5} & {\bf5} & {\bf0.00} & {\bf0.00} &  {\bf0.005} & 0.021 &  {0.262{\scriptsize[0.210]}} &  {1.334{\scriptsize[1.074]}} &  3.03 & 2.93 &  {\bf0.029{\scriptsize[0.025]}} &  {0.039{\scriptsize[0.019]}}\\
ZIP-LPCM Sup \text{Beta}(1,99)& {\bf5} & {\bf5} & 0.27 & {\bf0.00} &  0.023 & 0.029 &  {0.290{\scriptsize[0.229]}} &  {1.334{\scriptsize[1.074]}} &  {\bf3.02} & {\bf2.96} &  {0.084{\scriptsize[0.055]}} &  {\bf0.009{\scriptsize[0.002]}}\\
\midrule
Pois-LPCM Sup & {\bf5} & {\bf5} & 0.42 & {\bf0.00} & 0.369 & 0.025 & {0.384{\scriptsize[0.293]}} & {\bf1.322{\scriptsize[1.070]}} & 2.62 & 2.91 & -- & -- \\
Pois-LPCM unSup & 2 & {\bf5} & 1.53 & {\bf0.00} & 1.086 & 0.072 & {0.400{\scriptsize[0.320]}} & {1.333{\scriptsize[1.072]}} & 2.63 & {\bf2.96} & -- & -- \\
\midrule
\end{tabular}
\end{adjustbox}
\label{SS1table}
\end{table}

The results also highlight that the ZIP-LPCM implementations with higher prior mean of the probability of unusual zeros tend to overestimate $K$ on ZIP-LPCM networks.
For example, in scenario 1 the posterior clustering of the ``ZIP-LPCM Sup \text{Beta}(1,1)'' tends to split some of the groups into smaller subgroups, leading to $\hat{K}=7$ against $K^*=5$, even though such an implementation is supervised and thus uses some extra information.
On the contrary, the unsupervised Pois-LPCM underestimates $K$: the posterior clustering of the ``Pois-LPCM unSup'' merges several groups together and provides $\hat{K}=2$.
Considering that the Pois-LPCM is in fact an extreme case of the ZIP-LPCM wherein all the probability of unusual zeros is set to be zero, this overestimated/underestimated $K$ performance correspond to the fact that the probability of unusual zeros controls the sparsity of the network, where larger probability corresponds to sparser latent positions bringing more separated clusters, and vice-versa.

\begin{figure}[h!]
\centering
\includegraphics[scale=0.475]{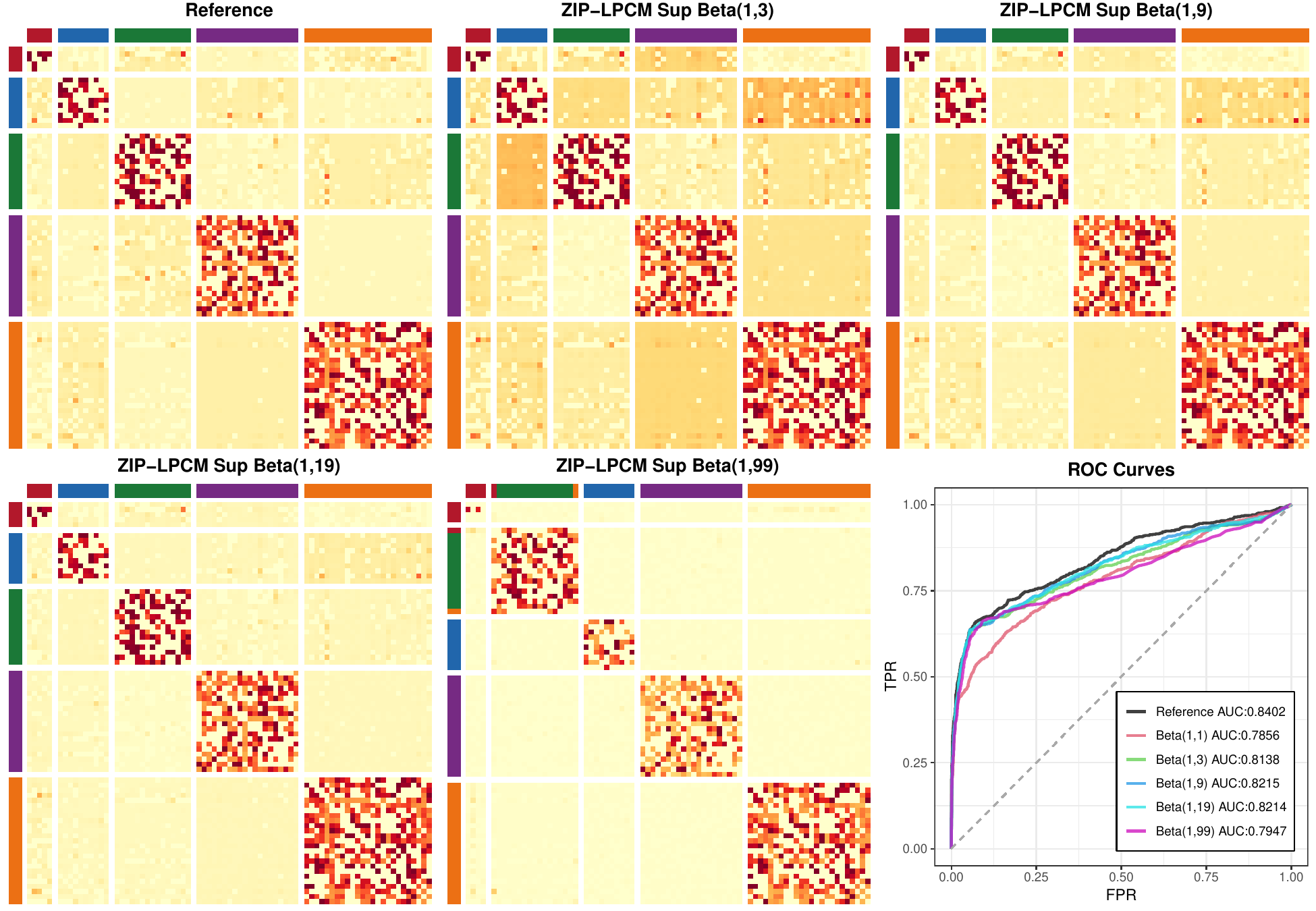}
\caption{Simulation study 1 scenario 1. Performance of the posterior mean $\hat{\boldsymbol{\nu}}$, which approximates the conditional probability in Eq.~\eqref{P_m0}.
The top-left plot is the heatmap of the reference values of Eq.~\eqref{P_m0}, obtained by leveraging the reference model parameters used for simulating the network, whereby darker entry colors correspond to higher values.
The other four heatmap plots describe $\hat{\boldsymbol{\nu}}$ as inferred by the corresponding priors indicated on top of the heatmap.
The rows and columns of the matrices are rearranged and separated according to $\hat{\boldsymbol{z}}$ while the side-bars indicate the true clustering of each individual.
The last plot shows the Receiver Operating Characteristic (ROC) curves for all the supervised ZIP-LPCM cases, where the reference $\boldsymbol{\nu}^*$ is the response variable.}
\label{SS1ROC}
\end{figure}

Each element $\nu_{ij}$ of the indicators of unusual zeros $\boldsymbol{\nu}$ is defined to be either 1 or 0.
Hence, the posterior mean of each $\nu_{ij}$ measures the proportion of the times that the corresponding $y_{ij}$ is inferred as an unusual zero.
In the case that $y_{ij}>0$, the posterior samples of $\nu_{ij}$ are all inferred to zero by default.
Denoting $\hat{\nu}_{ij}$ as the posterior mean of $\nu_{ij}$, then $\hat{\nu}_{ij}$ becomes an approximation of the conditional probability of the unusual zeros in Eq.~\eqref{P_m0} accounting for the posterior uncertainty of all other model parameters and latent variables.
The performance of $\hat{\nu}_{ij}$ for the supervised ZIP-LPCM cases from scenario 1 are shown in Figure~\ref{SS1ROC}. 
We note that the $\text{Beta}(1,1)$ case is excluded because its summarized clustering is not satisfactory.
The figure shows that the $\text{Beta}(1,9)$ case provide the best matching and the best ROC curves with respect to the references, while slightly tuning the $\beta_2$ provides comparable performance.
However, it is also interesting that the $\text{Beta}(1,99)$ prior also successfully and correctly prioritizes many zero interactions which are more likely to be unusual zeros compared to other zeros.
In fact, more non-zero interactions exist in a particular block, more accurate the pairwise distances between the nodes inside are, hence bringing more robust inference results of how likely the zero interactions in the block are unusual zeros.
Note here that in general smaller $p_{ij}$ does not imply smaller value of Eq.~\eqref{P_m0}: that also depends on $\beta$ and $\boldsymbol{U}$.
Hence, from a practical standpoint we suggest the $\text{Beta}(1,9)$ as the default prior setting for the probability of unusual zeros, whereas priors with smaller means can be considered when practitioners expect a more clustered network or prefer to learn more towards true zeros rather than unusual zeros.
By contrast, higher prior mean is encouraged if practitioners expect sparser network architecture bringing more subgroup features.
However, moderate tuning of the prior parameters does not significantly affect the overall performance as shown in both Figure~\ref{SS1ROC} and Table~\ref{SS1table}.

\subsection{Simulation study 2}
\label{SS2}

In this second simulation study, the network size $N$ and the clustering $\boldsymbol{z}^*$ of the synthetic networks are the same as those proposed in the first simulation study.
The networks are randomly simulated from the ZIP-SBM which can be obtained by removing the latent position parts in Eq.~\eqref{LPM} and in Eq.~\eqref{LPCM} from the ZIP-LPCM, and directly proposing a $\text{Pois}(\lambda_{z_iz_j})$ in Eq.~\eqref{Aug_ZI}.
The model parameter $\boldsymbol{\lambda}$ is a $K \times K$ matrix with entries $\{\lambda_{gh}:g,h=1,2,\dots,K\}$ being the Poisson rates defined for the interactions between any two occupied groups.

\begin{figure}[h!]
\centering
\includegraphics[scale=0.55]{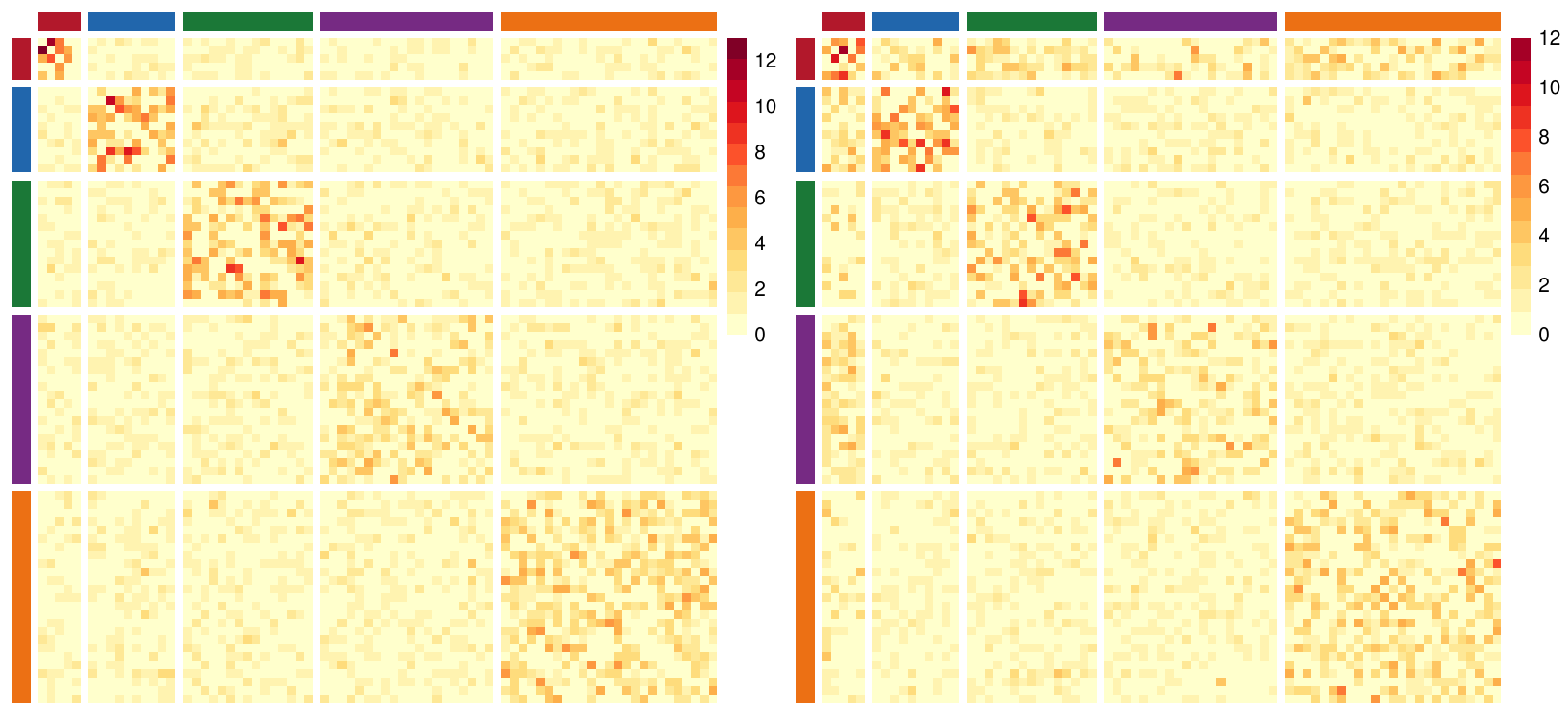}
\caption{Simulation study 2. Synthetic networks' adjacency matrix heatmap plots. Darker entries correspond to higher edge weights. The side-bars indicate the reference clustering $\boldsymbol{z}^*$. Left plot: scenario 1 network, generated from a ZIP-SBM without hubs. Right plot: scenario 2 network, generated from a ZIP-SBM with hubs.}
\label{SS2obsYadj}
\end{figure}

In scenario 1 of simulation study 2, the values of $\boldsymbol{\lambda}$ used for generating the network are indicated with $\boldsymbol{\lambda_1}$ in the equation shown below, whereas the probability of unusual zeros $\boldsymbol{P}$ is the same as the one in Section~\ref{SS1}.
\begin{equation*}
\resizebox{.975\hsize}{!}{
$\boldsymbol{\lambda_1}=\begin{pmatrix}7.0 & 0.5 & 0.5 & 0.5 & 0.5 \\ 0.5 & 4.5 & 0.5 & 0.5 & 0.5\\0.5 & 0.5 & 3.5 & 0.5 & 0.5 \\0.5 & 0.5 & 0.5 & 2.0 & 0.5 \\ 0.5 & 0.5 & 0.5 & 0.5 & 2.5\end{pmatrix},\boldsymbol{\lambda_2}=\begin{pmatrix}7.0 & 2.0 & 2.0 & 2.0 & 2.0 \\ 2.0 & 4.5 & 0.5 & 0.5 & 0.5\\2.0 & 0.5 & 3.5 & 0.5 & 0.5 \\2.0 & 0.5 & 0.5 & 2.0 & 0.5 \\ 2.0 & 0.5 & 0.5 & 0.5 & 2.5\end{pmatrix},\boldsymbol{P_2}=\begin{pmatrix}0.40 & 0.60 & 0.20 & 0.60 & 0.20 \\ 0.20 & 0.40 & 0.05 & 0.10 & 0.05\\0.60 & 0.10 & 0.40 & 0.05 & 0.10 \\0.20 & 0.05 & 0.10 & 0.40 & 0.05 \\ 0.60 & 0.10 & 0.05 & 0.10 & 0.40\end{pmatrix}$}.
\end{equation*}
The values indicated with $\boldsymbol{\lambda_2}$ and $\boldsymbol{P_2}$ describe instead the setup for the second scenario of this simulation study, whereby the interaction rate matrix $\boldsymbol{\lambda_2}$ includes a group of hubs, that is, nodes that have relatively high connection rates to all other groups.
The heatmap plots of the networks' adjacency matrices from both scenarios are shown in Figure~\ref{SS2obsYadj}.
In practice, we observe that the execution time of our approach is affected by the complexity of the observed networks, whereby all the simulation study 2 implementations require approximately 10\% less time compared to those in simulation study 1.

\begin{figure}[h!]
\centering
\includegraphics[scale=0.335]{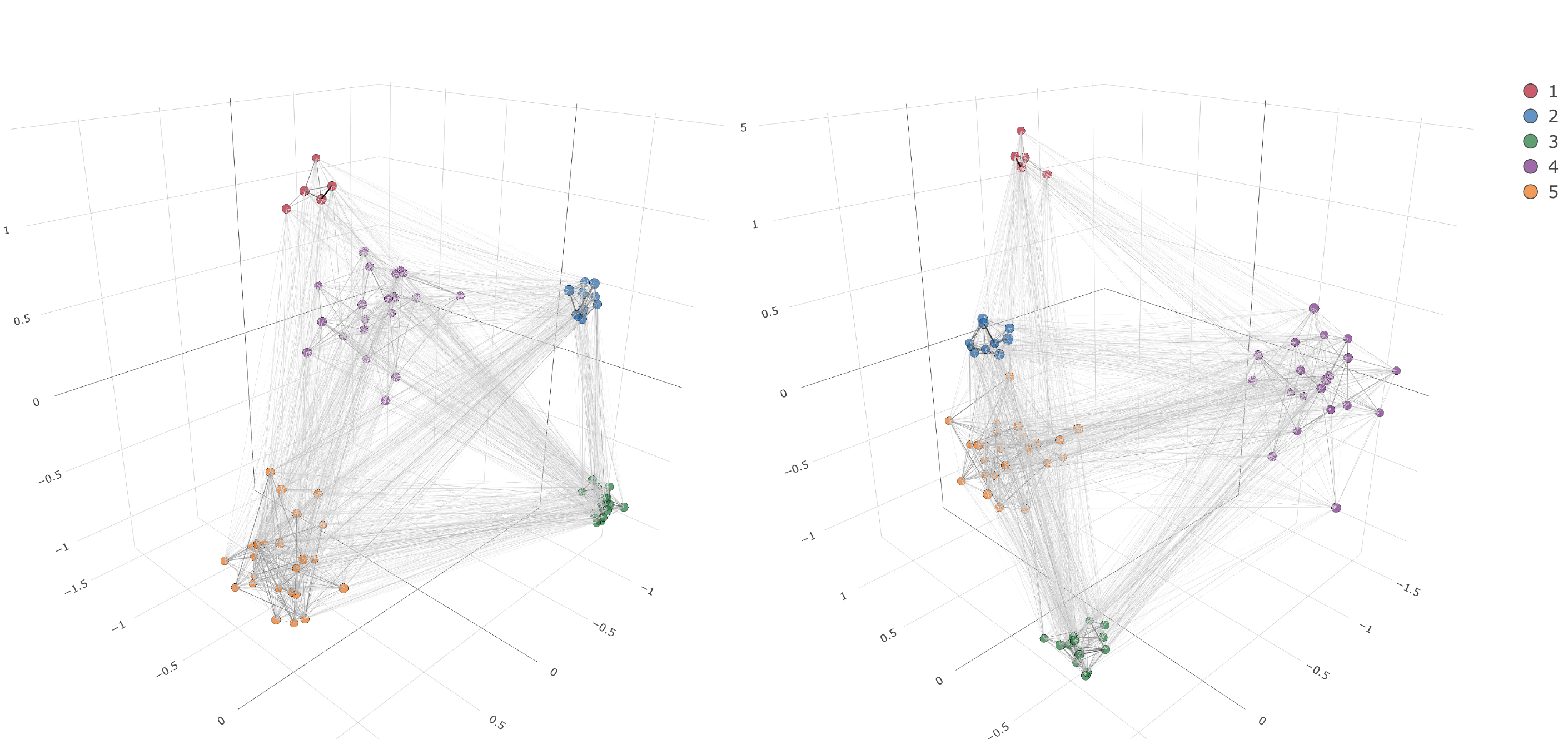}

\vspace{0.5em}

\includegraphics[scale=0.33]{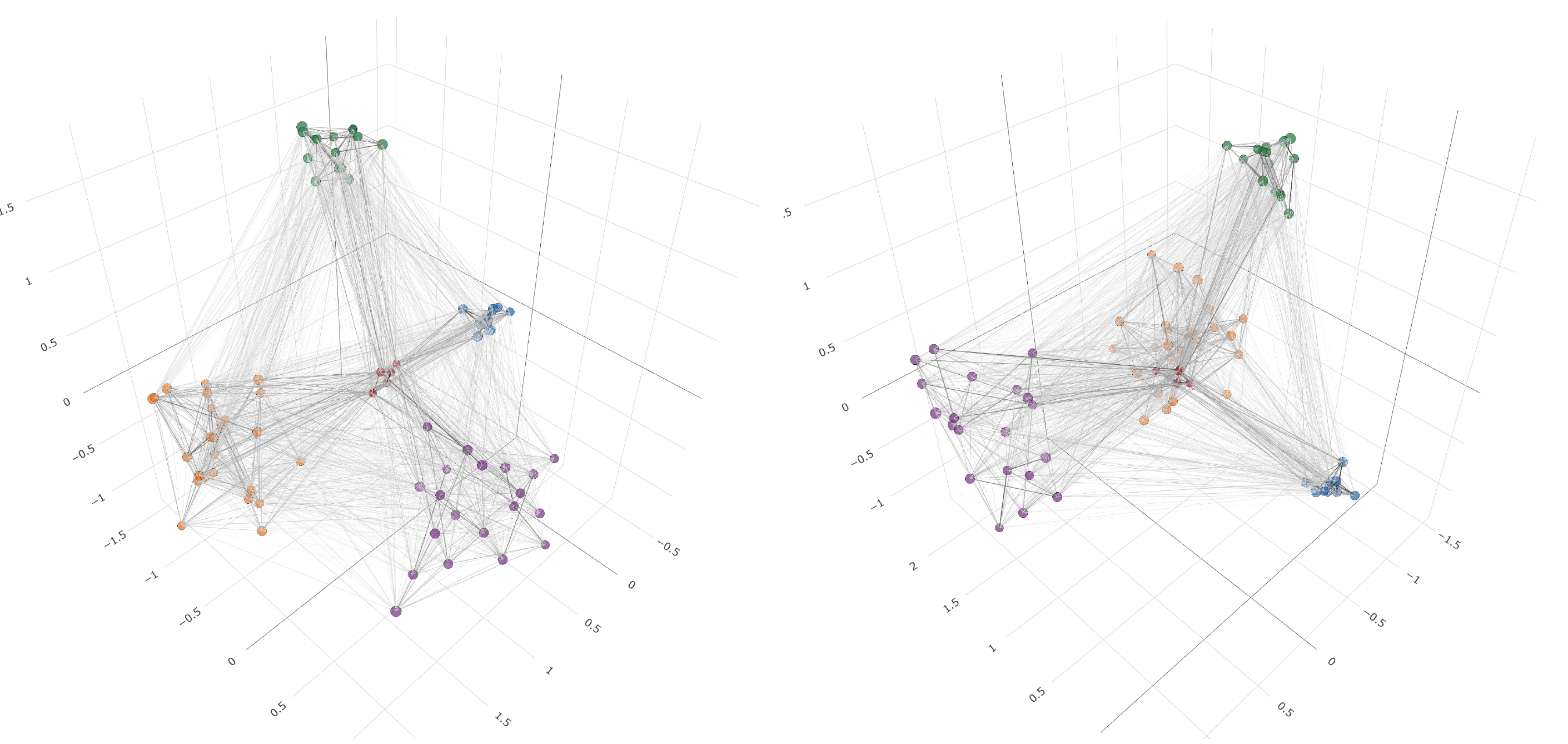}
\caption{Simulation study 2. The 1st and the 2nd rows illustrate the inferred point estimate $\hat{\boldsymbol{U}}$ obtained by ZIP-LPCM Sup Beta(1,9) implementations for Scenario 1 and Scenario 2, respectively. The 2nd column plots are rotated version of the 1st column plots where each inferred latent position rotated for $90\degree$ clockwise with respect to the vertical axis. Different node colors correspond to different inferred groups according to the corresponding $\hat{\boldsymbol{z}}$.
Node sizes are proportional to node betweenness while edge widths and colors are proportional to edge weights. }
\label{SS2_Obs_hatU}
\end{figure}
\setlength{\tabcolsep}{12pt}
\begin{table}[h!]
\renewcommand{\arraystretch}{0.9}
\centering
\caption{\footnotesize{Simulation study 2. Performance of eight different implementations where (\lowerromannumeral{1}) $\hat{K}$: the number of clusters in $\hat{\boldsymbol{z}}$; (\lowerromannumeral{2}) $\text{VI}(\hat{\boldsymbol{z}},\boldsymbol{z}^*)$: the VI distance between the point estimate $\hat{\boldsymbol{z}}$ and the true clustering $\boldsymbol{z}^*$; (\lowerromannumeral{3}) $\mathbbm{E}_{\boldsymbol{z}}[\text{VI}(\hat{\boldsymbol{z}},\boldsymbol{z}) \mid \boldsymbol{Y}]$: the minimized expected posterior VI loss of the clustering with respect to $\hat{\boldsymbol{z}}$. This statistic measures the uncertainty of the posterior clustering around the $\hat{\boldsymbol{z}}$; (\lowerromannumeral{4}) $\hat{\beta}$: the posterior mean of $\beta$; (\lowerromannumeral{5}) $\mathbbm{E}(\{|\hat{p}_{z_iz_j}-p^*_{z_iz_j}|\})${\scriptsize [\text{sd}]}: the mean of $\{|\hat{p}_{z_iz_j}-p^*_{z_iz_j}|: i,j=1,2,\dots,N; i>j\}$ with the corresponding standard deviation (sd) shown in the square bracket; (\lowerromannumeral{6}) $\mathbbm{E}(\{|\hat{\lambda}_{ij}-\lambda^*_{ij}|\})${\scriptsize [\text{sd}]}: the mean of $\{|\hat{\lambda}_{ij}-\lambda^*_{ij}|: i,j=1,2,\dots,N; i>j\}$ with the corresponding sd in the square bracket. More details are included in Section~\ref{SS2}. The best performance within each column excluding the $\hat{\beta}$ column are highlighted in bold font.}}
\begin{adjustbox}{width=1.02\textwidth,center=\textwidth}
\begin{tabular}[c]{c|cc|cc|cc|cc|cc|cc}
\multicolumn{1}{c}{} & \multicolumn{2}{c}{$\hat{K}$} & \multicolumn{2}{c}{$\text{VI}(\hat{\boldsymbol{z}},\boldsymbol{z}^*)$}  &  \multicolumn{2}{c}{$\mathbbm{E}_{\boldsymbol{z}}[\text{VI}(\hat{\boldsymbol{z}},\boldsymbol{z}) \mid \boldsymbol{Y}]$} &   \multicolumn{2}{c}{$\hat{\beta}$}  & \multicolumn{2}{c}{$\mathbbm{E}(\{|\hat{p}_{z_iz_j}-p^*_{z_iz_j}|\})${\scriptsize [\text{sd}]}} & \multicolumn{2}{c}{$\mathbbm{E}(\{|\hat{\lambda}_{ij}-\lambda^*_{ij}|\})${\scriptsize [\text{sd}]}} \\
\midrule
\textsc{Scenario} & 1 & 2 & 1 & 2 & 1 & 2 & 1 & 2&1&2&1& 2\\
\midrule
ZIP-LPCM Sup \text{Beta}(1,1)& {\bf5} & {\bf5} & {\bf0.00} & {\bf0.00} &  {\bf0.00} & 0.17 &  1.42 & 1.60 &  {0.076{\scriptsize[0.069]}} &  {0.092{\scriptsize[0.078]}} & {0.152{\scriptsize[0.261]}} & {0.245{\scriptsize[0.299]}}\\
ZIP-LPCM Sup \text{Beta}(1,3)& {\bf5} & {\bf5} & {\bf0.00} & {\bf0.00} &  {\bf0.00} & {\bf0.00} &  1.42 & 1.56 &  {0.063{\scriptsize[0.071]}} &  {0.073{\scriptsize[0.067]}} & {0.149{\scriptsize[0.259]}} & {0.243{\scriptsize[0.306]}}\\
\midrule
ZIP-LPCM Sup \text{Beta}(1,9)& {\bf5} & {\bf5} & {\bf0.00} & {\bf0.00} &  {\bf0.00} & {\bf0.00} &  1.40 & 1.55 &  {0.051{\scriptsize[0.048]}} &  {0.059{\scriptsize[0.061]}} & {0.149{\scriptsize[0.265]}} & {0.240{\scriptsize[0.318]}}\\
ZIP-LPCM unSup \text{Beta}(1,9)& {\bf5} & {\bf5} & {\bf0.00} & {\bf0.00} &  {\bf0.00} & {\bf0.00} &  1.42 & 1.55 &  {0.042{\scriptsize[0.038]}} &  {0.060{\scriptsize[0.058]}} & {0.145{\scriptsize[0.260]}} & {\bf0.239{\scriptsize[0.319]}}\\
\midrule
ZIP-LPCM Sup \text{Beta}(1,19)& {\bf5} & {\bf5} & {\bf0.00} & {\bf0.00} &  {\bf0.00} & {\bf0.00} &  1.39 & 1.54 &  {\bf0.037{\scriptsize[0.026]}} &  {\bf0.056{\scriptsize[0.071]}} & {\bf0.141{\scriptsize[0.267]}} & {0.239{\scriptsize[0.329]}}\\
ZIP-LPCM Sup \text{Beta}(1,99)& {\bf5} & {\bf5} & {\bf0.00} & 0.42 &  0.19 & 0.08&  1.37 & 1.53 &  {0.082{\scriptsize[0.049]}} &  {0.112{\scriptsize[0.126]}} & {0.151{\scriptsize[0.288]}} & {0.287{\scriptsize[0.413]}}\\
\midrule
Pois-LPCM Sup & {\bf5} & {\bf5} & {\bf0.00} & {\bf0.00} & 0.07 & 0.05 & 1.03 & 1.15 & -- & -- & {0.312{\scriptsize[0.511]}} & {0.461{\scriptsize[0.564]}}\\
Pois-LPCM unSup & {\bf5} & 4 & {\bf0.00} & 0.40 & 0.21 & 0.23& 1.02 & 1.16 & -- & -- & {0.312{\scriptsize[0.514]}} & {0.465{\scriptsize[0.573]}}\\
\midrule
\end{tabular}
\end{adjustbox}
\label{SS2table}
\end{table}
Figure~\ref{SS2_Obs_hatU} illustrates the inferred $\hat{\boldsymbol{U}}$ obtained by the supervised ZIP-LPCM with the default $\text{Beta}(1,9)$ prior setting, whereas Table~\ref{SS2table} illustrates the performance of all the implementations that we take into account in this second simulation study.
The latent positions show well-separated clusters when they are inferred for ZIP-SBM networks, with significant hubs exiting in the center for scenario 2. 
Based on the $\boldsymbol{\lambda}^*$s used for simulating the networks, it is also shown that smaller $\{\lambda^*_{gh}\}$ correspond to sparser inferred latent positions between and within the clusters.
The presence of the hubs leads to slightly more clustered latent positions in scenario 2, which in turn brings slightly higher $\hat{\beta}$ in scenario 2 as shown in Table~\ref{SS2table}.\\

In the ZIP-LPCM, the $\lambda_{ij}$ can be obtained based on $\beta$ and $\boldsymbol{U}$ following Eq.~\eqref{LPM} for each pair of nodes $i,j$: we use this to construct the posterior samples of $\lambda_{ij}$ for each ZIP-LPCM implementation.
The posterior mean of these samples forms the $\hat{\lambda}_{ij}$ which contributes to the statistic, $\mathbbm{E}(\{|\hat{\lambda}_{ij}-\lambda^*_{ij}|\})${\scriptsize [\text{sd}]}, shown in Table~\ref{SS2table}.
The corresponding reference parameter $\lambda_{ij}^*$ is obtained according to $\lambda^*_{z^*_iz^*_j}$.

In general, our ZIP-LPCM still performs reasonably well in situations where the data are generated using a ZIP-SBM.
However, we note that in this case the best performance is provided by the Beta(1,19) prior instead of the default Beta(1,9).
The Pois-LPCM fails to recover the true clustering of the scenario 2 network when the inference is unsupervised.
The same deteriorated performance is also observed for the supervised ZIP-LPCM with $\text{Beta}(1,99)$ prior, the performance of which is considered to be close to that of the Pois-LPCM.
After a careful analysis of the resulting posterior samples, we noticed that even though these two settings are able to recover the true clustering, it is easy for them to get stuck around some local posterior mode of the clustering, whereby the hubs are merged into another group.
We find that these simulation results highlight that the zero-inflation feature of the model becomes critical to efficiently characterize the group differences and thus recover the correct partitioning of the nodes.\\

\begin{figure}[h!]
\centering
\includegraphics[scale=0.475]{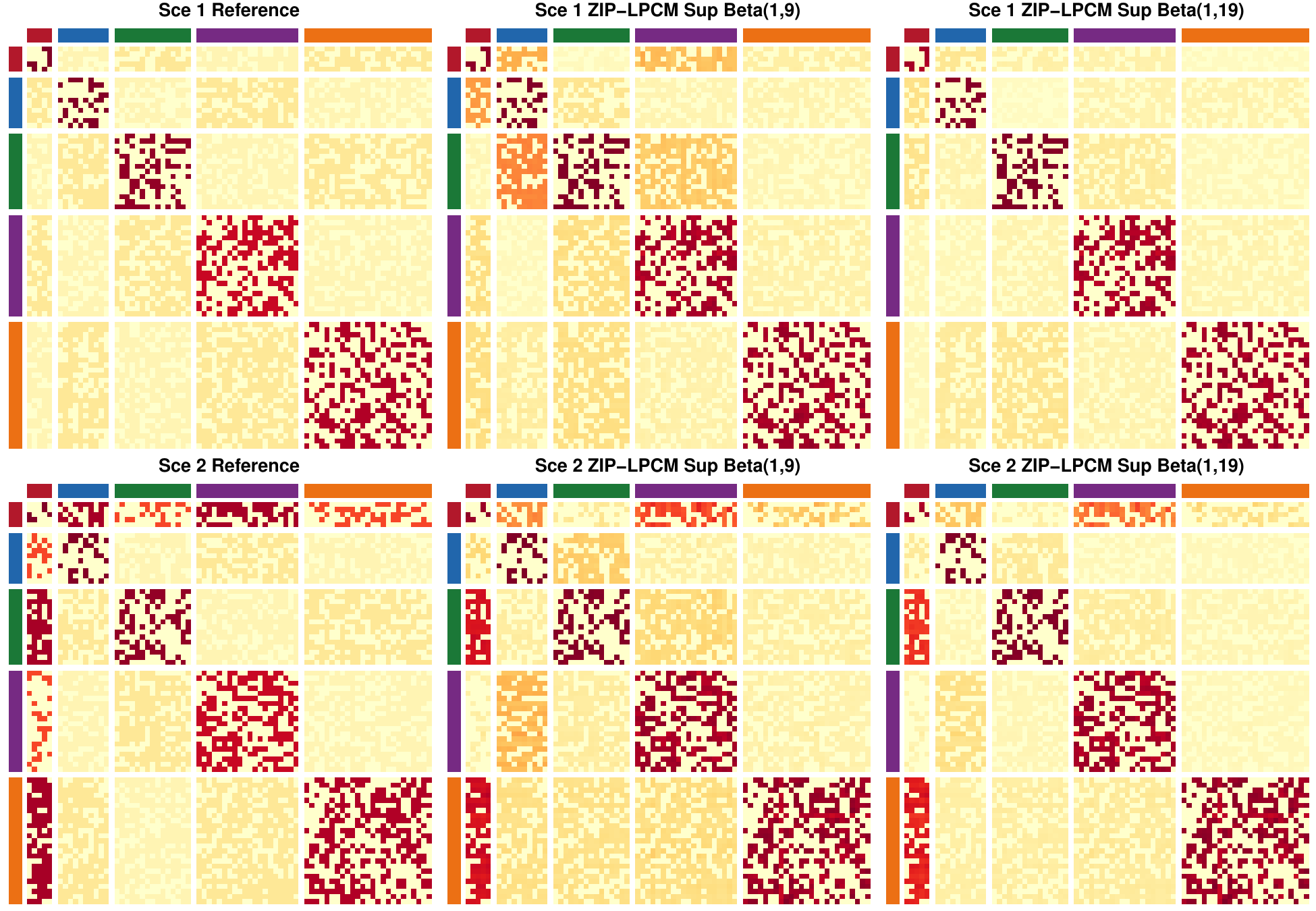}
\caption{Simulation study 2. Performance of $\hat{\boldsymbol{\nu}}$ based on the $\text{Beta}(1,9)$ and $\text{Beta}(1,19)$ prior settings. The 1st and 2nd rows, respectively, correspond to scenarios 1 and 2. Darker colors indicate higher (approximate) probability of unusual zero conditional on the fact that the corresponding observed interaction is a zero interaction. The rows and columns of the matrices are rearranged and separated according to $\hat{\boldsymbol{z}}$ while the side-bars indicate the true clustering of each individual.}
\label{SS2_hatnu}
\end{figure}

Figure~\ref{SS2_hatnu} further illustrates the performance of $\hat{\boldsymbol{\nu}}$ for the ZIP-LPCM.
Though the elements in the diagonal blocks are well approximated, the elements in the off-diagonal blocks are difficult to infer.
This shows that, in some cases, it is difficult for the ZIP-LPCM in the latent space to capture specific network architectures produced by those freely chosen model parameters of ZIP-SBM.
This is especially significant for the asymmetric patterns of the upper-diagonal and the lower-diagonal blocks shown in Figure~\ref{SS2_hatnu}.
However, we can still observe remarkably many elements in the off-diagonal blocks of $\hat{\boldsymbol{\nu}}$ from scenario 2 that have significantly higher values than others, especially for those blocks associated to the hubs.
The patterns of these elements generally agree with the corresponding reference patterns, thus successfully and correctly highlighting non-negligible many zero interactions that are more likely to be unusual zeros compared to others.\\

As a final note, we also applied the supervised ZIP-SBM with the Beta(1,9) prior and a Ga(1,1) Poisson rate prior, which extend the applications in \citet{lu2025zero} to directed networks, on both ZIP-SBM networks from the two different scenarios. We indicate these as ``ZIP-SBM Sup'' here.
Since the network data was generated from the ZIP-SBM, it is expected that the ``ZIP-SBM Sup'' implementations have natural advantages over ZIP-LPCM implementations in this case.
However, we note that, though the results of the ``ZIP-SBM Sup'' implementations are even better than the best ``ZIP-LPCM Sup Beta(1,19)'' cases illustrated in Table~\ref{SS2table} and Figure~\ref{SS2obsYadj}, the discrepancies in performance between them are not substantial.
Especially for the inferred approximate conditional probability of unusual zeros $\{\hat{\nu}_{ij}\}$, the mean absolute error of which between the ``ZIP-SBM Sup'' case and the ``ZIP-LPCM Sup Beta(1,19)'' case is $0.0403$ in scenario 1, and is $0.0301$ in scenario 2, while the corresponding standard deviation is $0.0320$ and $0.0756$, respectively.
Considering that the ZIP-SBM is not able to visualize networks, these deterioration can be fairly neglected compared to the gains in the ability of network visualization for the ZIP-LPCM.

\subsection{Simulation study 3}
\label{SS3}

In this third simulation study, we first replicate the simulation study 1 scenario 1 experiments for multiple times to check the robustness of the ZIP-LPCM performance.
We repeatedly draw 50 different networks with the same model settings introduced therein, and we leverage the same implementation settings for the inference.
The performance of this replication simulation study is shown as the top table of Table~\ref{SS3table}, wherein we use the same set of summary statistics as being used in Table~\ref{SS1table} for the performance exploration.
The median, 10\% and 90\% quantiles of the corresponding statistics for all the 50 replication experiments are provided.
The results show that all the 50 implementations successfully select the true number of groups $K^*=5$ for the 50 different networks, respectively, and provide similar performance of the distance matrix $\{d_{ij}\}$, the intercept $\beta$, and the individual-level probability of unusual zeros $\bm{p}$, which are as good as those illustrated in simulation study 1.
However, we observe that not all the implementations successfully recover the true clustering.
In fact, there are 60\% implementations which perfectly recover the true clustering; 32\% implementations misclassified 1 node; 6\% implementations misclassified 2 nodes, and there is only 1 output clustering which clusters two true groups together leading to the final 4-group clustering.

\setlength{\tabcolsep}{12pt}
\begin{table}[h!]
\renewcommand{\arraystretch}{0.9}
\centering
\caption{\footnotesize{Simulation study 3. Performance of the replication simulation study (top table) and the simulation studies for networks with larger network sizes (bottom table). The same set of summary statistics used in Table~\ref{SS1table} is also leveraged here for performance explorations. The values in each row of the top table below correspond to the median, 10\% and 90\% quantiles of the corresponding summary statistics for all 50 replicated implementations. The values in the bottom table below correspond to the output values of the corresponding summary statistics for the two simulation studies of larger networks.}}
\begin{adjustbox}{width=1.02\textwidth,center=\textwidth}
\begin{tabular}[c]{l|c|c|c|c|c|c}
\midrule
\multicolumn{1}{l}{50-Replications} & \multicolumn{1}{c}{$\hat{K}$} & \multicolumn{1}{c}{$\text{VI}(\hat{\boldsymbol{z}},\boldsymbol{z}^*)$}  &  \multicolumn{1}{c}{$\mathbbm{E}_{\boldsymbol{z}}[\text{VI}(\hat{\boldsymbol{z}},\boldsymbol{z}) \mid \boldsymbol{Y}]$}  & \multicolumn{1}{c}{$\mathbbm{E}(\{|\hat{d}_{ij}-d^*_{ij}|\})${\scriptsize [\text{sd}]}} &   \multicolumn{1}{c}{$\hat{\beta}$}  & \multicolumn{1}{c}{$\mathbbm{E}(\{|\hat{p}_{z_iz_j}-p^*_{z_iz_j}|\})${\scriptsize [\text{sd}]}}  \\
\midrule
10\% quantile   & 5 & 0.00 & 0.0058 &  {0.2486{\scriptsize[0.1978]}} & 2.9164 &  {0.0280{\scriptsize[0.0207]}}\\
Median            & 5 & 0.00 & 0.0354 &  {0.2818{\scriptsize[0.2323]}} & 2.9704 &  {0.0345{\scriptsize[0.0310]}}\\
90\% quantile   & 5 & 0.15 & 0.1160 &  {0.3249{\scriptsize[0.2725]}} & 3.0188 &  {0.0469{\scriptsize[0.0458]}}\\
\midrule
\end{tabular}
\end{adjustbox}

\begin{adjustbox}{width=1.02\textwidth,center=\textwidth}
\begin{tabular}[c]{c|c|c|c|c|c|c}
\midrule
\multicolumn{1}{c}{} & \multicolumn{1}{c}{$\hat{K}$} & \multicolumn{1}{c}{$\text{VI}(\hat{\boldsymbol{z}},\boldsymbol{z}^*)$}  &  \multicolumn{1}{c}{$\mathbbm{E}_{\boldsymbol{z}}[\text{VI}(\hat{\boldsymbol{z}},\boldsymbol{z}) \mid \boldsymbol{Y}]$}  & \multicolumn{1}{c}{$\mathbbm{E}(\{|\hat{d}_{ij}-d^*_{ij}|\})${\scriptsize [\text{sd}]}} &   \multicolumn{1}{c}{$\hat{\beta}$}  & \multicolumn{1}{c}{$\mathbbm{E}(\{|\hat{p}_{z_iz_j}-p^*_{z_iz_j}|\})${\scriptsize [\text{sd}]}}  \\
\midrule
ZIP-LPCM N150  & 5 & 0.00 & 0.0042 &  {0.2924{\scriptsize[0.2920]}} & 2.9536 &  {0.0237{\scriptsize[0.0189]}}\\
ZIP-LPCM N225  & 5 & 0.00 & 0.0119 &  {0.2197{\scriptsize[0.1983]}} & 3.0153 &  {0.0182{\scriptsize[0.0241]}}\\
\midrule
\end{tabular}
\end{adjustbox}
\label{SS3table}
\end{table}

\begin{figure}[h!]
\centering
\includegraphics[scale=0.335]{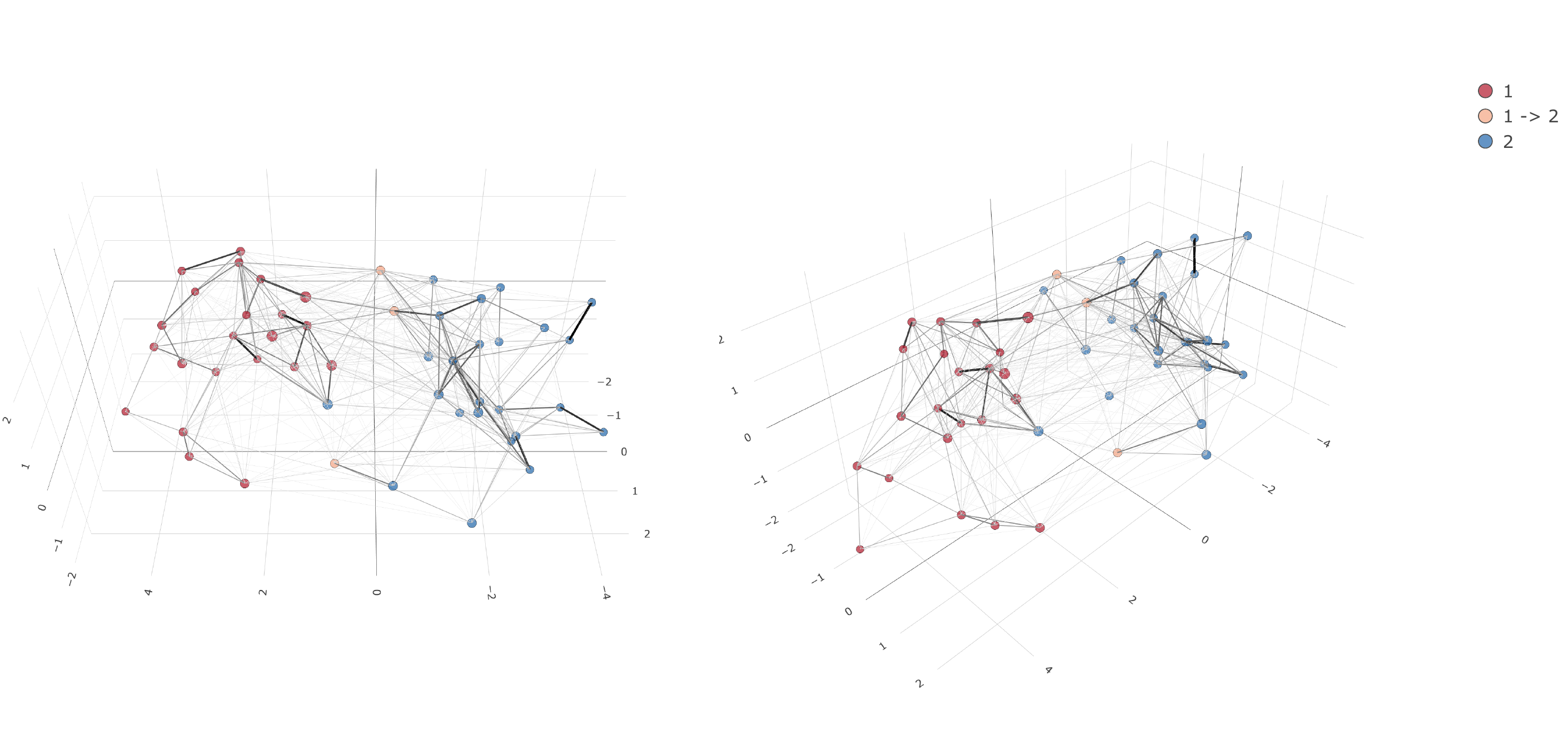}
\caption{Simulation study 3. A two-group synthetic network example of showing possible misclassifications of nodes' clustering. The 2nd plot is the rotated plot of the 1st latent positions' plot that is rotated for $45\degree$ anti-clockwise with respect to the vertical axis. The latent positions are those used for generating the network, whereas the dark red (group 1) and dark blue (group 2) nodes correspond to the nodes which are inferred to have true clustering. The three light red nodes are the ones which should belong to group 1 according to the reference clustering $\bm{z}^*$ but they are instead inferred to be misclassified to group 2 based on their network structure. No misclassification observed for group 2 nodes' clustering. Node sizes are proportional to node betweenness. Edge widths and colors are proportional to edge weights.}
\label{SS2_K2_3d_Plot}
\end{figure}

The above observed misclassifications are due to the model assumption that the clustering is highly dependent on the latent positions $\bm{U}$.
Figure~\ref{SS2_K2_3d_Plot} provides a more challenging simulation study to explain this situation.
The synthetic network shown in the figure has $N=50$ nodes, which are evenly assigned to $K=2$ groups.
The network is randomly generated from a ZIP-LPCM with settings: $\beta=3, \bm{\mu}=[(2,0,0)^T,(-2,0,0)^T], \bm{\tau} = (0.75,0.75)$, and the unusual zero probability is set to be $0.6$ and $0.4$, respectively, for diagonal and off-diagonal elements.
These settings make the two groups of the network in the latent space be close to each other, forming some mixed nodes in the middle that are challenging to classify as shown in Figure~\ref{SS2_K2_3d_Plot}.
By treating the reference clustering $\bm{z}^*$ with 15 nodes' clustering contaminated (randomly reallocated) as the node attributes $\bm{c}$, our supervised inference approach misclassified three nodes which should belong to group 1 (dark reds) but which are inferred to be in group 2 (dark blues) according to the reference clustering used for generating such a network.
However, this ``misclassification'' might not be wrong based on the network architecture.
In fact, the complete likelihood value evaluated based on the reference values of those model parameters and variables used for generating the network is -3114.734, whereas that of our inferred output is -2771.003, indicating a better fitting to such a synthetic network.
Considering also the fact that 98\% of the replication simulation studies provide either true clustering or the clustering that is very close to the reference clustering, our proposed method is shown to work well in this robustness checking experiment.

We also experiment on two more synthetic networks with larger network sizes in this simulation study 3.
The first network has double of the number of nodes of the simulation study 1 networks, that is, $N=150$, whereas the second network has triple of the number, that is, $N=225$.
Both networks are randomly generated from the ZIP-LPCMs that have settings which are the same as those proposed in simulation study 1 scenario 1, except that the latent space group centers are scaled to be larger and are 1.25 and 1.375 of the $\bm{\mu}^*$ therein, respectively, in order to maintain similar or less network sparsity.
The supervised inference is applied on both networks where 40 and 60 nodes' clustering is contaminated in the reference clustering to, respectively, form the corresponding node attributes.
The inference results on such two networks are shown as the bottom table of Table~\ref{SS3table}, where the performance which are comparable as those illustrated in Sections~\ref{SS1} and \ref{SS2} are observed.
However, though it is shown that our good performance can be generalized to larger networks, we found that the practical implementations are more challenging, where the 12000-iteration executions of our approach for the $N=150$ network take approximately 43 mins to finish.
The execution time spent by fitting the $N=225$ network is much longer and is around 1.2 hours, even when a k-means algorithm is used to initialize the partition with a 25-group solution.
The convergence of posterior samples of the $N=225$ network simulation study are shown to require more than 4000 iterations, which is significantly longer than all our previous experiments and thus 4000-iteration burn-in is instead proposed in this case.

%----------------------------------------------------------------------------------------------------------------------------------------------------------------------------------------------------------------------------------------------------------------------------------------
%----------------------------------------------------------------------------------------------------------------------------------------------------------------------------------------------------------------------------------------------------------------------------------------
%----------------------------------------------------------------------------------------------------------------------------------------------------------------------------------------------------------------------------------------------------------------------------------------
%----------------------------------------------------------------------------------------------------------------------------------------------------------------------------------------------------------------------------------------------------------------------------------------

\section{Real data applications}
\label{RDA}

In this section, we illustrate the ZIP-LPCM performance on four different real networks.
The experiment priors and proposal variance settings are similar to those applied in the simulation studies in Section~\ref{SS} with the default prior setting of the probability of unusual zeros being $\text{Beta}(1,9)$.
Since the real networks have different network sizes and complexity, a conservative $60,000$ iterations are used for each real network, with the first $30,000$ iterations treated as burn-in.
We find that this leads to satisfactory mixing and convergence in all the applications considered.

In this section, we introduce an extra summary statistic to help with the illustration of the results.
Recall that the posterior mean $\hat{\nu}_{ij}$ approximates Eq.~\eqref{P_m0}, which is the conditional probability of $y_{ij}$ being an unusual zero provided that $y_{ij}$ is observed as a zero.
However, once an observed $y_{ij}=0$ is inferred as an unusual zero, the corresponding missing weight $x_{ij}\sim \text{Pois}(\lambda_{ij})$ can be assumed following Eq.~\eqref{Aug_X}, where $\lambda_{ij}=\text{exp}(\beta-||\boldsymbol{u}_i-\boldsymbol{u}_j||)$.
This allows us to construct the conditional probability of an observed zero interaction that should actually be a non-zero interaction by using the product of $\text{P}(x_{ij}>0|\beta, \boldsymbol{u}_i, \boldsymbol{u}_j)=1-f_{\text{Pois}}(0|\lambda_{ij})$ and Eq.~\eqref{P_m0}, confirmed by:
$$
\text{P}(x_{ij}>0|y_{ij}=0,\beta, \boldsymbol{u}_i, \boldsymbol{u}_j, p_{z_iz_j})=\text{P}(x_{ij}>0|\beta, \boldsymbol{u}_i, \boldsymbol{u}_j)\text{P}(\nu_{ij}=1|y_{ij}=0,\beta, \boldsymbol{u}_i, \boldsymbol{u}_j, p_{z_iz_j}).
$$
The above probability can be approximated by $\hat{\text{P}}(x_{ij}>0|y_{ij}=0,\dots)=[1-f_{\text{Pois}}(0|\hat{\lambda}_{ij})]\hat{\nu}_{ij}$ accounting for the uncertainty of the posterior samples.
From a practical standpoint, this statistic is usually the one that the practitioners are more interested in, compared to Eq.~\eqref{P_m0}. 
Here, the $\hat{\lambda}_{ij}$ can be obtained by the posterior mean of Eq.~\eqref{LPM} accounting for the uncertainty of posterior samples of $\beta$ and $\boldsymbol{U}$.

%----------------------------------------------------------------------------------------------------------------------------------------------------------------------------------------------------------------------------------------------------------------------------------------
%----------------------------------------------------------------------------------------------------------------------------------------------------------------------------------------------------------------------------------------------------------------------------------------

\subsection{Sampson monks network}

The social interactions among a group of $18$ monks were recorded by \citet{sampson1968novitiate} during his stay in a monastery.
A political ``crisis in the cloister'' occurred in that period, leading to the expulsion of four monks and the voluntary departure of several others.
The interactions were recorded as follows: each monk was asked at three different time points whether they had positive relations to each of other monks and to rank the three monks they were closest to.
The dataset has been studied on many previous works, including \citet{hoff2002latent}.
We aggregated the three time points leading to a directed non-negative discrete weighted social network describing different levels of friendships between the monks.

\begin{figure}[h!]
\centering
\includegraphics[scale=0.485]{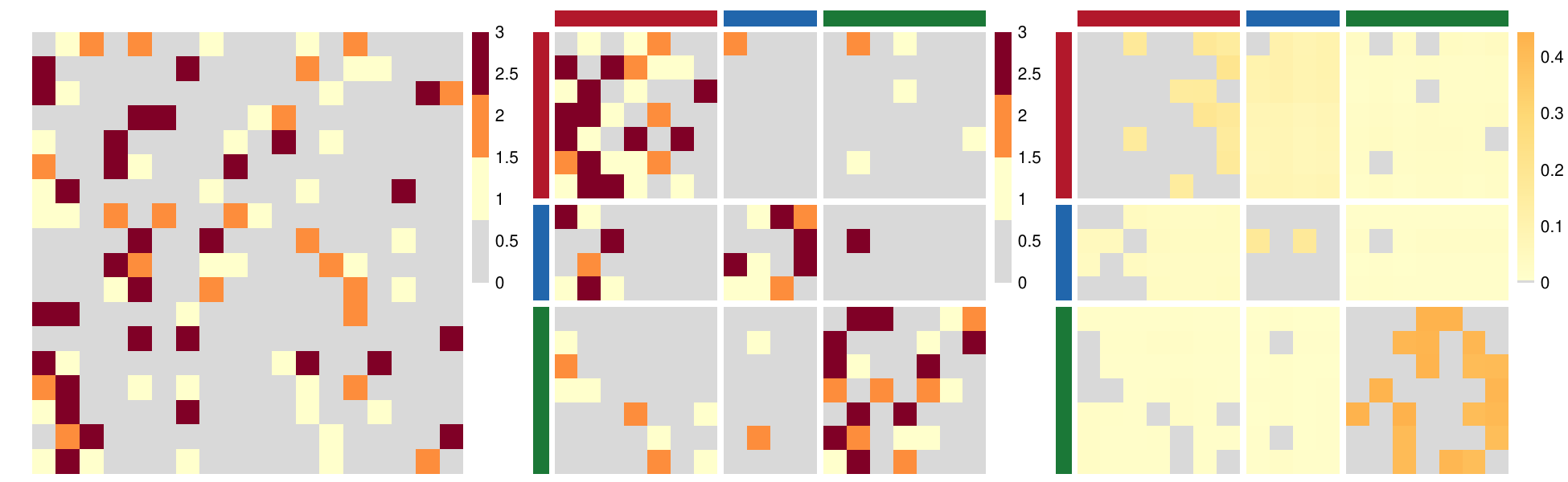}
\caption{The heatmap plots for the Sampson monks real network where the grays are used for zero values in order to highlight other non-zero elements.
Left plot: original observed adjacency matrix, $\boldsymbol{Y}$. Middle plot: plot of $\boldsymbol{Y}$ where the rows and columns of the matrices are rearranged and separated according to $\boldsymbol{z}^*$. The different colors in the side-bars of this and the last plot correspond to different reference clustering of each individual. Right plot: inferred $\hat{\text{P}}(x_{ij}>0|y_{ij}=0,\dots)$.}
\label{RDA_SampMonks_heatmaps}
\end{figure}

The observed adjacency matrix $\boldsymbol{Y}$ is shown as the first plot of Figure~\ref{RDA_SampMonks_heatmaps}, within which the non-negative discrete values of each entry ranges from 0 to 3.
Each monk was classified by Sampson to be within one of the three groups: ``Turks'', ``Outcasts'' or ``Loyal'', and such a clustering is treated as the reference clustering $\boldsymbol{z}^*$ for this network.
The plot of $\boldsymbol{Y}$ rearranged based on $\boldsymbol{z}^*$ is shown as the second plot of Figure~\ref{RDA_SampMonks_heatmaps}, where we use red, blue and green colors to represent the three reference groups, respectively.
An extra true-or-false variable ``Cloisterville'' from \citet{de2018exploratory} was also added to this dataset to indicate whether or not each monk attended the minor seminary of ``Cloisterville'' before coming to the monastery.
In our experiments, we leverage the combination of the $\boldsymbol{z}^*$ and the ``Cloisterville'' information as the exogenous node attributes indicated with $\boldsymbol{c}$, to implement the supervised version of the ZIP-LPCM on this network.
For example, the monks in the ``Turks'' group are separated into two different levels in $\boldsymbol{c}$ depending on whether each of them attended the minor seminary or not.
Similar for the other two groups leading to a total of $C=6$ levels in $\boldsymbol{c}$.

\begin{figure}[h!]
\centering
\includegraphics[scale=0.335]{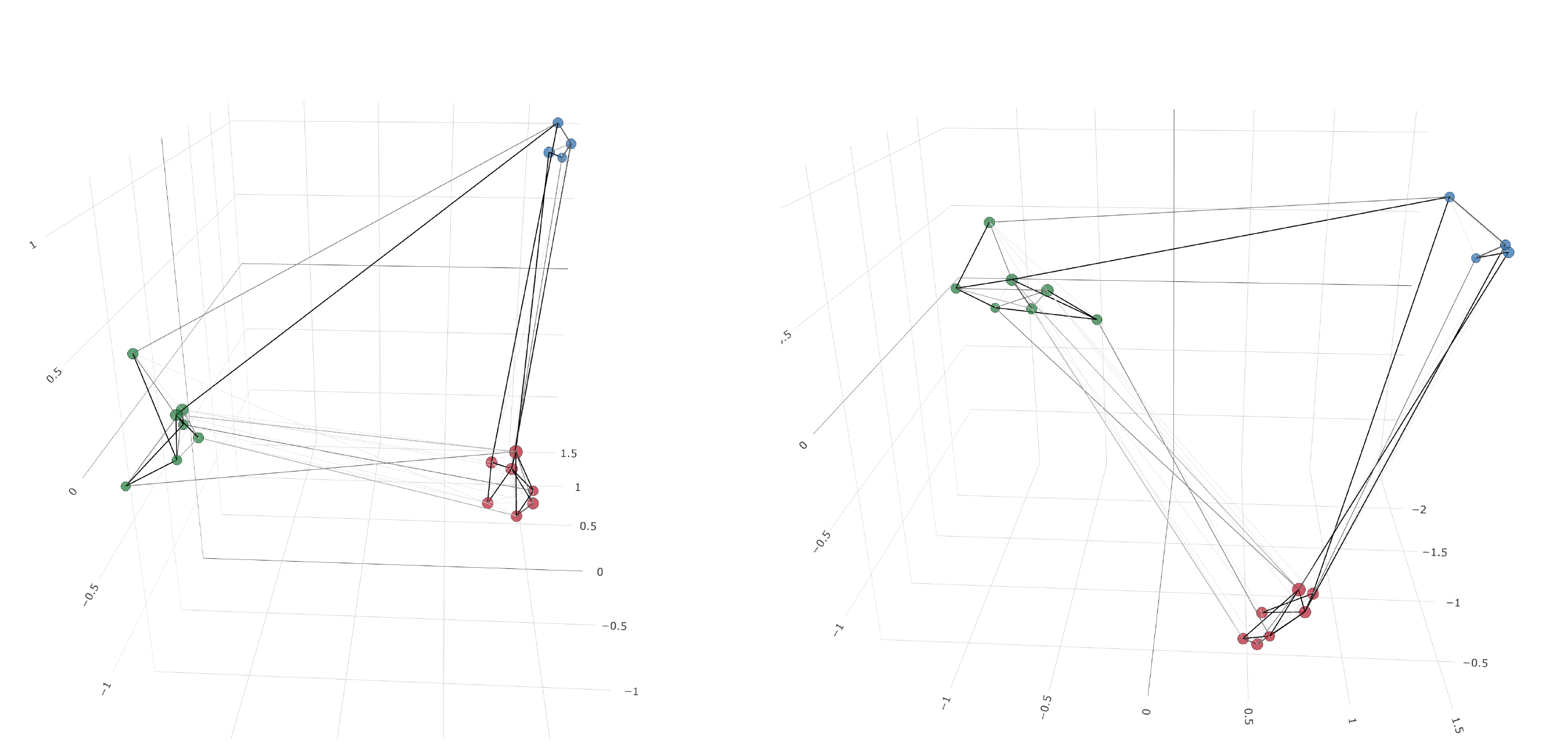}
\caption{The Sampson monks real network. Left plot: inferred latent positions, $\hat{\boldsymbol{U}}$. The three inferred groups in $\hat{\boldsymbol{z}}$ are distinguished by different colors, and perfectly agree with the reference clustering. Node sizes are proportional to node betweenness. Edge widths and colors are proportional to edge weights. Right plot: rotated version of the latent positions shown in the left plot where the whole latent space is rotated by $90\degree$ clockwise with respect to the vertical axis.}
\label{RDA_SampMonks_hat_U_Y}
\end{figure}

The inferred latent positions, $\hat{\boldsymbol{U}}$, shown in Figure~\ref{RDA_SampMonks_hat_U_Y} illustrate a typical SBM structure which resembles the ones shown in our simulation study 2 in Section~\ref{SS2}.
The inferred clustering $\hat{\boldsymbol{z}}$ perfectly agrees with the reference clustering $\boldsymbol{z}^*$, even consider the more informative exogenous node attributes $\boldsymbol{c}$ during inference.
It is shown that the exogenous ``Cloisterville'' information does not bring more subgroup features to the clustering of this network under the ZIP-LPCM.
Our unsupervised ZIP-LPCM implementation, which is not detailed here, also returns the same inferred clustering but with higher uncertainty of the posterior clustering.

The red ``Turks'' group is shown to be more tightly clustered than the blue ``Outcasts'' group, which is itself more clustered than the green ``Loyal'' group.
According to the analysis of the conditional probability of observed zeros that should actually be non-zero interactions shown in the last plot of Figure~\ref{RDA_SampMonks_heatmaps}, the within-group zero interactions of the ``Loyal'' group are more likely to be non-zeros compared to other zero interactions, though the corresponding inferred probability is around 0.4 which is not particularly high.
The number of zero interactions from ``Turks'' to ``Outcasts'' are shown to be significantly less than that in the reverse direction, leading to slightly higher conditional probability of those zero-interactions being unusual zeros.
However, considering the far latent distance between the two groups illustrated in Figure~\ref{RDA_SampMonks_hat_U_Y}, the corresponding conditional probability of being non-zeros shown in the last plot of Figure~\ref{RDA_SampMonks_heatmaps} remains relatively small and negligible.

%----------------------------------------------------------------------------------------------------------------------------------------------------------------------------------------------------------------------------------------------------------------------------------------
%----------------------------------------------------------------------------------------------------------------------------------------------------------------------------------------------------------------------------------------------------------------------------------------

\subsection{Windsurfers network}

This is an undirected non-negative discrete weighted network \citep{freeman1998exploring,freeman1988human,konect} recording the interpersonal contacts between $43$ windsurfers in southern California during the fall of 1986.
The values of interaction weights range from 0 to 47 recorded in the adjacency matrix.
As the network size is larger than the Sampson monks network, the $\text{Beta}(1,19)$ prior of the probability of unusual zeros is proposed to encourage more clustering.
Further, no reference clustering and exogenous node attributes are available for this network, so the unsupervised ZIP-LPCM is used.

\begin{figure}[h!]
\centering
\includegraphics[scale=0.4]{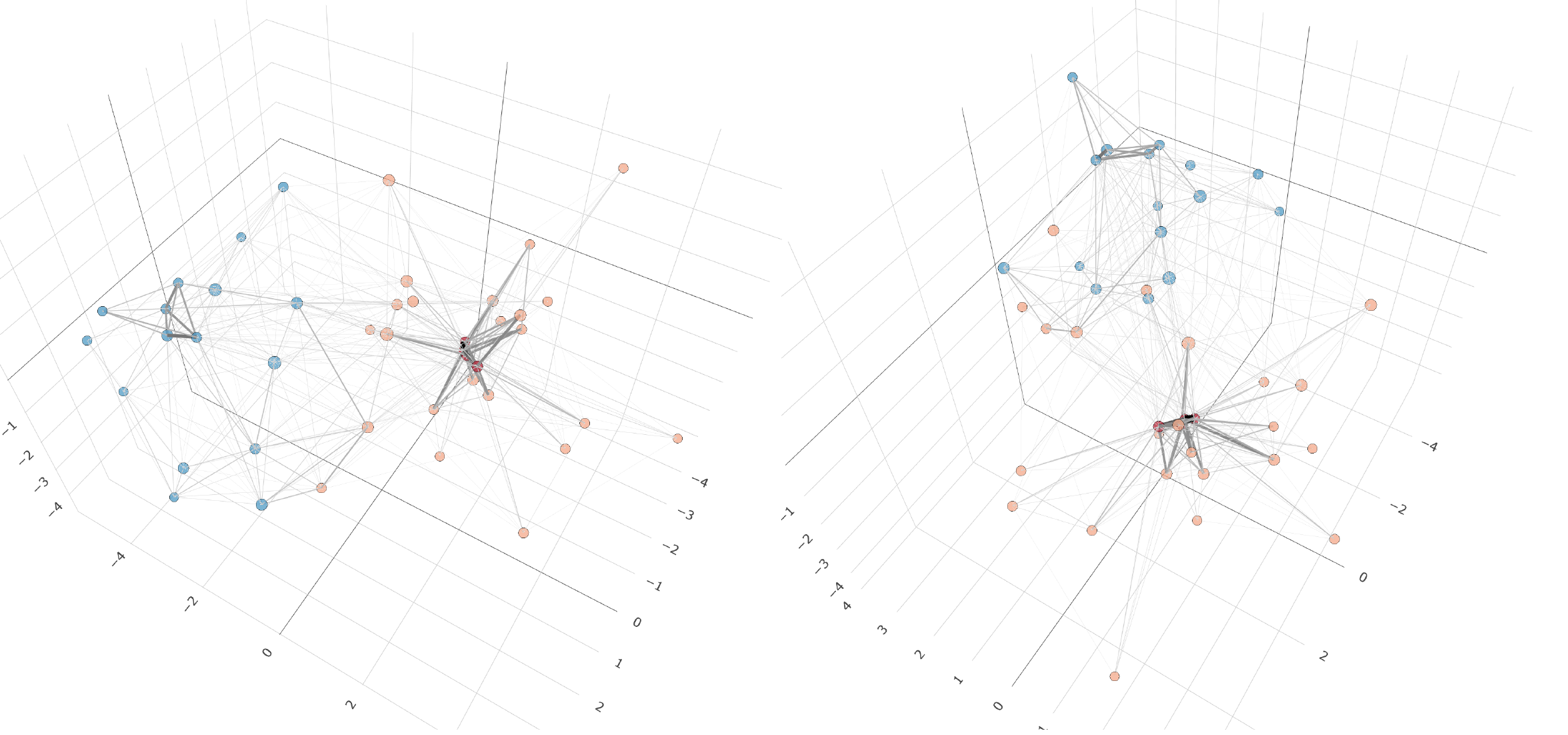}
\caption{Windsurfers real network. Left plot: inferred latent positions, $\hat{\boldsymbol{U}}$. The three inferred groups in $\hat{\boldsymbol{z}}$ are distinguished by different colors. Node sizes are proportional to node betweeness. Edge widths and colors are proportional to edge weights. Right plot: rotated version of the latent positions shown in the left plot where the whole latent space is rotated by $90\degree$ clockwise with respect to the vertical axis.}
\label{RDA_Windsurfers_hat_U_Y}
\end{figure}

Figure~\ref{RDA_Windsurfers_hat_U_Y} shows that the whole network may generally be split into two main groups.
However, it is interesting that our inferred clustering $\hat{\boldsymbol{z}}$ returns $\hat{K}=3$ inferred groups, one of which is likely to be a core group since it sits at the center of another group.
\begin{figure}[h!]
\centering
\includegraphics[scale=0.485]{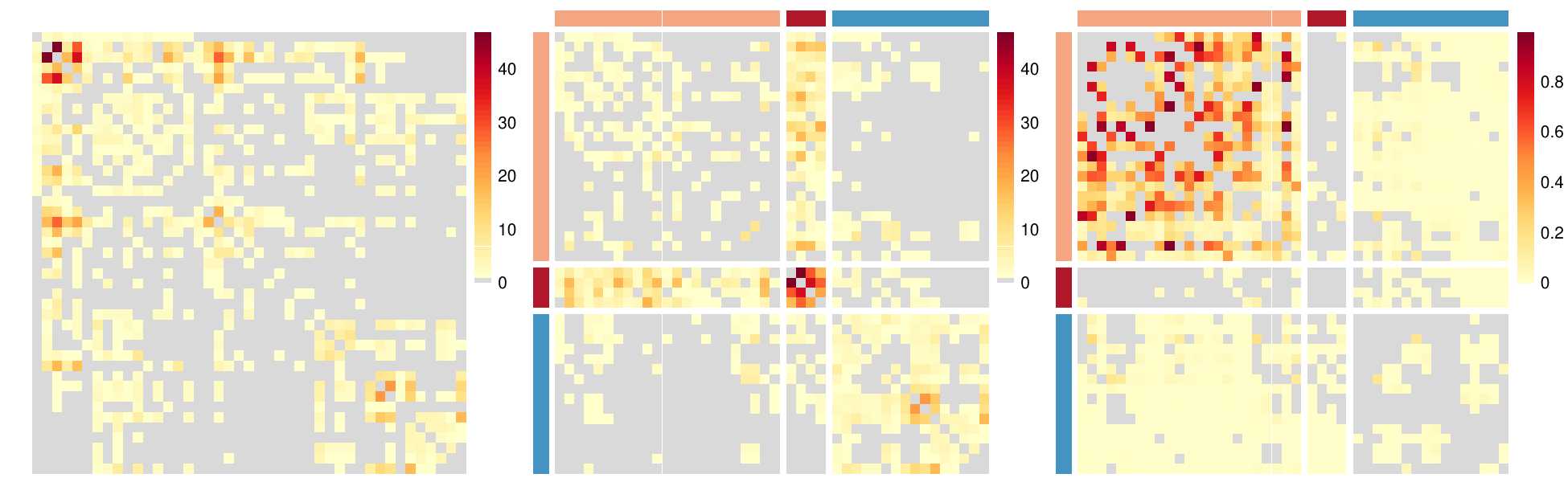}
\caption{The heatmap plots for the windsurfers real network where the grays are used for zero values in order to highlight other non-zero elements.
Left plot: original observed adjacency matrix, $\boldsymbol{Y}$. Middle plot: plot of $\boldsymbol{Y}$ where the rows and columns of the matrices are rearranged and separated according to $\hat{\boldsymbol{z}}$. The different colors on the side-bars correspond to different inferred clustering of each individual. Right plot: inferred $\hat{\text{P}}(x_{ij}>0|y_{ij}=0,\dots)$.}
\label{RDA_Windsurfers_heatmaps}
\end{figure}
We use dark red color to represent the core group while the light red and light blue colors are used, respectively, for the corresponding non-core group and another group.
In combination with the middle plot of Figure~\ref{RDA_Windsurfers_heatmaps}, it is shown that there are four windsurfers clustered into the red core group: they frequently interact with each other thus leading to very strong connections.
This core group also actively interacts with all other windsurfers from the red non-core group, within which a few windsurfers seem to be the ``close friends'' of the core windsurfers.
These findings indicate that the members from the core group may have a central or leader role or generally high reputation across the whole network.\\

It can also be observed in Figure~\ref{RDA_Windsurfers_hat_U_Y} that four blue windsurfers are also well connected with each other and tend to form a core of the blue group, but their connection patterns tend to be significantly weaker than the red core group, and thus our model prefers not to distinguish them from the whole blue group.
The zero interactions which are more likely to be non-zeros are mainly those within the red non-core group: some of them have very high probability to be non-zero interactions.
This result indicates that either the corresponding interpersonal contact data was lost, or that those pairs of windsurfers might have other forms of interactions which were not recorded in this dataset.

%----------------------------------------------------------------------------------------------------------------------------------------------------------------------------------------------------------------------------------------------------------------------------------------
%----------------------------------------------------------------------------------------------------------------------------------------------------------------------------------------------------------------------------------------------------------------------------------------

\subsection{Train bombing network}

This dataset \citep{hayes2006connecting,konect} aims at reconstructing the contacts between suspected terrorists involved in the train bombing of Madrid on 11th March, 2004.
The corresponding undirected network records 64 suspects, and the pair-wise interaction strengths, which range from 0 to 4, indicate the aggregation of friendship and co-participating in training camps or previous attacks.
For this dataset, we propose a $\text{Beta}(1,9)$ prior and we use the unsupervised setting.

\begin{figure}[h!]
\centering
\includegraphics[scale=0.35]{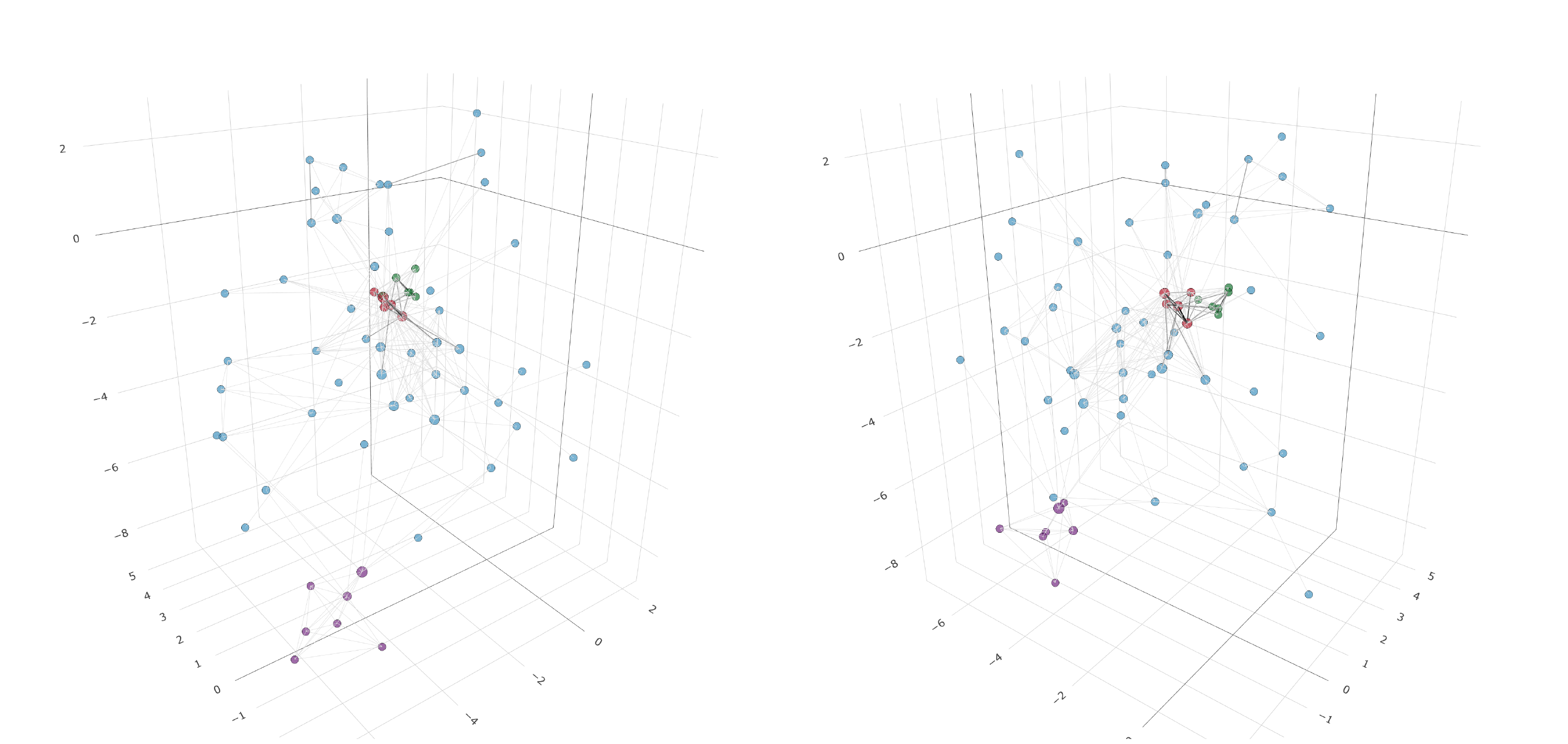}
\caption{Train bombing real network. Left plot: inferred latent positions, $\hat{\boldsymbol{U}}$. The inferred clustering distinguished by different colors. Node sizes are proportional to node betweenness. Edge widths and colors are proportional to edge weights. Right plot: rotated version of the latent positions shown in the left plot where the whole latent space is rotated by $90\degree$ clockwise with respect to the vertical axis.}
\label{RDA_TrainBombing_hat_U_Y}
\end{figure}

\begin{figure}[h!]
\centering
\includegraphics[scale=0.485]{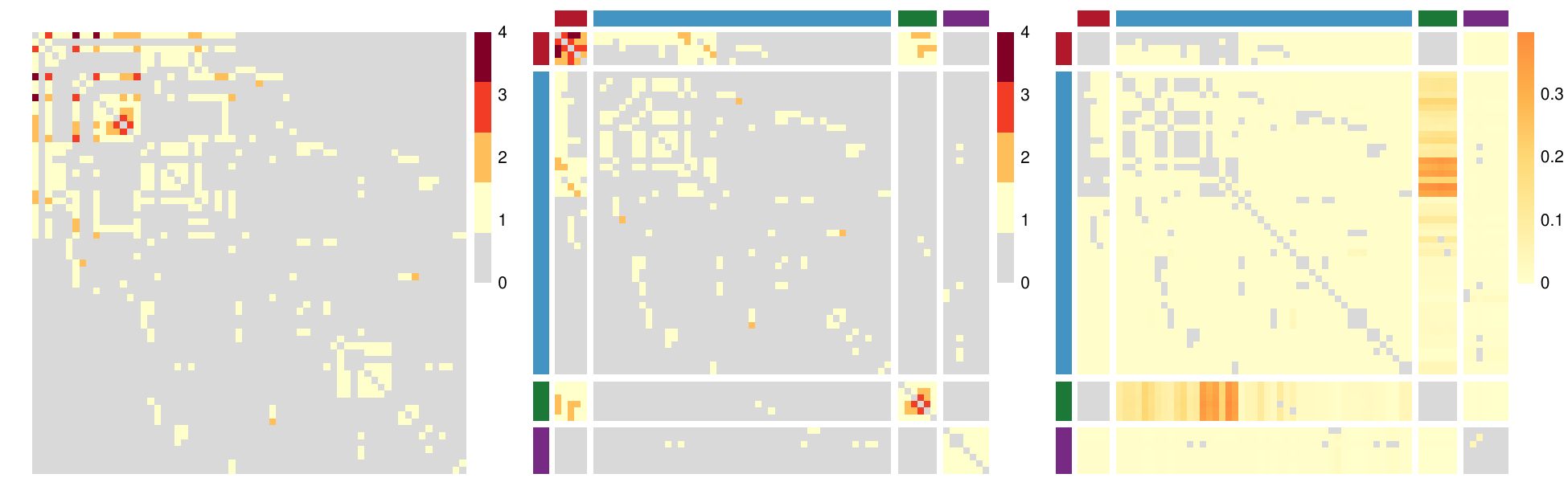}
\caption{The heatmap plots for the train bombing real network where the grays are used for zero values in order to highlight other non-zero elements.
Left plot: original observed adjacency matrix, $\boldsymbol{Y}$. Middle plot: the plot of $\boldsymbol{Y}$ where the rows and columns of the matrices are rearranged and separated according to $\hat{\boldsymbol{z}}$. The different colors in the side-bars of this and the last plot correspond to different inferred clustering of each individual. Right plot: inferred $\hat{\text{P}}(x_{ij}>0|y_{ij}=0,\dots)$.}
\label{RDA_TrainBombing_heatmaps}
\end{figure}

The inferred latent positions and clustering shown in Figure~\ref{RDA_TrainBombing_hat_U_Y} illustrate that a small red group containing five suspects forms the core of the whole network: strong connections between each pair of members inside this core group can be observed.
The big blue group contains the non-core suspects and, based on the middle plot of Figure~\ref{RDA_TrainBombing_heatmaps}, some of them seem to be more ``central'' than others in light of the significantly stronger connections with each other and with the core nodes.
By contrast, the other ``non-central'' blue members do not interact with the core nodes and only interact with these ``central'' blue members.
Though this ``central'' and ``non-central'' feature of the inferred blue group can be illustrated by the plots, our inference results suggest not to split them into two separate groups.\\

It is interesting that we also observe a small green group which contains five suspects who interact strongly with each other.
This special group is shown to only interact well with the core red group and has very few connections with the non-core blue group.
According to the corresponding inferred latent positions shown in Figure~\ref{RDA_TrainBombing_hat_U_Y}, these features are unusual and in agreement with the results shown in the last plot of Figure~\ref{RDA_TrainBombing_heatmaps} where there are significantly many zero interactions, which are more likely to be non-zeros compared to other zero interactions, between this special green group and the non-core blue group.
Finally, a small purple group seems to form a separate small network by themselves and only has few weak connections with the blue group.

%----------------------------------------------------------------------------------------------------------------------------------------------------------------------------------------------------------------------------------------------------------------------------------------
%----------------------------------------------------------------------------------------------------------------------------------------------------------------------------------------------------------------------------------------------------------------------------------------

\subsection{Summit co-attendance criminality network}
\label{RDA_Criminal}

The last real network that we propose in this paper was obtained from \url{https://sites.google.com/site/ucinetsoftware/datasets/covert-networks}, and was previously analyzed in a number of works, including \citet{legramanti2022extended} and \citet{lu2025zero}.
This is an undirected network recording the co-attendance of summits for 84 criminal suspects operating in the north of Italy. The edge weights represent the number of summits that any two individuals co-attended, and the maximum edge weight is 13.
We consider a ground-truth partition of the nodes as in \citet{lu2025zero}, which we indicate with $\boldsymbol{z}^*$. This reference clustering is made of 10 groups, which describe the roles and affiliations of the suspects.

The network also contains additional exogenous node attributes, which we indicate with $\boldsymbol{c}$, and use for the supervised implementations. The additional node information is also a partition with $C=7$ clusters.
Since these partitions contain more levels compared to the ones explored in the previous sections, we propose to apply a $\text{Beta}(1,6)$ as the prior of the probability of the unusual zeros, so that we encourage the network sparsity, thus bringing more subgroup features.\\

\begin{figure}[h!]
\centering
\includegraphics[scale=0.335]{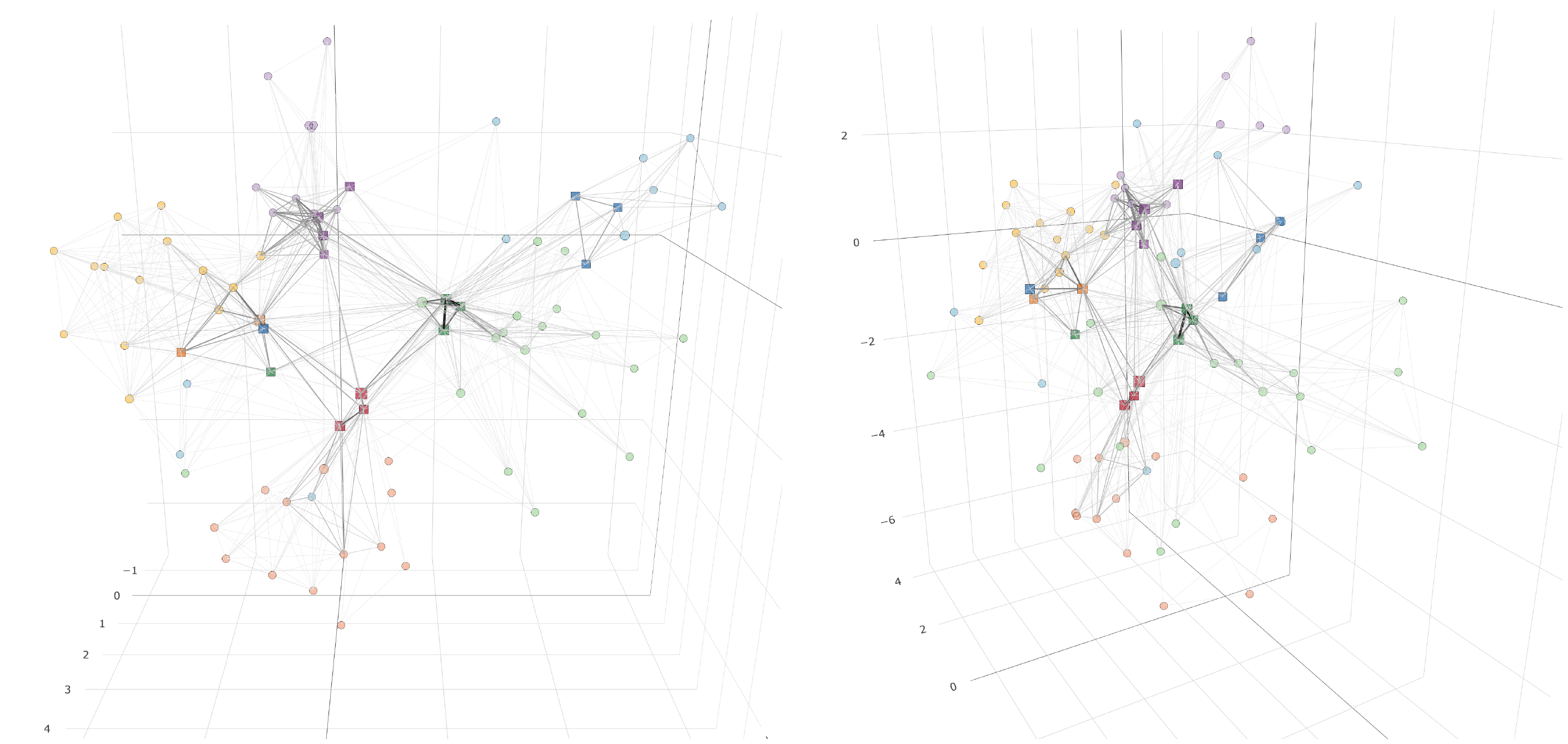}

\vspace{0.5em}

\includegraphics[scale=0.335]{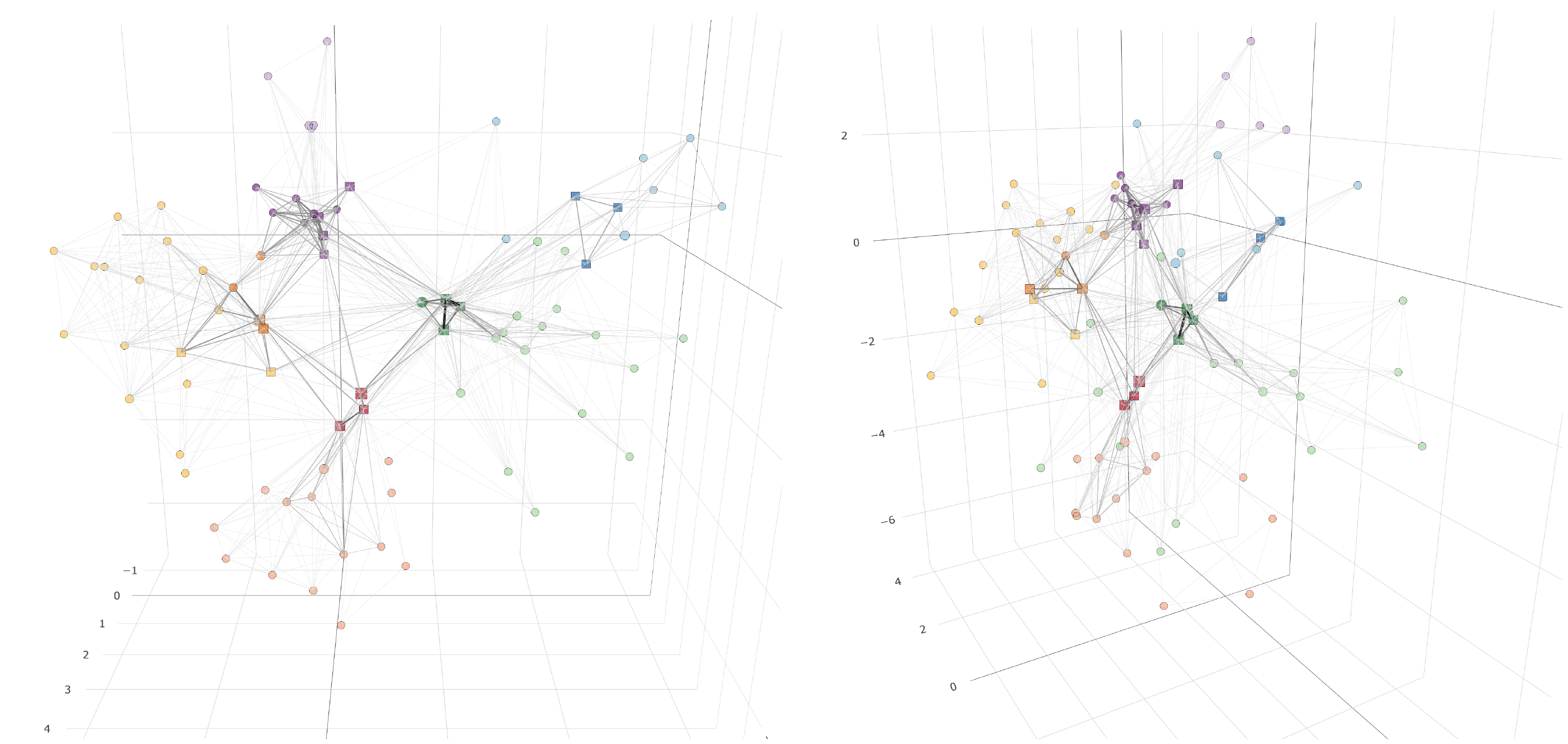}
\caption{Summit co-attendance criminality network. The 1st row plots illustrate the inferred latent positions, $\hat{\boldsymbol{U}}$, along with the different dark or light node colors indicating the reference clustering $\boldsymbol{z}^*$. Darker square nodes indicate a more central role in the organization. The same $\hat{\boldsymbol{U}}$ is shown in the 2nd row plots but the node colors denote instead the inferred partition $\hat{\boldsymbol{z}}$. The 2nd column plots are the rotated version of the latent positions shown in the 1st column plots where the whole latent space is rotated by $60\degree$ clockwise with respect to the vertical axis. Node sizes are proportional to node betweenness. Edge widths and colors are proportional to edge weights.}
\label{RDA_CriminalSummit_hat_U}
\end{figure}

\begin{figure}[h!]
\centering
\includegraphics[scale=0.385]{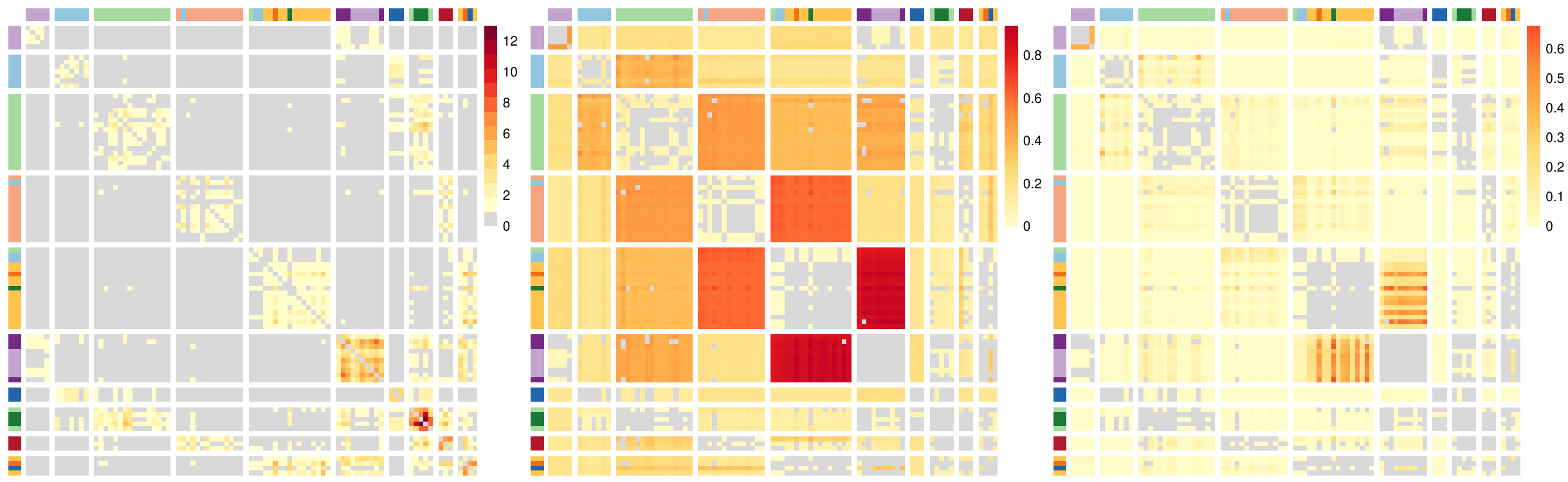}
\caption{The output heatmap plots for the summit co-attendance criminality network where the grays are used for zero values in order to highlight other non-zero elements. The different colors in the side-bars correspond to the role-affiliation information, $\boldsymbol{z}^*$. The rows and columns of the matrices are rearranged and separated according to $\hat{\boldsymbol{z}}$, where the inferred groups containing central nodes are placed at the bottom while the others are placed at the top. Left plot: adjacency matrix $\boldsymbol{Y}$. Middle plot: inferred $\hat{\boldsymbol{\nu}}$. Right plot: inferred $\hat{\text{P}}(x_{ij}>0|y_{ij}=0,\dots)$.}
\label{RDA_CriminalSummit_heatmaps}
\end{figure}

Figure~\ref{RDA_CriminalSummit_hat_U} illustrates the inferred latent positions plotted along with a set of different node colors and shapes.
The clustering in the 1st row plots is the reference clustering, $\boldsymbol{z}^*$, where different node colors correspond to different affiliations.
Darker squares indicate individuals with more central roles within the organization.
Figure~\ref{RDA_CriminalSummit_heatmaps} shows the heatmaps for this network, with sidebars indicating $\boldsymbol{z}^*$, and row and column gaps indicating $\hat{\boldsymbol{z}}$.
We note that there is a fairly strong agreement between the ground truth partition and the inferred one.

Combining both Figure~\ref{RDA_CriminalSummit_hat_U} and \ref{RDA_CriminalSummit_heatmaps}, significant core-periphery structure can be observed.
Based on the inferred latent positions, the core groups, consisting mainly of the criminal suspects with central roles shown as darker groups in the 2nd row plots of Figure~\ref{RDA_CriminalSummit_hat_U}, mostly sit at the center of the network, while the non-core groups are at the periphery.
The core groups generally show a flat structure as they roughly lie within a flat plane, whereas the periphery non-core groups are more tridimensional and more pervasive.
But the members of the non-core groups mainly have connections within their own group or with the corresponding core groups, indicating a lack of interactions between the non-core groups for this network.
This is further confirmed by visualizing the first plot of Figure~\ref{RDA_CriminalSummit_heatmaps}.
The members within each core group are densely connected with each other, but significant sparsity between the cores can be observed, as exemplified by those relatively sparse between-group interactions of the cores at the center of the latent positions in Figure~\ref{RDA_CriminalSummit_hat_U}.
However, most of the corresponding zeros within and between these core groups are not likely to be non-zeros according to the right plot of Figure~\ref{RDA_CriminalSummit_heatmaps}.\\

It is worthwhile to note that our inferred clustering, $\hat{\boldsymbol{z}}$, is similar to the inferred clustering (denoted as $\boldsymbol{z}^\dagger$ here) obtained by ZIP-SBM in \citet{lu2025zero}.
However, one key difference between $\hat{\boldsymbol{z}}$ and $\boldsymbol{z}^\dagger$ is relevant to some of the central nodes, including two orange nodes, one blue and one green, whose inferred latent positions are close to each other and are near the center of the network as shown in Figure~\ref{RDA_CriminalSummit_hat_U}.

%----------------------------------------------------------------------------------------------------------------------------------------------------------------------------------------------------------------------------------------------------------------------------------------
%----------------------------------------------------------------------------------------------------------------------------------------------------------------------------------------------------------------------------------------------------------------------------------------

\subsection{Comparisons to inference on 2-dimensional latent space}
\label{2d_Comparison}

\begin{figure}[h!]
\centering
\includegraphics[scale=1.0]{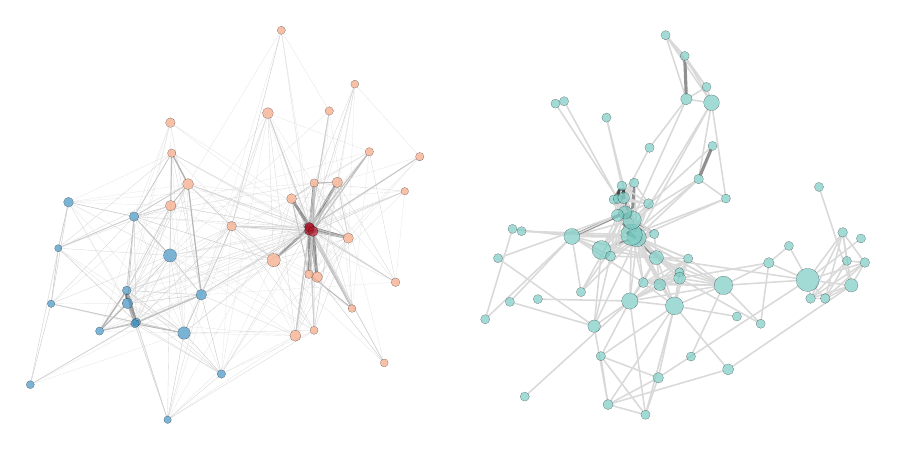}

\vspace{0.5em}

\includegraphics[scale=1.0]{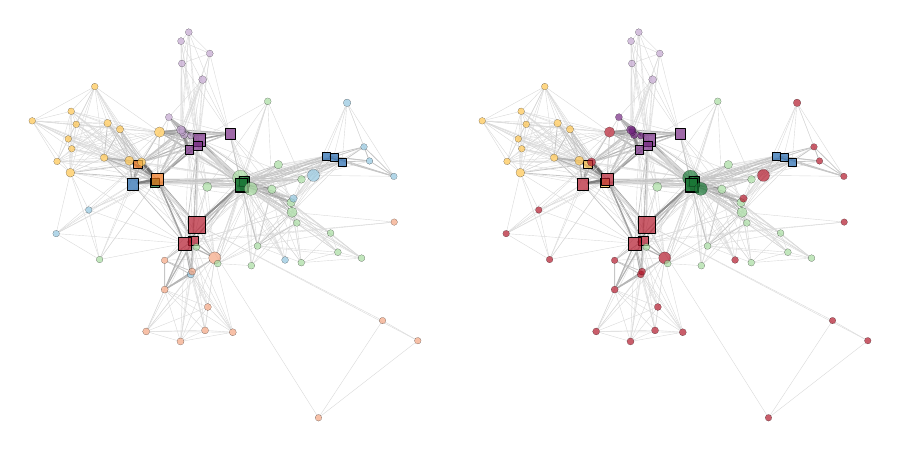}
\caption{The windsurfers, train bombing and summit co-attendance criminality network in 2-dimensional latent space. The plot settings regarding node sizes, colors, types and edge colors, widths are similar to those applied in the previous subsections. Top-left: inferred 2-d latent positions' plot of the windsurfers network. Different node colors correspond to different inferred groups in $\hat{\boldsymbol{z}}$. Top-right: inferred 2-d latent positions' plot of the train bombing network where the inferred clustering has only 1 group. Bottom plots: inferred 2-d latent positions of the summit co-attendance criminality network, where the different node colors in the left plot indicate the reference clustering $\boldsymbol{z}^*$, while those in the right plot correspond to different inferred groups in $\hat{\boldsymbol{z}}$.}
\label{RDA_2d_Plots}
\end{figure}

We finish this section of real data applications by comparing the inference output obtained in a 3-dimensional latent space to the output obtained in a 2-dimensional latent space for the real datasets.
The analyses are performed in an analogous way, with the obvious exception that the dimension of the latent positions is instead set as $d=2$.
Figure~\ref{RDA_2d_Plots} illustrates the inferred 2-dimensional latent positions for the real networks.
In summary, these results show that the plot of the 2-d inferred latent positions is roughly like a screenshot of the corresponding 3-d inferred latent positions from a specific angle, for example, if we compare the top-left plot of Figure~\ref{RDA_2d_Plots}, which illustrates the 2-dimensional $\hat{\boldsymbol{U}}$ of the windsurfers network, to the left plot of Figure~\ref{RDA_Windsurfers_hat_U_Y}.
Similarly, the bottom-left plot of Figure~\ref{RDA_2d_Plots}, which shows the 2-dimensional $\hat{\boldsymbol{U}}$ and the reference clustering $\boldsymbol{z}^*$ of the criminality network, generally follow the pattern which is similar to that of the top-left plot of Figure~\ref{RDA_CriminalSummit_hat_U} which is a screenshot of the 3-dimensional $\hat{\boldsymbol{U}}$ of this network.

However, this does not mean that the third dimension is redundant, actually, there is good evidence that a 2-dimensional latent space is not sufficient to characterize the architectures of these complex networks.
For example, in the summit co-attendance criminality network, the green nodes are distributed pervasively and tend to overlap with red and blue nodes in 2-d latent space as shown in the bottom-left plot of Figure~\ref{RDA_2d_Plots}.
In comparison, this group of nodes can be well separated along the extra 3rd dimension in 3-d latent space according to Figure~\ref{RDA_CriminalSummit_hat_U}.
Furthermore, it is also shown that the networks are generally inferred to be more aggregated in 2-d space compared to the corresponding 3-d cases, making it harder to well infer the network clustering.

Similarly for the train bombing network, the 2-dimensional $\hat{\boldsymbol{U}}$ and the corresponding $\hat{\boldsymbol{z}}$ are plotted at top-right of Figure~\ref{RDA_2d_Plots}.
The core red group and the special green group are well inferred in the 3-d space shown in Figure~\ref{RDA_TrainBombing_hat_U_Y}, but can no longer be well separated and be identified in the 2-d space, bringing a trivial inferred clustering which only has one singleton group.
Our experiments show that the illustrated unsatisfactory clustering of the summit co-attendance criminality network and the train bombing network inferred in 2-d space also brings unreasonable performance of inferred probability of unusual zeros.
On the other hand, we point out that the inference performance of the windsurfers network and the Sampson monks network in 2-d space are shown to be comparable with those in 3-d space.
Similar inferred latent positions' patterns and inferred clustering are observed in both 2-d and 3-d space, though the inferred probability of unusual zeros is generally higher in 2-d cases due to the more aggregated patterns of the 2-d latent positions.
The plot of inferred 2-d latent positions of the Sampson monks network, which is not included in Figure~\ref{RDA_2d_Plots}, is similar to the plots shown in Figure~\ref{RDA_SampMonks_hat_U_Y}.
We refer to \url{https://github.com/Chaoyi-Lu/ZIP-LPCM} for more details of all the 2-d implementations for the four real networks.

The above findings show that 3-dimensional latent positions are capable of giving a more nuanced model-based representation compared to 2-d latent spaces.
The choice of the number of latent space dimensions remains a challenging problem. According to our results, 3-dimensional latent spaces can allow for a better model fit while still maintaining visualizations that are easy to interpret. Thus, our work emphasizes once more the criticality of this model-choice research question within the scope of latent space modeling.

%----------------------------------------------------------------------------------------------------------------------------------------------------------------------------------------------------------------------------------------------------------------------------------------
%----------------------------------------------------------------------------------------------------------------------------------------------------------------------------------------------------------------------------------------------------------------------------------------
%----------------------------------------------------------------------------------------------------------------------------------------------------------------------------------------------------------------------------------------------------------------------------------------
%----------------------------------------------------------------------------------------------------------------------------------------------------------------------------------------------------------------------------------------------------------------------------------------

\section{Conclusion and discussion}
\label{Conclusion}

This paper describes an original zero-inflated Poisson latent position cluster model which leverages several recent ideas from the literature on computational statistics to analyze non-negative weighted networks with unusual zeros or missing data.
Our methodology combines zero-inflated Poisson distributions, clustering (via mixture of finite mixtures), and optional nodes attributes to be used within the clustering structure, when available.
As regards the inferential procedure, our novel approach relies on a partially collapsed sampler which features a new truncated absorb-eject move, and leads to an automatic and computationally efficient selection of the optimal number of groups.
One fundamental output of our proposed procedure is that it provides new model-based visualizations of the complex network data using a 3-dimensional latent space.

The results that we obtain using this methodology include the latent space visualizations, the clustering of the nodes, and the detection of unusual zeros, that is, missing or non-reported data.
As we demonstrate via various examples on simulated datasets, the model has great flexibility and can generalize well to a variety of data patterns. In addition, the inferential procedure is scalable and is able to discover the accurate estimates for the model parameters and model structure.
Applications on various real networks show that we are able to uncover multiple complex and interesting architectures for the social networks, which provide new perspectives on the analyses of these datasets.

Future work will focus on computational scalability, since we are interested in estimating networks with much larger sizes that are common in real world.
Moreover, instead of our proposed partially collapsed MCMC approach, the variational Bayesian methods \citep[for example, ][]{salter2013variational,gollini2016joint,marchello2024deep} are also viable approaches for inference tasks, and it will be interesting to compare the performance of these two widely-used classes of inference methods.
As motivated by \citet{legramanti2022extended}, instead of a mixture of finite mixtures clustering prior, one could compare different forms of Gibbs-type priors \citep{gnedin2006exchangeable} to complement our framework.
If exogenous node attributes are not categorical, one may consider a more generalized structure of the clustering prior, as suggested by \citet{shen2024bayesian}, which can accommodate continuous or categorical or mixed type attributes.
Another viable modification is to replace Eq.~\eqref{LPM}, which is one of the simplest forms to link between the $\mathbbm{E}(y_{ij})$ and the corresponding latent positions, with a more sophisticated choice, for example, considering a different metric between pair-wise nodes on a different latent geometric space \citep{smith2019geometry}. Similarly, count data distributions beyond the Poisson may also be considered.
Finally, following \citet{sewell2015latent}, a natural and ambitious extension for this network model would be to consider a dynamic setting, for the interactions, the clustering, and the missing data specifications.

\section*{Funding Statement}
The Insight Research Ireland Centre for Data Analytics is supported by Science Foundation Ireland under Grant Number 12/RC/2289$\_$P2.

\section*{Acknowledgements}
The authors would like to gratefully thank the anonymous reviewers for the constructive comments to help improve this article.

\section*{Competing Interests}
None.

\section*{Code and Data}
The application code and data for the simulation studies and the real data applications shown in Sections 4 and 5 is available at \url{https://github.com/Chaoyi-Lu/ZIP-LPCM}.

\bibliographystyle{apalike}
\bibliography{ZIP_LPCM_MFM_Final.bib}

@article{rastelli2018choosing,
  title={Choosing the number of groups in a latent stochastic blockmodel for dynamic networks},
  author={Rastelli, Riccardo and Latouche, Pierre and Friel, Nial},
  journal={Network Science},
  volume={6},
  number={4},
  pages={469--493},
  year={2018},
  publisher={Cambridge University Press}
}

@article{lambert1992zero,
  title={Zero-inflated Poisson regression, with an application to defects in manufacturing},
  author={Lambert, Diane},
  journal={Technometrics},
  volume={34},
  number={1},
  pages={1--14},
  year={1992},
  publisher={Taylor \& Francis}
}

@article{tanner1987calculation,
  title={The calculation of posterior distributions by data augmentation},
  author={Tanner, Martin A and Wong, Wing Hung},
  journal={Journal of the American Statistical Association},
  volume={82},
  number={398},
  pages={528--540},
  year={1987},
  publisher={Taylor \& Francis}
}

@article{ghosh2006bayesian,
  title={Bayesian analysis of zero-inflated regression models},
  author={Ghosh, Sujit K and Mukhopadhyay, Pabak and Lu, Jye-Chyi JC},
  journal={Journal of Statistical Planning and Inference},
  volume={136},
  number={4},
  pages={1360--1375},
  year={2006},
  publisher={Elsevier}
}

@article{sewell2016latent,
  title={Latent space models for dynamic networks with weighted edges},
  author={Sewell, Daniel K and Chen, Yuguo},
  journal={Social Networks},
  volume={44},
  pages={105--116},
  year={2016},
  publisher={Elsevier}
}

@article{hoff2002latent,
  title={Latent space approaches to social network analysis},
  author={Hoff, Peter D and Raftery, Adrian E and Handcock, Mark S},
  journal={Journal of the American Statistical Association},
  volume={97},
  number={460},
  pages={1090--1098},
  year={2002},
  publisher={Taylor \& Francis}
}

@article{handcock2007model,
  title={Model-based clustering for social networks},
  author={Handcock, Mark S and Raftery, Adrian E and Tantrum, Jeremy M},
  journal={Journal of the Royal Statistical Society Series A: Statistics in Society},
  volume={170},
  number={2},
  pages={301--354},
  year={2007},
  publisher={Oxford University Press}
}

@article{geng2019probabilistic,
  title={Probabilistic community detection with unknown number of communities},
  author={Geng, Junxian and Bhattacharya, Anirban and Pati, Debdeep},
  journal={Journal of the American Statistical Association},
  volume={114},
  number={526},
  pages={893--905},
  year={2019},
  publisher={Taylor \& Francis}
}

@article{miller2018mixture,
  title={Mixture models with a prior on the number of components},
  author={Miller, Jeffrey W and Harrison, Matthew T},
  journal={Journal of the American Statistical Association},
  volume={113},
  number={521},
  pages={340--356},
  year={2018},
  publisher={Taylor \& Francis}
}

@article{rastelli2018optimal,
  title={Optimal Bayesian estimators for latent variable cluster models},
  author={Rastelli, Riccardo and Friel, Nial},
  journal={Statistics and Computing},
  volume={28},
  pages={1169--1186},
  year={2018},
  publisher={Springer}
}

@article{legramanti2022extended,
  title={Extended stochastic block models with application to criminal networks},
  author={Legramanti, Sirio and Rigon, Tommaso and Durante, Daniele and Dunson, David B},
  journal={The Annals of Applied Statistics},
  volume={16},
  number={4},
  pages={2369},
  year={2022},
  publisher={NIH Public Access}
}

@article{muller2011product,
  title={A product partition model with regression on covariates},
  author={M{\"u}ller, Peter and Quintana, Fernando and Rosner, Gary L},
  journal={Journal of Computational and Graphical Statistics},
  volume={20},
  number={1},
  pages={260--278},
  year={2011},
  publisher={Taylor \& Francis}
}

@article{ryan2017bayesian,
  title={Bayesian model selection for the latent position cluster model for social networks},
  author={Ryan, Caitriona and Wyse, Jason and Friel, Nial},
  journal={Network Science},
  volume={5},
  number={1},
  pages={70--91},
  year={2017},
  publisher={Cambridge University Press}
}

@article{van2008partially,
  title={Partially collapsed Gibbs samplers: Theory and methods},
  author={Van Dyk, David A and Park, Taeyoung},
  journal={Journal of the American Statistical Association},
  volume={103},
  number={482},
  pages={790--796},
  year={2008},
  publisher={Taylor \& Francis}
}

@article{park2009partially,
  title={Partially collapsed Gibbs samplers: Illustrations and applications},
  author={Park, Taeyoung and Van Dyk, David A},
  journal={Journal of Computational and Graphical Statistics},
  volume={18},
  number={2},
  pages={283--305},
  year={2009},
  publisher={Taylor \& Francis}
}

@article{nobile2007bayesian,
  title={Bayesian finite mixtures with an unknown number of components: The allocation sampler},
  author={Nobile, Agostino and Fearnside, Alastair T},
  journal={Statistics and Computing},
  volume={17},
  pages={147--162},
  year={2007},
  publisher={Springer}
}

@article{gnedin2009characterizations,
  title={Characterizations of exchangeable partitions and random discrete distributions by deletion properties},
  author={Gnedin, Alexander and Haulk, Chris and Pitman, Jim},
  journal={arXiv preprint arXiv:0909.3642},
  year={2009}
}

@book{pitman2006combinatorial,
  title={Combinatorial stochastic processes: Ecole d'et{\'e} de probabilit{\'e}s de saint-flour xxxii-2002},
  author={Pitman, Jim},
  year={2006},
  publisher={Springer}
}

@article{nobile2005bayesian,
  title={Bayesian finite mixtures: A note on prior specification and posterior computation (Technical report)},
  author={Nobile, A},
  journal={University of Glasgow},
  year={2005}
}

@article{meilua2007comparing,
  title={Comparing clusterings—an information based distance},
  author={Meil{\u{a}}, Marina},
  journal={Journal of Multivariate Analysis},
  volume={98},
  number={5},
  pages={873--895},
  year={2007},
  publisher={Elsevier}
}

@article{wade2018bayesian,
  title={Bayesian cluster analysis: Point estimation and credible balls (with discussion)},
  author={Wade, Sara and Ghahramani, Zoubin},
  year={2018}
}

@book{borg2005modern,
  title={Modern multidimensional scaling: Theory and applications},
  author={Borg, Ingwer and Groenen, Patrick JF},
  year={2005},
  publisher={Springer Science \& Business Media}
}

@article{gower1966some,
  title={Some distance properties of latent root and vector methods used in multivariate analysis},
  author={Gower, John C},
  journal={Biometrika},
  volume={53},
  number={3-4},
  pages={325--338},
  year={1966},
  publisher={Oxford University Press}
}

@book{sampson1968novitiate,
  title={A novitiate in a period of change: An experimental and case study of social relationships},
  author={Sampson, Samuel Franklin},
  year={1968},
  publisher={Cornell University}
}

@book{de2018exploratory,
  title={Exploratory social network analysis with Pajek: Revised and expanded edition for updated software},
  author={De Nooy, Wouter and Mrvar, Andrej and Batagelj, Vladimir},
  volume={46},
  year={2018},
  publisher={Cambridge University Press}
}

@article{sewell2015latent,
  title={Latent space models for dynamic networks},
  author={Sewell, Daniel K and Chen, Yuguo},
  journal={Journal of the American Statistical Association},
  volume={110},
  number={512},
  pages={1646--1657},
  year={2015},
  publisher={Taylor \& Francis}
}

@article{salter2012review,
  title={Review of statistical network analysis: models, algorithms, and software},
  author={Salter-Townshend, M and White, A and Gollini, I and Murphy, TB},
  journal={Statistical Analysis and Data Mining},
  volume={5},
  number={4},
  pages={243--264},
  year={2012},
  publisher={John Wiley \& Sons, Inc. New York, NY, USA}
}

@article{rastelli2016properties,
  title={Properties of latent variable network models},
  author={Rastelli, Riccardo and Friel, Nial and Raftery, Adrian E},
  journal={Network Science},
  volume={4},
  number={4},
  pages={407--432},
  year={2016},
  publisher={Cambridge University Press}
}

@article{gnedin2006exchangeable,
  title={Exchangeable Gibbs partitions and Stirling triangles},
  author={Gnedin, Alexander and Pitman, Jim},
  journal={Journal of Mathematical Sciences},
  volume={138},
  pages={5674--5685},
  year={2006},
  publisher={Springer}
}

@article{hall2000zero,
  title={Zero-inflated Poisson and binomial regression with random effects: a case study},
  author={Hall, Daniel B},
  journal={Biometrics},
  volume={56},
  number={4},
  pages={1030--1039},
  year={2000},
  publisher={Wiley Online Library}
}

@article{ridout2001score,
  title={A score test for testing a zero-inflated Poisson regression model against zero-inflated negative binomial alternatives},
  author={Ridout, Martin and Hinde, John and Dem{\'e}trio, Clarice GB},
  journal={Biometrics},
  volume={57},
  number={1},
  pages={219--223},
  year={2001},
  publisher={Wiley Online Library}
}

@article{lemonte2019zero,
  title={Zero-inflated Bell regression models for count data},
  author={Lemonte, Artur J and Moreno-Arenas, Germ{\'a}n and Castellares, Fredy},
  journal={Journal of Applied Statistics},
  year={2019},
  publisher={Taylor \& Francis}
}

@phdthesis{ma2024statistical,
  title={Statistical latent space models for international classification of diseases (ICD) codes},
  author={Ma, Cheng},
  year={2024}
}

@inproceedings{hoff2003random,
  title={Random effects models for network data},
  author={Hoff, Peter D},
  booktitle={Dynamic Social Network Modeling and Analysis: Workshop Summary and Papers},
  year={2003},
  organization={Citeseer}
}

@article{ma2020universal,
  title={Universal latent space model fitting for large networks with edge covariates},
  author={Ma, Zhuang and Ma, Zongming and Yuan, Hongsong},
  journal={Journal of Machine Learning Research},
  volume={21},
  number={4},
  pages={1--67},
  year={2020}
}

@article{salter2013variational,
  title={Variational Bayesian inference for the latent position cluster model for network data},
  author={Salter-Townshend, Michael and Murphy, Thomas Brendan},
  journal={Computational Statistics \& Data Analysis},
  volume={57},
  number={1},
  pages={661--671},
  year={2013},
  publisher={Elsevier}
}

@article{mcdaid2012clustering,
  title={Clustering in networks with the collapsed stochastic block model},
  author={McDaid, Aaron F and Murphy, Thomas Brendan and Friel, Nial and Hurley, Neil J},
  journal={arXiv preprint arXiv:1203.3083},
  year={2012}
}

@article{wyse2012block,
  title={Block clustering with collapsed latent block models},
  author={Wyse, Jason and Friel, Nial},
  journal={Statistics and Computing},
  volume={22},
  pages={415--428},
  year={2012},
  publisher={Springer}
}

@inbook{LPNM_2023,
author = {Kaur, H. and Rastelli, R. and Friel, N.},
title = {Latent Position Network Models},
booktitle = {The Sage Handbook of Social Network Analysis},
publisher = {SAGE Publications},
year = {2023},
chapter = {36},
pages = {526-541},
isbn = {9781529614671}
}

@article{freeman1998exploring,
  title={Exploring social structure using dynamic three-dimensional color images},
  author={Freeman, Linton C and Webster, Cynthia M and Kirke, Deirdre M},
  journal={Social Networks},
  volume={20},
  number={2},
  pages={109--118},
  year={1998},
  publisher={Elsevier}
}

@article{freeman1988human,
  title={On human social intelligence},
  author={Freeman, Linton C and Freeman, Sue C and Michaelson, Alaina G},
  journal={Journal of Social and Biological Structures},
  volume={11},
  number={4},
  pages={415--425},
  year={1988},
  publisher={Elsevier}
}

@inproceedings{konect,
	title = {{KONECT} -- {The} {Koblenz} {Network} {Collection}},
	author = {Jérôme Kunegis},
	year = {2013},
	booktitle = {Proc. Int. Conf. on World Wide Web Companion},
	pages = {1343--1350},
	url = {http://dl.acm.org/citation.cfm?id=2488173},
	url_presentation = {https://www.slideshare.net/kunegis/presentationwow},
	url_web = {http://konect.cc/},
	url_citations = {https://scholar.google.com/scholar?cites=7174338004474749050},
}

@article{hayes2006connecting,
  title={Connecting the dots},
  author={Hayes, Brian},
  journal={American Scientist},
  volume={94},
  number={5},
  pages={400--404},
  year={2006}
}

@article{lu2025zero,
  title={Zero-inflated stochastic block modeling of efficiency-security trade-offs in weighted criminal networks},
  author={Lu, Chaoyi and Durante, Daniele and Friel, Nial},
    journal = {Journal of the Royal Statistical Society Series A: Statistics in Society},
    pages = {qnaf029},
    year = {2025},
    issn = {0964-1998},
    doi = {10.1093/jrsssa/qnaf029},
}

@article{gollini2016joint,
  title={Joint modeling of multiple network views},
  author={Gollini, Isabella and Murphy, Thomas Brendan},
  journal={Journal of Computational and Graphical Statistics},
  volume={25},
  number={1},
  pages={246--265},
  year={2016},
  publisher={Taylor \& Francis}
}

@article{marchello2024deep,
  title={A Deep Dynamic Latent Block Model for Co-clustering of Zero-Inflated Data Matrices},
  author={Marchello, Giulia and Corneli, Marco and Bouveyron, Charles},
  journal={Journal of Computational and Graphical Statistics},
  volume={33},
  number={4},
  pages={1224--1239},
  year={2024},
  publisher={Taylor \& Francis}
}

@article{govaert2003clustering,
  title={Clustering with block mixture models},
  author={Govaert, G{\'e}rard and Nadif, Mohamed},
  journal={Pattern Recognition},
  volume={36},
  number={2},
  pages={463--473},
  year={2003},
  publisher={Elsevier}
}

@article{smith2019geometry,
  title={The geometry of continuous latent space models for network data},
  author={Smith, Anna L and Asta, Dena M and Calder, Catherine A},
  journal={Statistical Science: A Review Journal of the Institute of Mathematical Statistics},
  volume={34},
  number={3},
  pages={428},
  year={2019}
}

@article{shen2024bayesian,
  title={Bayesian community detection for networks with covariates},
  author={Shen, Luyi and Amini, Arash and Josephs, Nathaniel and Lin, Lizhen},
  journal={Bayesian Analysis},
  volume={1},
  number={1},
  pages={1--28},
  year={2024},
  publisher={International Society for Bayesian Analysis}
}

@article{krivitsky2009representing,
  title={Representing degree distributions, clustering, and homophily in social networks with latent cluster random effects models},
  author={Krivitsky, Pavel N and Handcock, Mark S and Raftery, Adrian E and Hoff, Peter D},
  journal={Social networks},
  volume={31},
  number={3},
  pages={204--213},
  year={2009},
  publisher={Elsevier}
}

\end{document}